\documentclass{aa}  
\usepackage{txfonts}
\usepackage{orcidlink}
\usepackage{amsmath}
\usepackage{makecell}
\usepackage{graphicx}
\usepackage{longtable}
\usepackage{wasysym}
\usepackage{booktabs}
\usepackage{multicol}
\usepackage{multirow}
\usepackage{array}
\usepackage{color}
\usepackage[dvipsnames]{xcolor}
\usepackage{ulem}
\usepackage{pdflscape}
\usepackage{caption}
\usepackage{placeins}
\usepackage{dblfloatfix}
\usepackage{tabularx}
\usepackage{threeparttable}

\usepackage{array,makecell} % array for >{...}p{..}, makecell for wrapped headers
\newcolumntype{L}[1]{>{\raggedright\arraybackslash}p{#1}}   % left, wrapping
\newcolumntype{C}[1]{>{\centering\arraybackslash}p{#1}}     % centered, wrapping
\begin{document}

   \title{An H$\alpha$ survey of infrared bow-shocks around OB-type stars}

   \author{Kuljeet Singh Saddal%\orcidlink{0009-0006-6176-1796}
   \inst{1,2} \corrauth{kuljeet.saddal@gmail.com}
          \and
          Michaela Kraus %\orcidlink{0000-0002-4502-6330}
          \inst{1} \email{michaela.kraus@asu.cas.cz}
          \and
          Dieter H. Nickeler%\orcidlink{0000-0001-5165-6331}
          \inst{1} \email{dieter.nickeler@asu.cas.cz}
          \and
          Tiina Liimets%\orcidlink{0000-0003-2196-9091}
          \inst{3} \email{sinopesky@gmail.com}
          \and
          Sergey Karpov%\orcidlink{0000-0003-0035-651X}
          \inst{4} \email{karpov.sv@gmail.com}
          \and \\
          Olga V. Maryeva%\orcidlink{0000-0003-1442-4755}
          \inst{1} \email{olga.maryeva@gmail.com}
          }

   \institute{Astronomical Institute, Czech Academy of Science, Fri\v{c}ova 298, 251 65 Ond\v{r}ejov, Czech Republic
%              \email{, michaela.kraus@asu.cas.cz}
         \and
             Astronomical Institute, Faculty of Mathematics and Physics, Charles University, V Hole\v{s}ovick\'{a}ch 2, 182 00 Prague, Czech Republic
        \and
             Tartu Observatory, University of Tartu, Observatooriumi 1, 61602 T\~oravere, Estonia 
        \and   
             Institute of Physics of the  Czech Academy of Sciences, CZ-182 21 Prague 8, Czech Republic\\
             }

   \date{Received xxx; accepted xxx}

% \abstract{}{}{}{}{} 
% 5 {} token are mandatory
 
  \abstract
   {Bow shocks around massive stars are commonly identified in mid-infrared surveys, where emission from heated dust traces the interaction between stellar winds and the surrounding interstellar medium (ISM). In contrast, systematic optical investigations of these structures remain scarce with only a few possible detections reported so far.}
   {This survey aims to investigate whether bow shocks identified in mid-infrared observations around OB stars also produce detectable emission in the H$\alpha$ recombination line, confirm prior detections, compare the morphology of bow shocks detected in H$\alpha$ and mid-infrared wavelengths, and assess the role of stellar radiation pressure in contributing to the support of these structures.}
   {We performed a targeted H$\alpha$ imaging survey of 78 infrared bow-shock candidates compiled from several literature catalogues. Observations were obtained with multiple ground-based facilities and complemented with archival mid-infrared images from the \textit{Spitzer Space Telescope} and the \textit{Wide-field Infrared Survey Explorer}. For the detected bow structures,
   geometrical parameters such as spatial extent, width, and stand-off distance were measured, and the bows were classified based on the parameters of their central stars, the measured stand-off distances and adopted plausible ISM conditions.}
   {Clear arc-shaped H$\alpha$ nebulae are detected in 15 objects, with one additional bow shock discovered serendipitously in a wide-field image. A further 35 objects exhibit diffuse H$\alpha$ emission associated with complex background structures, while 28 show no detectable H$\alpha$ emission. The H$\alpha$ and infrared morphologies generally trace the same large-scale structures, although most of the infrared arcs lie slightly closer to the star and appear broader than their optical counterparts. We find that several systems are consistent with radiation-supported bow-shocks, bow waves or dust waves, while others remain compatible with classical wind-supported bow shocks.}
   {This survey demonstrates that a fraction of infrared bow-shock candidates around OB stars also exhibit detectable ionised gas structures in H$\alpha$, and both stellar wind momentum and radiation pressure can contribute to shaping these structures. 
   No correlation is found between the detection of an H$\alpha$ bow shock and the stellar parameters, and the overlap in objects observed in H$\alpha$ and in the radio regime was too small for firm conclusions other than that stars with no H$\alpha$ emission also show no radio emission.
   Future spectroscopic observations will be required to determine the physical conditions and kinematics of the ionised gas and to further constrain the nature of these bow-shock systems.}


   \maketitle
   \nolinenumbers
%
%-------------------------------------------------------------------

\section{Introduction}

Massive stars significantly influence the interstellar medium (ISM) through their winds, radiation fields, and ultimately through their explosion as supernovae. Thereby, they inject substantial amounts of energy and momentum throughout their evolution and
enrich the surrounding gas with chemically processed material. The interaction with the ISM can lead to the formation of characteristic gas and dust structures such as H\,{\sc ii} regions, shells or bow shocks. Bow shocks around massive stars arise when a supersonic relative motion develops between the stellar wind and the surrounding ISM. Supersonic motions can have different origins. In the runaway star scenario, the star itself moves with a high velocity generating wind--ISM interaction fronts aligned with their direction of motion. Such high-velocity stars may have been ejected from multiple systems or young stellar clusters through dynamical interactions, or by the disruption of binary systems, for example during a supernova explosion \citep{poveda1967,blaauw1961,pflamm2010,fujii2011}. On the other hand, non-runaway OB stars with bow shocks exist as well. In this so-called `weather vane' scenario introduced by \citet{2008ApJ...689..242P}, the bow-shock morphology is shaped by large-scale external flows, such as expanding H\,{\sc ii} regions, whose impinging momentum flux determines the shock orientation and structure.

Early analytical descriptions of wind–ISM interactions established the ram-pressure balance framework and the resulting bow-shock geometry \citep{baranov1971,wilkin1996,wilkin2000}. More recently, multidimensional simulations have examined bow-shock evolution under increasingly realistic conditions, including radiative heating and cooling, photoionization, and evolving wind properties \citep{Meyer2014,Meyer2016,mohamed2012,acreman2016,Mackey2016,Green2019}, while magnetohydrodynamical (MHD) studies have further explored the role of magnetized stellar winds and ambient interstellar fields in shaping the shock structure \citep{Meyer2017,Meyer2021,green2022,Mackey2021,2025A&A...696A..91M,2025ApJ...980..239D}. 

An alternative interpretation attributes the observed arcs around massive stars to radiation-pressure–driven bow waves, where stellar photons transfer sufficient momentum to decelerate and deflect the dust grains \citep{van1988}. This mechanism has been investigated in detail in the context of late O-type stars with weak winds embedded in dusty photoevaporation flows within H\,{\sc ii} regions, where radiation pressure can dominate the dust dynamics and support arc-shaped structures distinct from classical wind-supported bow shocks \citep{ochsendorf2014_1,ochsendorf2014_2,ochsendorf2015}. Building on this picture, \citet{henney2019_1,henney2019_2,henney2019_3} established a unified theoretical and observational framework for stellar bow shells around luminous stars. In  
their first paper, \citet{henney2019_1} developed a semi-analytic model in the strong gas–grain coupling limit, balancing the ram pressure of the ambient stream against stellar wind and radiation pressure. This framework yields three regimes defined by optical depth of the shocked shell and dominant internal support: wind-supported bow shocks (WBS), radiation-supported bow waves (RBW), and radiation-supported bow shocks (RBS). Radiation pressure becomes significant in dense environments with high shell optical depth, for weak-wind stars with low wind momentum, and at low relative velocities between the star and the surrounding ISM. \citet{henney2019_2} investigated the breakdown of the strong gas–grain coupling limit and showed that radiation pressure can decouple dust from the gas upstream of the hydrodynamic bow shock. This process introduces a fourth regime, in which a distinct dust wave (DW) forms exterior to the gas shell.  Their study also included the influence of different magnetic field orientations on the structure and existence of these dust waves. The third paper in the series introduced a novel diagnostic method which uses mid-infrared data to distinguish between wind and radiation-supported bow structures and to refine stellar wind mass-loss estimates \citep{henney2019_3}. Applying this method to a sample of observed bows, they identified four strong and three marginal RBW candidates but no strong evidence for decoupled dust waves, and suggested that stellar wind mass-loss rates derived from dust emissivities of the infrared bow shocks, as proposed by \citet{kobulnicky2018}, should be revised downward by approximately a factor of two.

From the observational point of view, the first bow shocks around massive stars were identified based on emission in optical nebular lines \citep{gull1979}, and pronounced structures detected on infrared images from the \textit{Infrared Astronomical Satellite} (IRAS) mission by \citet{van1988} and \citet{van1995}. These early surveys culminated in the catalogue of 21 candidate bow shocks and four bubbles by \citet{noriega1997}. Wide-field mid-infrared surveys subsequently expanded the Galactic census of stellar bow shocks, with the Extensive stellar BOw Shock Survey (E-BOSS) catalogue compiled from \textit{Wide-field Infrared Survey Explorer} (WISE) and 
\textit{Midcourse Space Experiment} (MSX) data by \citet{peri2012,peri2015}, and a systematic visual inspection of WISE and \textit{Spitzer Space Telescope} images yielding over 600 additional candidates \citep{kobulnicky2016,kobulnicky2017}. More recently, the combination of mid-infrared imaging with \textit{Gaia} DR3 astrometry has led to the identification of new runaway massive stars and associated bow-shock structures \citep{carretero2023,carretero2025}. Radio studies have examined infrared-selected bow shocks to identify thermal and non-thermal counterparts and to trace the shocked gas \citep{benaglia2010,benaglia2021,moutzouri2022,moutzouri2025,eijnden2022,eijnden2022_2,eijnden2024}, while X-ray investigations have searched for high-energy emission associated with particle acceleration in these systems \citep{lopez2012,terada2012,toala2016,becker2017,binder2019}. While extensive work has been carried out at infrared, radio, and X-ray wavelengths, optical investigations of massive-star bow shocks remain sparse, with the only dedicated H$\alpha$ survey conducted by \citet{brown2005} and detailed spectroscopy largely restricted to isolated cases such as the red supergiant IRC–10414 \citep{gvaramadze2014}. This lack of systematic H$\alpha$ studies motivated us to perform a targeted survey presented in this work.

The paper is structured as follows. Section~\ref{sec:targetselection} describes the target selection criteria adopted for this survey. Section~\ref{sec:observations} presents the H$\alpha$ observations obtained with various ground-based facilities, together with the archival mid-infrared imaging from \textit{Spitzer} and WISE (bands 2–4) used for comparison. The derived bow-shock parameters and the proposed morphological classification scheme are presented in Section~\ref{sec:results}. In Section~\ref{sec:discussion}, the results are discussed in the context of previous H$\alpha$, infrared, and radio surveys, and Section~\ref{sec:conclusions} summarises our conclusions and outlines prospects for future work.

\section{Target selection}
\label{sec:targetselection}

Bow shocks around runaway stars are efficiently identified in the infrared because dust swept into the shocked layer is heated by stellar radiation and gas compression, producing bright emission at $12-24$\,$\mu$m. We therefore prioritised targets with established mid-infrared arcs and assembled our sample from four literature sources.

The first infrared catalogue of bow-shock nebulae around early-type stars was compiled by \citet{van1988}, who conducted a visual survey of arcuate or bubble-like structures using the $60$\,$\mu$m all-sky survey of IRAS. Their search identified 15 objects with arc-like (or ring-like) morphologies, listed in their Table~1. From this sample we selected the 13 objects classified as OB or Wolf--Rayet stars, excluding the fourth and fifth entries whose spectral types are unknown. We assign the designations BS01--BS15 according to the order in which the objects appear in their table, omitting BS04 and BS05. One additional target was not included in our observational sample: BS03: HD\,192163, a Wolf--Rayet star for which the H$\alpha$ morphology in the Isaac Newton Telescope Photometric H$\alpha$ Survey \citep[IPHAS;][]{2005MNRAS.362..753D} indicates that the emission structure is not associated with a bow-shock nebula. 
In total, 12 targets from this catalogue were observed.

In a subsequent study, \citet{van1995} extended their earlier work by compiling a catalogue of 188 OB-type runaway stars compiled from the surveys of \citet{garmany1982}, \citet{cruz1974}, and \citet{stone1979}. Using IRAS data, they constructed 60\,$\mu$m excess maps using the 100\,$\mu$m images as reference, and identified 58 candidate bow shocks, of which 25 exhibit well-resolved arc-like structures. From this sample, \citet{brown2005} searched for objects with bow shock structures in H$\alpha$ targeting 37 O-type candidates using data from the two H$\alpha$ allsky surveys \textit{Southern H-Alpha Sky Survey Atlas} \citep[SHASSA,][]{2001PASP..113.1326G} and \textit{Virginia Tech Spectral-Line Survey} \citep[VTSS,][]{dennison1998}. Their methodology relied on the analysis of radial surface-brightness profiles as a more robust alternative to visual inspection, and on  
the geometry of the arc-like emission to constrain ambient ISM

\begin{landscape}
{\tiny
\begin{table}[t]
\caption{Target Selection Table. Spectral types are adopted from the SIMBAD database unless otherwise indicated. Proper motions are taken from \textit{Gaia} DR3, and distances are derived from \textit{Gaia} DR3 parallaxes, except where alternative sources are specified.}
\label{tab:TargetSelection}
\centering
\footnotesize
\begin{tabular}{lllcccc|llcccc}
\hline \hline
ID & Alt. ID & Star & Sp. T. & $\mu_{RA}$ & $\mu_{DEC}$  & d & ID & Star & Sp. T. & $\mu_{RA}$ & $\mu_{DEC}$  & d \\
  &  &  &  & (mas yr$^{-1}$) & (mas yr$^{-1}$)  & (kpc) &  &  & & (mas yr$^{-1}$) & (mas yr$^{-1}$) & (kpc) \\
 \hline
EB01 & K366 & HD\,2083  & B0.2V  & $ $7.90 & $ $0.17  & 0.40 $\pm$ 0.01 & K007 & G\,003.8417\,-01.0440 & B2-3V$^{1}$ & $ $1.93 & $-$1.87 & 1.22 $\pm$ 0.02 \\
EB02 & BS08 & HD\,2905  & BC0.7I & $ $3.13 & $-$1.83  & 0.94 $\pm$ 0.14 & K014 & G\,006.2977\,-00.2012 & - & $-$0.68 & $-$1.94  & 3.03 $\pm$ 0.73 \\
EB03 &      & HD\,15629 & O4.5V  & $-$0.67 & $-$0.66 & 2.06 $\pm$ 0.08 & K048 & HD\,166996 & B2I & $ $0.11 & $-$1.76  & 1.83 $\pm$ 0.14 \\
EB04 & K373 & HD\,21856 & B1V    & $-$8.89 & $ $1.67  & 0.42 $\pm$ 0.01 & K051 & NGC\,6618\,258 & O9.5V & $ $0.34 & $-$1.60 & 1.42 $\pm$ 0.04 \\
EB05 & BS10 & HD\,22928 & B5III  & $ $29.04 & $-$23.36  & 0.16 $\pm$ 0.01 & K052 & NGC 6618 326 & O7:V & $ $0.96 & $ $0.34  &  2.28 $\pm$ 1.30 \\
EB06 & BS09; K370 & HD\,30614 & O9I    & $ $0.22 & $ $7.19  & 1.69 $\pm$ 0.38 & K053 & G\,015.1032\,-00.6489 & B2V & $ $0.12 & $-$1.97  & 1.87 $\pm$ 0.24 \\
EB07 & K375 & HD\,34078 & O9.5V  & $-$4.75 & $ $43.54  & 0.39 $\pm$ 0.01 & K063 & BD\,-13\,4934 & B1V & $ $0.37 & $-$1.72 & 1.76 $\pm$ 0.06 \\
EB08 & K378 & HD\,36512 & O9.7V  & $-$0.70 & $-$4.88  & 0.41 $\pm$ 0.02 & K065 & BD\,-13\,4937 & B1.5V & $ $0.15 & $-$1.55  & 1.81 $\pm$ 0.06 \\
EB09 &      & HD\,37032 & B0.5V  & $ $0.25 & $-$2.46 & 2.35 $\pm$ 0.07 & K067 & NGC\,6611\,584 & O9V & $ $0.19 & $-$1.48  & 1.86 $\pm$ 0.06 \\
EB10 & K374 & HD\,41161 & O8V    & $-$2.51 & $-$1.28 & 1.49 $\pm$ 0.01 & K072 & G\,018.2660\,-00.2988 & O6V-O5V & $ $0.07 & $-$1.47 & 4.14 $\pm$ 0.30 \\
EB11 & BS14 & HD\,42933 & B1/2III & $-$3.75 & $ $8.09  & 0.46 $\pm$ 0.03 & K101 & G\,023.1100\,+00.5458 & - & $ $0.24 & $-$2.29  & 2.27 $\pm$ 0.12 \\
EB12 &      & HD\,47432 & O9.7I & $-$0.06  & $-$1.43  & 1.69 $\pm$ 0.01 & K148 & G\,030.3745\,+00.0252 & -  & - & - & - \\
EB13 & BB3; K376 & HD\,48099 & O5V $+$ O9V & $ $1.04 & $ $2.75  & 2.75 $\pm$ 0.04 & K150 &TYC\,5118-279-1 & B1V$^{1}$ & 0.61 & $-$5.80 & 0.61 $\pm$ 0.01 \\
EB14 & K381 & HD\,54662 & O6.5V $+$ 07.5V & $-$2.54 & $ $3.13 & 1.38 $\pm$ 0.01 & K159 & G\,031.6770\,+00.1775 & -  & - & - & - \\
EB15 &      & HD\,64315 & O5.5V $+$ O7V & $-$3.37 & $ $2.43  & 5.00 $\pm$ 0.80$^{3}$  & K262 & G\,049.4683\,-00.2527& - & - & -  & - \\
EB16 &      & HD\,110879 & B2V $+$ B3V & $-$37.43 & $-$10.58  & 0.10 $\pm$ 0.01 & K330 & HD\,229159 & B1I$ ^{1} $ & $-$2.01 & $-$5.80 & 1.71 $\pm$ 0.05 \\
EB17 &      & HD\,130298 & O6.5III & $-$6.49 & $-$1.20 & 2.55 $\pm$ 0.10 & K345 & HD\,199021 & B0V & $-$1.06 & $-$3.67  & 0.80 $\pm$ 0.01 \\
EB18 & K544 & HD\,136003 & B1I & $-$7.73 & $-$9.21  & 1.07 $\pm$ 0.03 & K346 & G\,083.9246\,-00.6778 & B1III$^{1}$ & $-$2.74 & $-$4.05  & 3.68 $\pm$ 0.20 \\
EB19 &      & HD\,329905 & O9I & $-$5.16 & $-$2.97  & 4.66 $\pm$ 0.34 & K347 & BD\,+42\,3914 & Be & $-$1.38$^{2}$  & $-$3.86$^{2}$   & 0.35 $\pm$ 0.12$^{2}$  \\
EB20 & BS15; K679 & HD\,143275 & B0.3IV & $-$10.21$^{2}$ & $-$35.41$^{2}$ & 0.15 $\pm$ 0.02$^{2}$  & K351 & LS\,III\,+55\,26 & B2I & $-$2.64 & $-$2.17  & 9.33 $\pm$ 1.51 \\
EB21 & BB7; BS01 & HD\,149757 & O9.2IV & $ $10.46 & $ $24.74  & 0.13 $\pm$ 0.01 & K359 & HD\,215806 & B0V$ ^{1} $ & $-$2.49 & $-$2.23  & 2.64 $\pm$ 0.09 \\
EB22 & K575 & HD\,150898 & B0I & $-$5.58  & $-$20.96  & 1.14 $\pm$ 0.01 & K379 & HD\,51756 & O9.7IV & $-$2.32$^{2}$ & $ $2.11$^{2}$  & 1.70 $\pm$ 1.00$^{4}$ \\
EB23 & K054 & HD\,165319 & O9.7I & $-$2.41 & $-$1.20  & 1.46 $\pm$ 0.01 & K383 & CD\,-26\,5136 & O6.5I & $-$3.69 & $ $4.52  & 6.27 $\pm$ 0.53 \\
EB24 &      & HD\,175514 & O7V $+$ ... &  $-$0.69 & $ $0.50 & 1.46 $\pm$ 0.05 & K385 & CD\,-41\,4637 & O6I & $-$4.91 & $ $6.92  & 2.35 $\pm$ 0.07 \\
EB25 &      & HD\,188001 & O7.5I & $-$0.24 & $-$10.12 & 1.93 $\pm$ 0.13 & K386 & HD\,77581 & B0.5I & $-$4.82 & $ $9.28  & 2.01 $\pm$ 0.06 \\
EB26 & K343 & HD\,195592 & O9.7I & $-$2.50 & $ $1.06  & 1.78 $\pm$ 0.05 & K388 & HD\,75860 & B1.5I & $-$6.68 & $ $4.79  & 2.29 $\pm$ 0.22 \\
EB27 & K344 & BD\,+43\,3654 & O4I & $-$2.59 & $ $0.73  & 1.72 $\pm$ 0.04 & K389 & HD\,76031 & B1I & $-$6.14 & $ $4.03  & 1.81 $\pm$ 0.08 \\
EB28 &      & BD+63\,1964 & B0I & $-$6.74 & $-$0.72  & 2.51 $\pm$ 0.10 & K391 & HD 77207 & B7I & $-$5.36 & $ $3.68  & 1.81 $\pm$ 0.05 \\
BB1 &    & HD\,17505 & O6.5III $+$ O8V & $-$0.69 & $-$0.81  & 2.41 $\pm$ 0.11 & K399 & G\,284.0765\,-00.4323 & O6V & $-$5.41 & $ $3.06  & 8.03 $\pm$ 3.41 \\
BB2 &     & HD\,24431 & O9III $+$ B1.5V & $-$1.77 & $-$0.05  & 0.96 $\pm$ 0.05 & K404 & G\,286.0498\,-01.65836 & - & $-$5.95 & $ $3.45  & 2.96 $\pm$ 0.10 \\
BB4 & BS13 & HD\,57061 & O9II & $-$2.31$^{2}$ & $ $5.02$^{2}$  & 0.92 $\pm$ 0.47$^{2}$ & K489 & G\,311.1657\,+00.0448 & - & $-$3.47 & $-$1.76  & 3.81 $\pm$ 0.37 \\
BB5 &    & HD\,92206 & O6.5V $+$ O6.5V & $-$7.29 & $ $2.81 & 2.58 $\pm$ 0.14 & K506 & HD 126593 & B1/2I & $-$5.26 & $-$2.91  & 2.55 $\pm$ 0.15 \\
BB6 &    & HD\,135240 & O8V & $-$4.09 & $-$3.76  & 0.68 $\pm$ 0.06 & K509 & G\,315.3719\,+00.6043	 & - & $-$4.99 & $-$1.53  & 4.56 $\pm$ 0.70 \\
BB8 &    & HD\,158186 & O9.5V & $ $1.76 & $-$1.34  & 1.12 $\pm$ 0.07 & K534 & G\,319.7182\,-00.7290 & - & $-$4.19 & $-$2.73  & 2.68 $\pm$ 0.38 \\
BS02 & K154 & HD\,171491 & B3IV & $-$2.69 & $-$5.95  & 0.65 $\pm$ 0.02  & K542 & G\,322.1986\,+00.5183  & - &$-$5.94 & $-$4.66 & 2.93 $\pm$ 0.86 \\
BS06 &    & HD\,203467 & B3IV & $ $5.47 & $ $5.92  & 0.40 $\pm$ 0.01 & K634 & HD\,152756 & B3/5I & $-$0.44 & $-$2.88  & 1.60 $\pm$ 0.04 \\
BS07 &    & HD\,210839 & O6.5I & $-$7.02 & $-$10.81  & 0.85 $\pm$ 0.06 & K681 & G\,350.6032\,+01.0332 & B3I & $-$0.23 & $-$2.12  & 1.88 $\pm$ 0.06 \\
BS11 &    & BD\,+09\,879 & O8III $+$ B0.5V & $ $2.90 & $-$3.18 & 0.39 $\pm$ 0.06 & K692 & G\,353.4162\,+00.4482 & O7.5 & $ $1.49 & $-$2.27  & 1.79 $\pm$ 0.06 \\
BS12 &    & HD\,50896 & WN4 & $-$4.32 & $ $3.10  & 1.53 $\pm$ 0.09 & K705 & CD\,-29\,13925 & B0.5I & $-$1.12 & $-$2.60  & 2.35 $\pm$ 0.10 \\
\hline
\end{tabular}
\tablebib{(1) \cite{patten2025} (2) \cite{hipparcos} (3) \cite{lorenzo2017} (4) \cite{papics2011}.}
\end{table}
}
\end{landscape}
\noindent properties. Eight bow-shock detections were reported in their Table~1. We selected these eight objects for deeper and (where appropriate) higher-resolution follow-up observations and refer to them hereafter with the prefix “BB”.

The E-BOSS catalogue was compiled in two stages, namely the initial release \cite{peri2012} and the extended release \cite{peri2015}. The first catalogue drew from two sources: (i) 56 OB stars previously listed as bow-shock candidates by \cite{noriega1997}, and (ii) 244 O–B2 runaway stars extracted from \cite{tetzlaff2010}, among which 17 objects were common to both groups. The extended catalogue incorporated additional material by including 234 B3–B5 stars also selected from \cite{tetzlaff2010},
42 runaways identified in the Galactic O Star Catalog \cite{maiz2004}, and 10 further runaway stars from \cite{hoogerwerf2001}, together with 45 objects previously reported in the literature or discovered by the authors. In both releases, a bow shock was defined as a comma- or arc-shaped structure located upstream of the stellar motion and brightest in the 12–24\,$\mu$m bands. A total of 28 bow-shock candidates were identified in the first release, and 45 additional candidates in the second, as listed in \cite[Table~8]{peri2015}. In the present work, we select the first 28 candidates listed in that table, retaining the same identifiers (EB01–EB28). Four objects from this subset (EB03, EB07, EB09 and EB28) were not observed during our dedicated observing runs, but available IPHAS images have been used to complete the sample.
The authors noted that their inspection of SHASSA and VTSS H$\alpha$ maps did not yield any convincing evidence of bow-shock morphology.

Finally, the largest catalogue to date of 709 infrared bow shocks was compiled by \citet{kobulnicky2016}, through a systematic examination of the 24\,$\mu$m band of \textit{Multiband Imaging Photometer for Spitzer} (MIPS) and WISE band~4 surveys of the Galactic plane for symmetric mid-infrared arc-like nebulae. Of the 709 identified candidates, 660 are new detections. 
From this catalogue, we selected a total of 39 candidates, with a primary focus on evolved massive stars according to their classification in SIMBAD\footnote{We note that many objects in that catalogue lack a classification in SIMBAD.}. The remaining targets were selected based on a manual inspection of the infrared morphologies, with preference given to objects displaying well-defined and isolated arc-like structures and that were suitable for follow-up H$\alpha$ imaging observations. Additional practical considerations included target visibility from the observing sites and the angular size relative to the available telescope fields of view.
We denote each selected candidate with the prefix “K”, followed by its catalogue index number.

After merging the above lists and removing duplicates, our study comprises of 78 unique targets listed in Table~\ref{tab:TargetSelection}, which were observed in our imaging survey.

%--------------------------------------------------------------------

\section{Observations and data reduction}
\label{sec:observations}

\subsection{H$\alpha$ Imaging}

Imaging observations were carried out over a period of four years beginning in 2020, with four objects observed in early 2026, using a combination of northern and southern facilities with different fields of view to adequately cover the spatial extent of the nebulae identified in the infrared surveys. The campaign made use of the Danish 1.54-m telescope at La Silla (DK154) and the robotic T17 and T31 telescopes of the iTelescope network located at Siding Spring Observatory in Australia for small and large objects on the southern sky, and of the Perek 2-m telescope and the Small Binocular Telescope (SBT) at Ond\v{r}ejov Observatory for small and large objects on the northern sky.

The DK154 is equipped with a 2048 $\times$ 2048 CCD camera that delivers a pixel scale of 0\farcs396 per pixel and covers a square field of view of 13\farcm7 $\times$ 13\farcm7. Observations were obtained through the ESO H$\alpha$ filter No. 693, which has a central wavelength of 656.23\,nm and a bandpass of 6.21\,nm.

The T31 telescope employs a 3056 $\times$ 3056 pixel CCD with a sampling of 1\farcs1 per pixel and a field of view of 55\farcm9 by 55\farcm9. The T17 telescope uses a 4788 $\times$ 3194 CCD that provides a pixel scale of 0\farcs531 per pixel and a field of 42\farcm4 $\times$ 28\farcm3. Both telescopes are equipped with an Astrodon H$\alpha$ narrowband filter with a central wavelength of 656.3\,nm and a full width at half maximum (FWHM) of 5\,nm.

Two CCD systems were used on the Perek 2-m telescope during the time span of the program. The older CCD camera delivered a sampling of 0\farcs404 per pixel and a field measuring 7\farcm31 by 4\farcm94. The newer Perek CCD operates with a 1056 × 1027 detector that provides a pixel scale of 0\farcs385 per pixel and a square field of view of 6\farcm3 $\times$ 6\farcm3. Both instruments were used with a narrow-band H$\alpha$ filter of 3\,nm bandwidth.

The SBT is equipped with a 4096 × 4096 CCD detector. This instrument offers a very wide field of view of 3\fdg5 $\times$ 3\fdg5 and a sampling of 3\farcs14 per pixel, making it well suited for detecting large-scale nebular structures \citep{2019AN....340..633S}. The H$\alpha$ filter has a bandwidth of 3\,nm.

We used a baseline exposure time of $900$\,s with our main survey facility DK154 for typical seeing conditions ($1\arcsec - 2\arcsec$). If no discernable structure was detected, the object was classified as non-detection. Otherwise, the exposure time was increased for better S/N values. For the supplementary facilities, we applied the same strategy. Considering the different sizes of the telescope collecting areas, the different pixel sizes, and the different seeing conditions, we estimated that the $900$\,s exposure with the DK154 translates to $\sim 3600$\,s with the T31 and to $\sim 900$\,s with the SBT (due to the usually very bad seeing conditions under which the telescope was available for our purposes). Therefore, the sensitivity was very similar with these facilities. For each of them, the faintest detection, when scaled to the baseline exposure times, corresponds to a surface brightness value of $\sim 10$\,Rayleigh (see Sect.~\ref{sec:calibration}), which we consider as the limiting value for our survey. Due to its very small FOV, the Perek telescope was not an ideal facility for our survey and was used only occasionally for a few small bow-shock nebulae in the northern sky. The same sensitivity would have been reached with a $\sim 600$\,s exposure. However, we used $3600$\,s as baseline for (non-)detection, which is four times deeper than with the other facilities.

In all cases, the total exposure time for each target was divided into a series of shorter individual integrations. This approach prevents excessive saturation of bright stars in the field and also mitigates guiding instabilities that are commonly encountered during long single exposures. 
The full observational log is presented in Table \ref{Tab:observation_log}, which lists for each target the host star, equatorial coordinates, date of observation, telescope used, total exposure time, number of combined frames, measured seeing, and the $V$ band magnitude of the source.

\begin{figure*}
   \centering
\includegraphics[width=0.49\hsize]{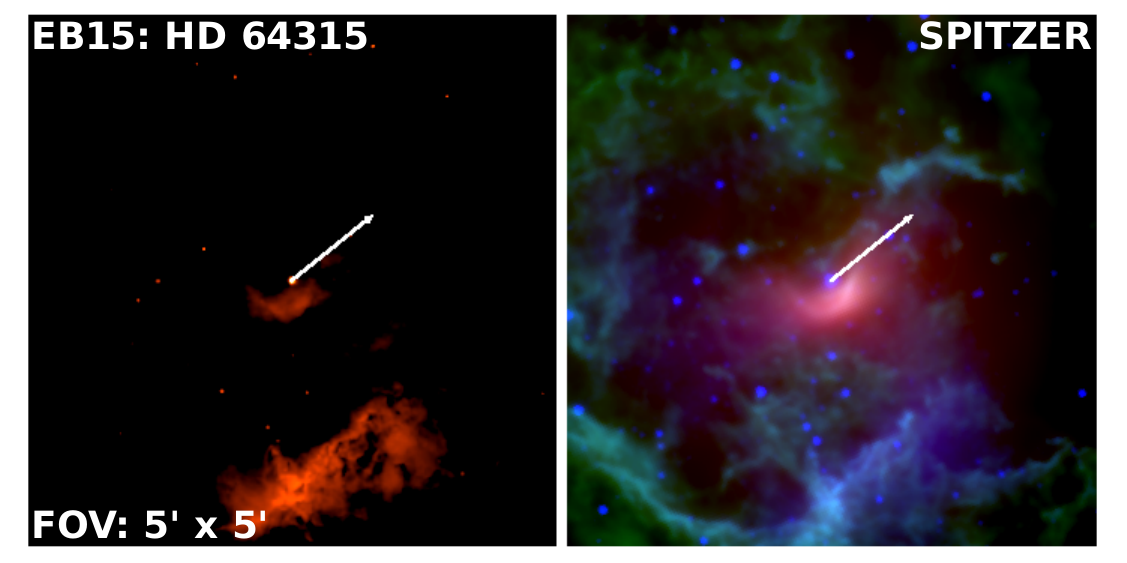} 
\includegraphics[width=0.49\hsize]{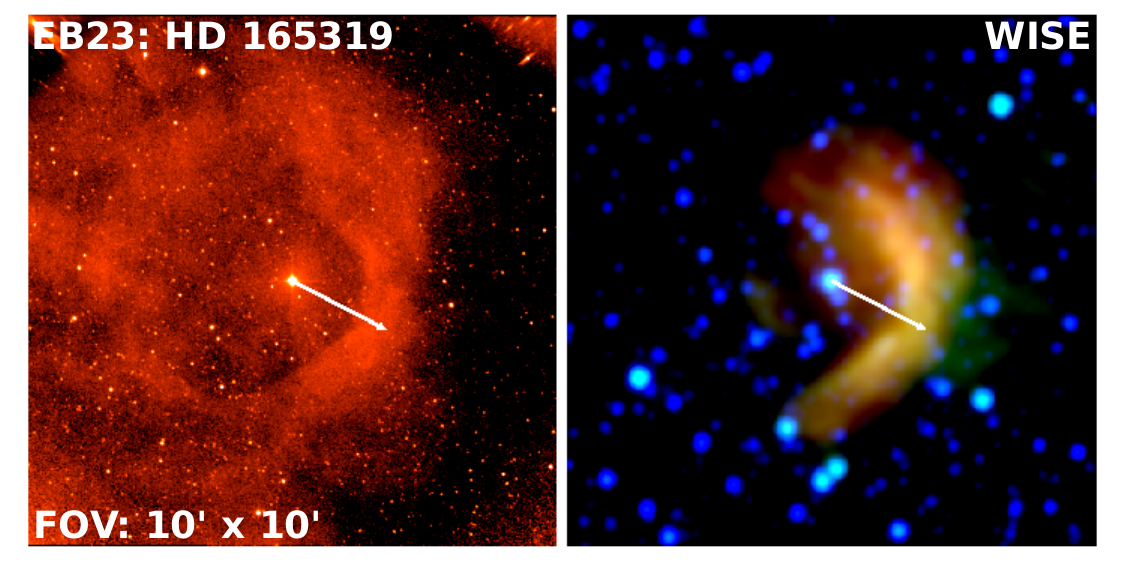}
\includegraphics[width=0.49\hsize]{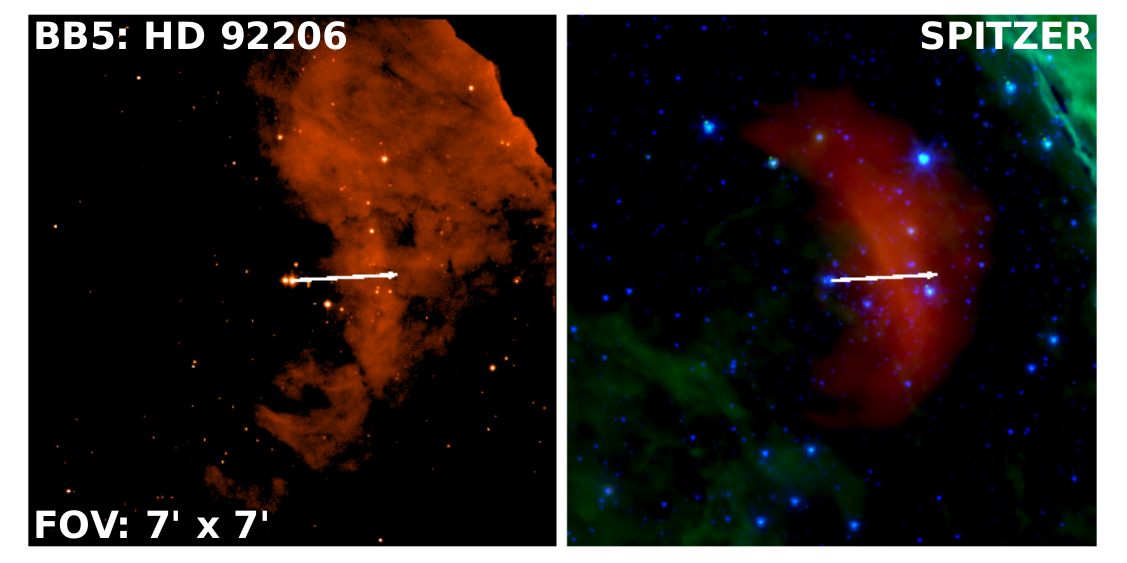} 
\includegraphics[width=0.49\hsize]{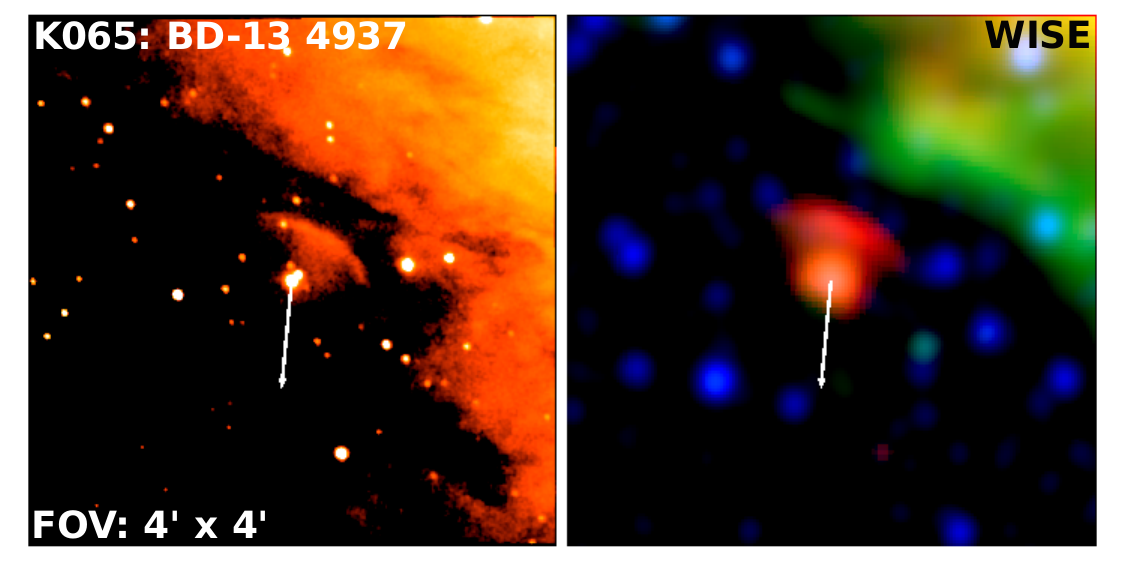} 
\includegraphics[width=0.49\hsize]{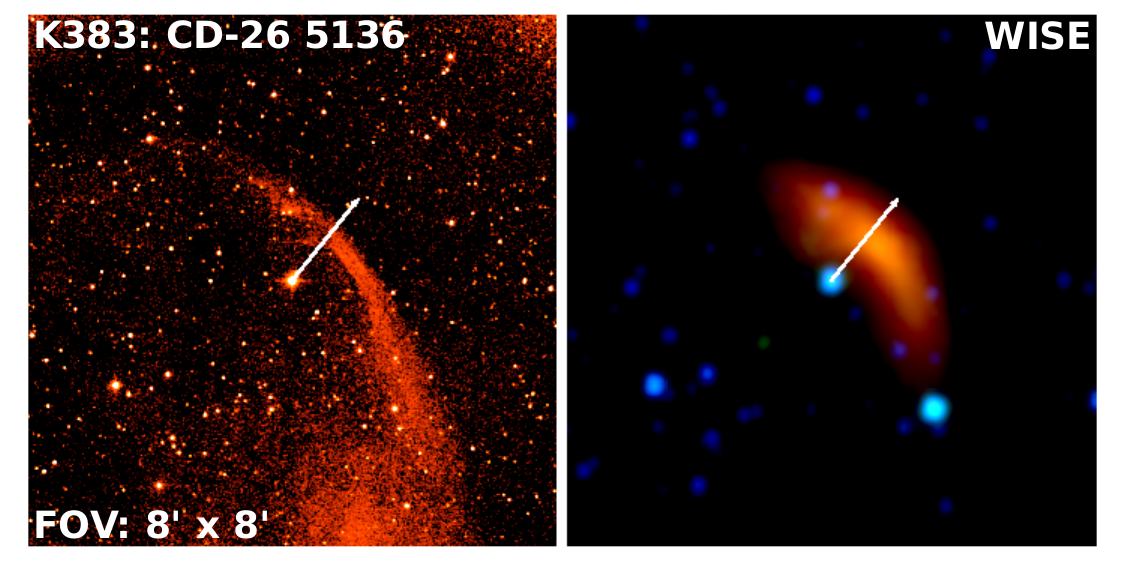}
\includegraphics[width=0.49\linewidth]{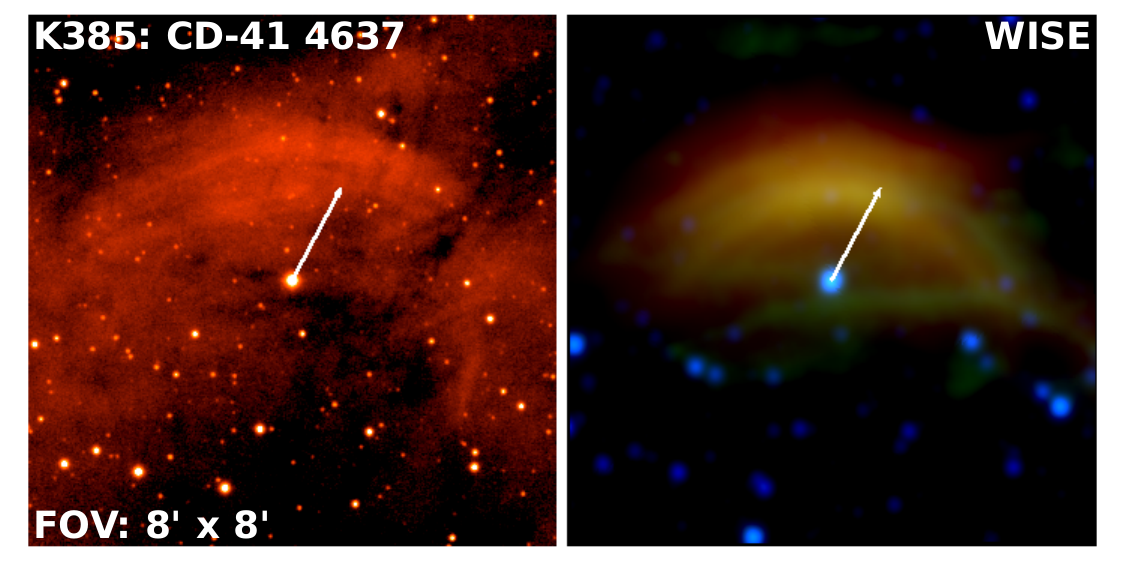} 
\includegraphics[width=0.49\linewidth]{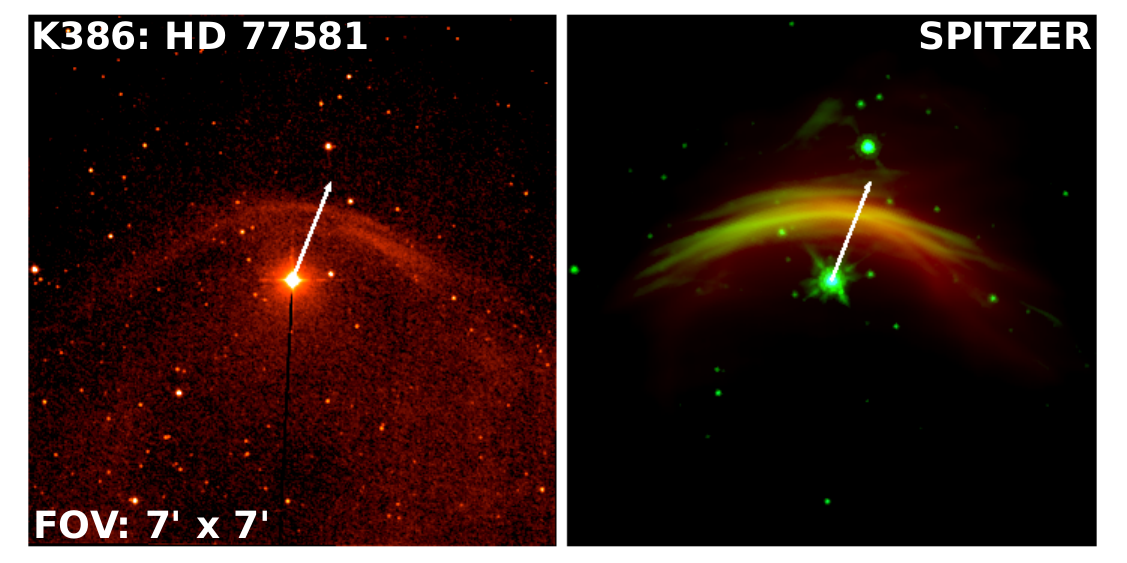} 
\includegraphics[width=0.49\linewidth]{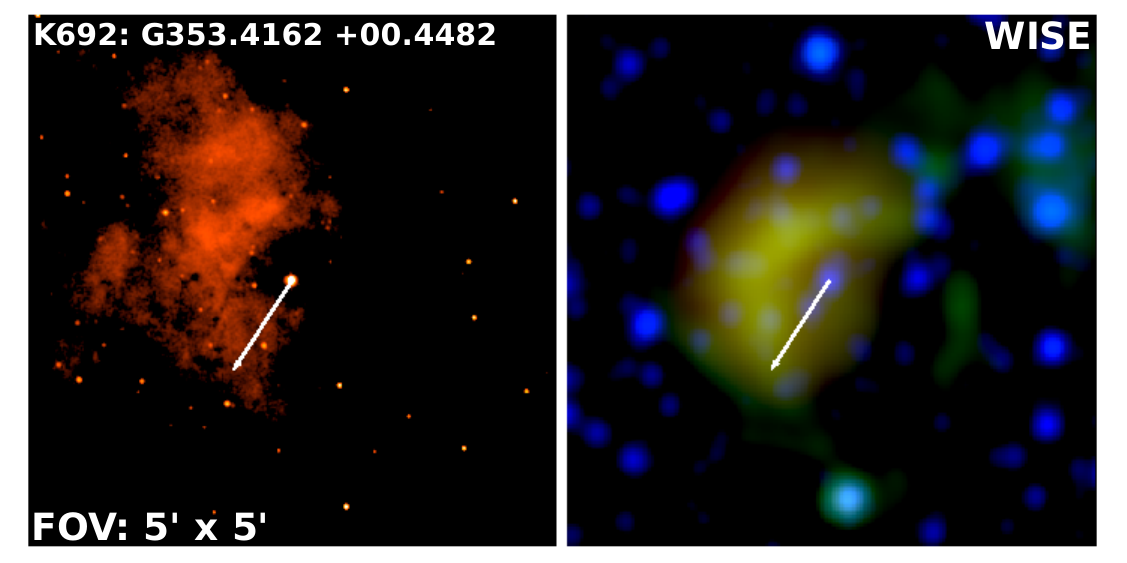}
\includegraphics[width=0.49\hsize]{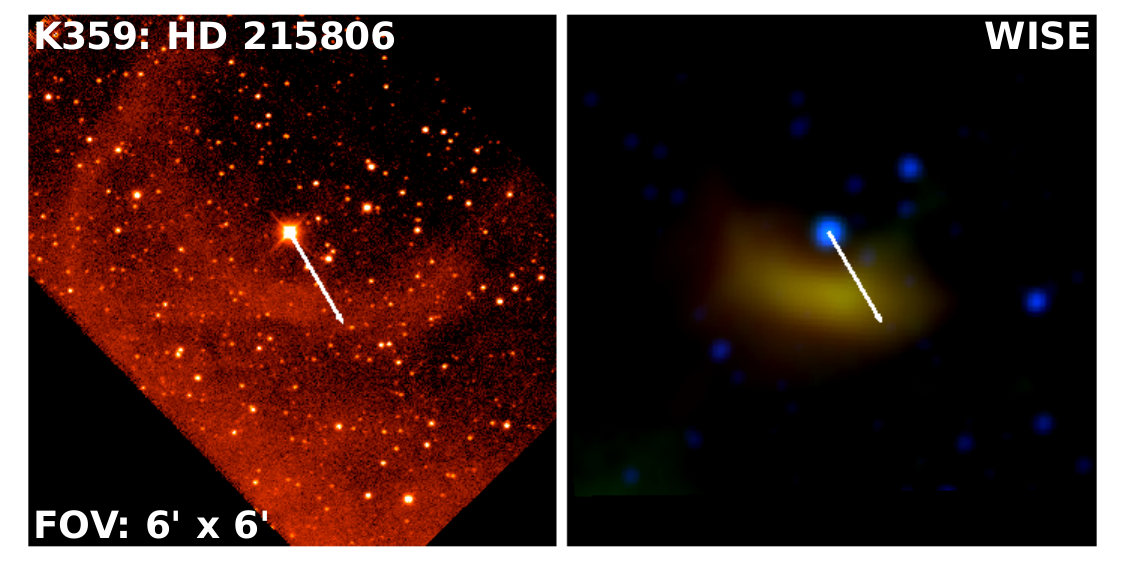}
\includegraphics[width=0.49\hsize]{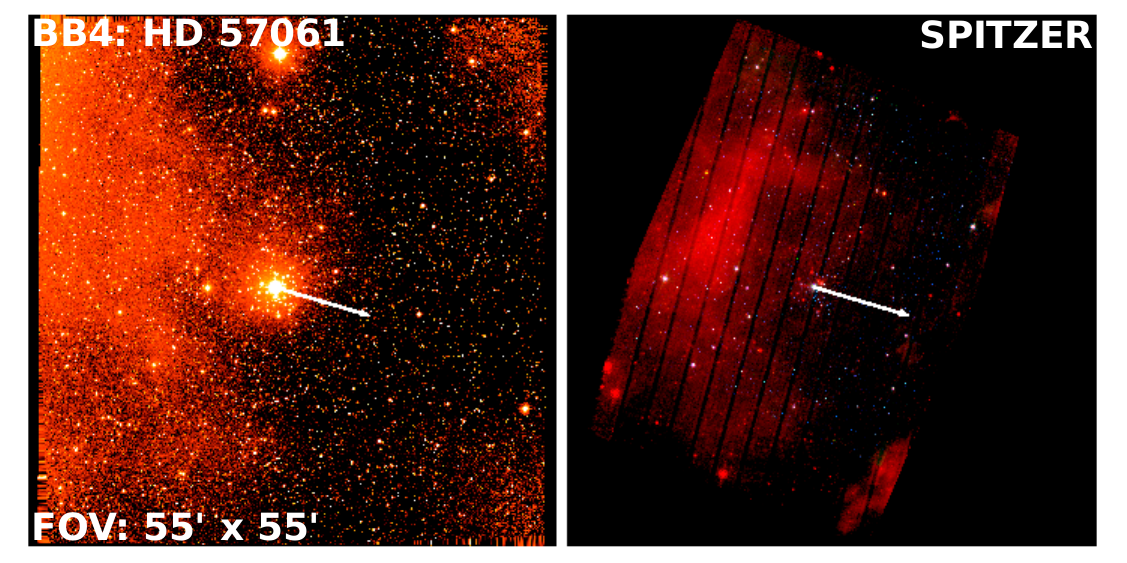}
\caption{Detections of H$\alpha$ bow shocks with DK154, except for K359 (Perek) and BB4 (T31). The proper-motion vector (not to scale) corresponds to 20\% of the image field of view. Left: H$\alpha$ image obtained from our survey. Right: three-colour mid-infrared image of the corresponding nebula, constructed from archival data. Blue: WISE W2 (4.6\,$\mu$m) or IRAC 4.5\,$\mu$m; green: WISE W3 (12\,$\mu$m) or IRAC 8\,$\mu$m; red: WISE W4 (22\,$\mu$m) or MIPS 24\,$\mu$m. Images are in equatorial coordinates with north up and east to the left.}
  \label{Detections1}%
\end{figure*}
\begin{figure*}[ht!]
\ContinuedFloat
   \centering
\includegraphics[width=0.49\hsize]{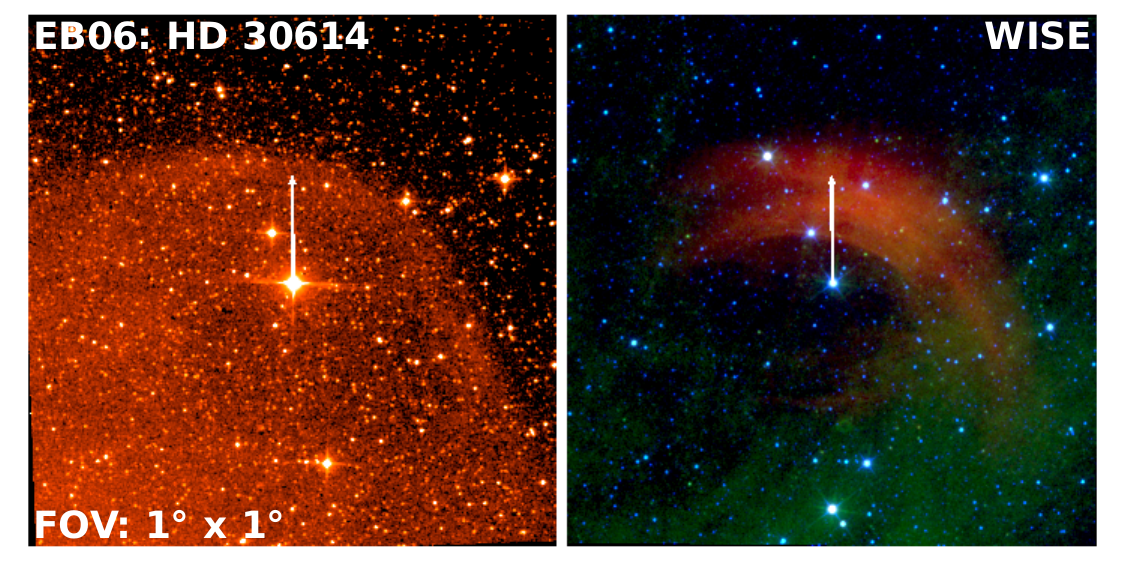}
\includegraphics[width=0.49\hsize]{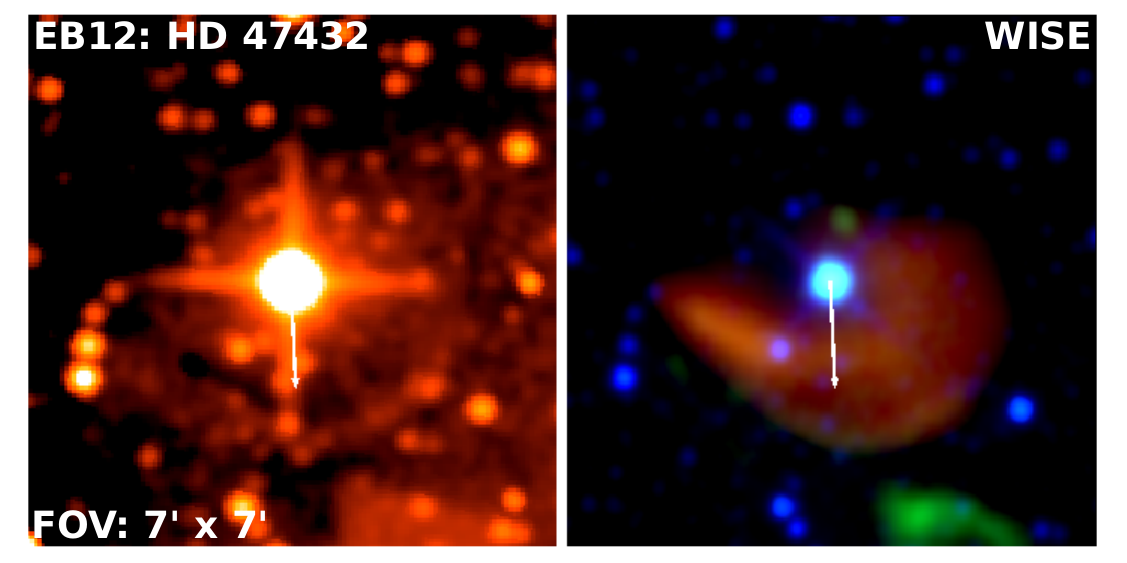}
\includegraphics[width=0.49\hsize]{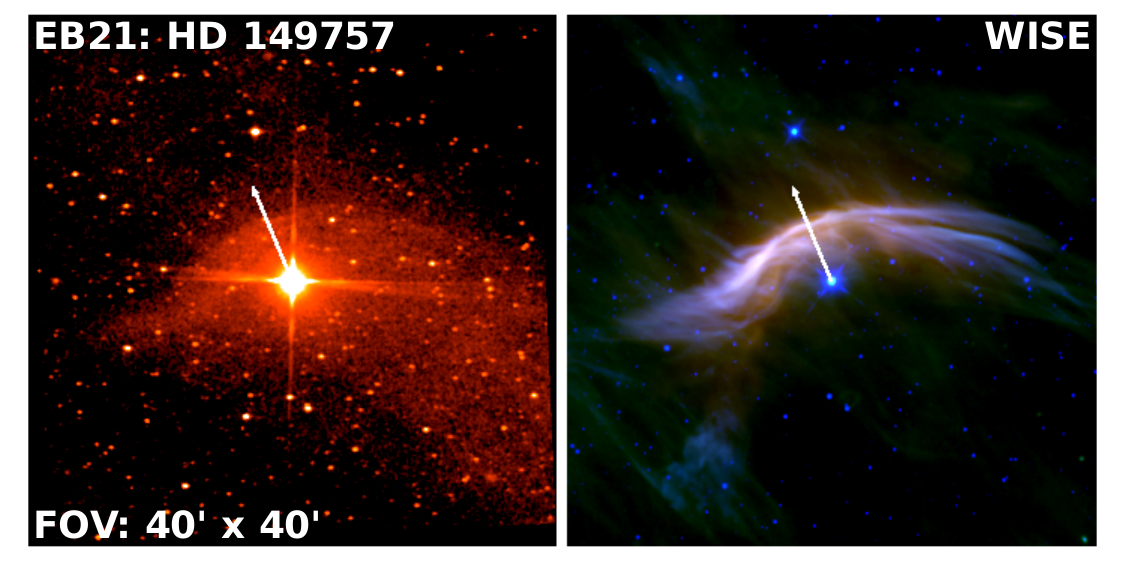}
\includegraphics[width=0.49\hsize]{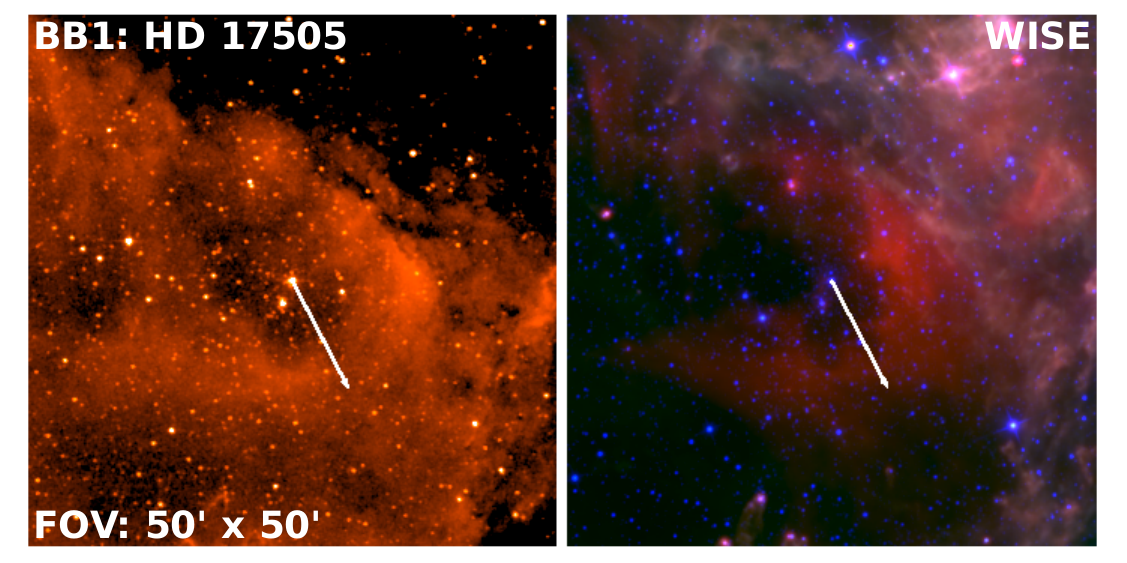}
\includegraphics[width=0.49\hsize]{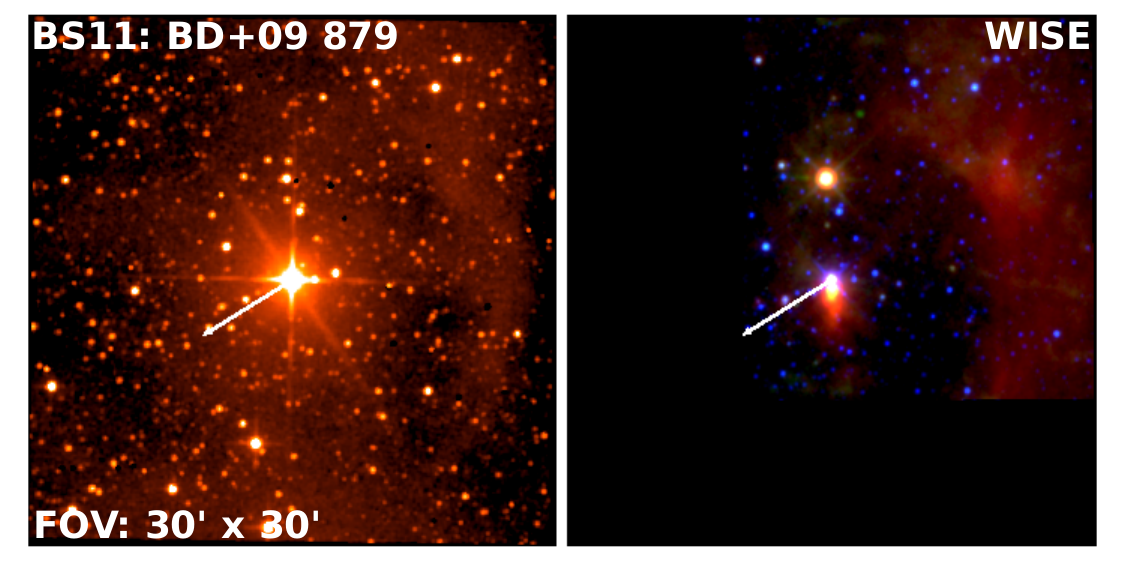}
\includegraphics[width=0.49\linewidth]{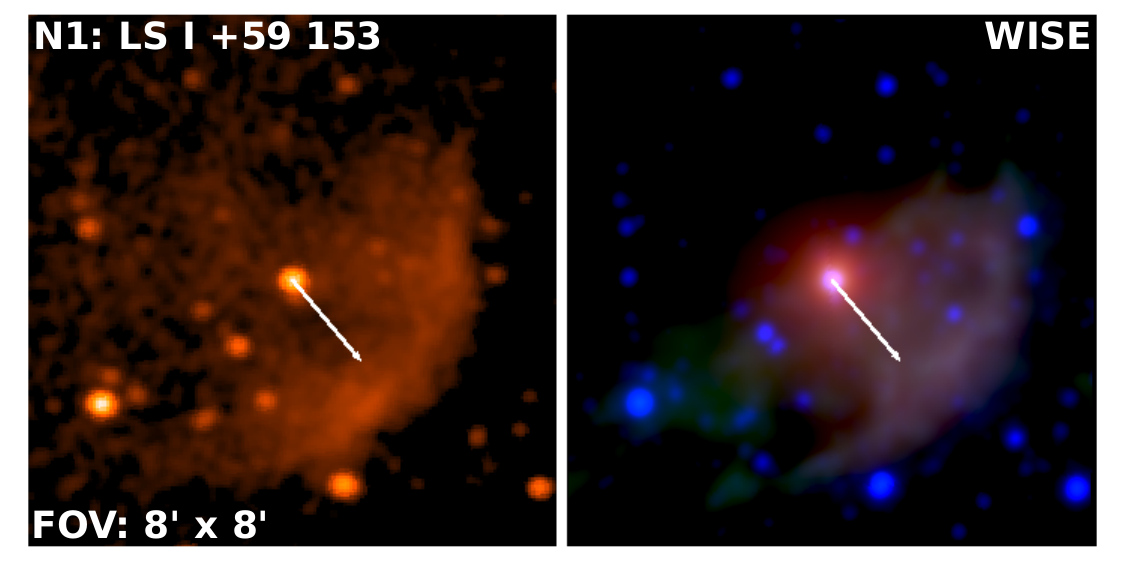}
   \caption{continued. Images were taken with the SBT.}
  \label{Detections2}
\end{figure*}

The data reduction was carried out using standard procedures within IRAF\footnote{IRAF was written at the National  Optical Astronomy Observatory, which was operated by the Association of Universities for Research in Astronomy (AURA) under cooperative agreement with the National Science Foundation.} \citep{1986SPIE..627..733T, 1993ASPC...52..173T}. This included subtraction of the bias level and division by normalised flat fields to correct for pixel-to-pixel sensitivity variations. Individual exposures were then aligned and median-combined to remove cosmic rays and to increase the resulting signal-to-noise ratio. Astrometry was performed on all images. 

\begin{figure}[t!]
    \centering
    \includegraphics[width=0.75\linewidth]{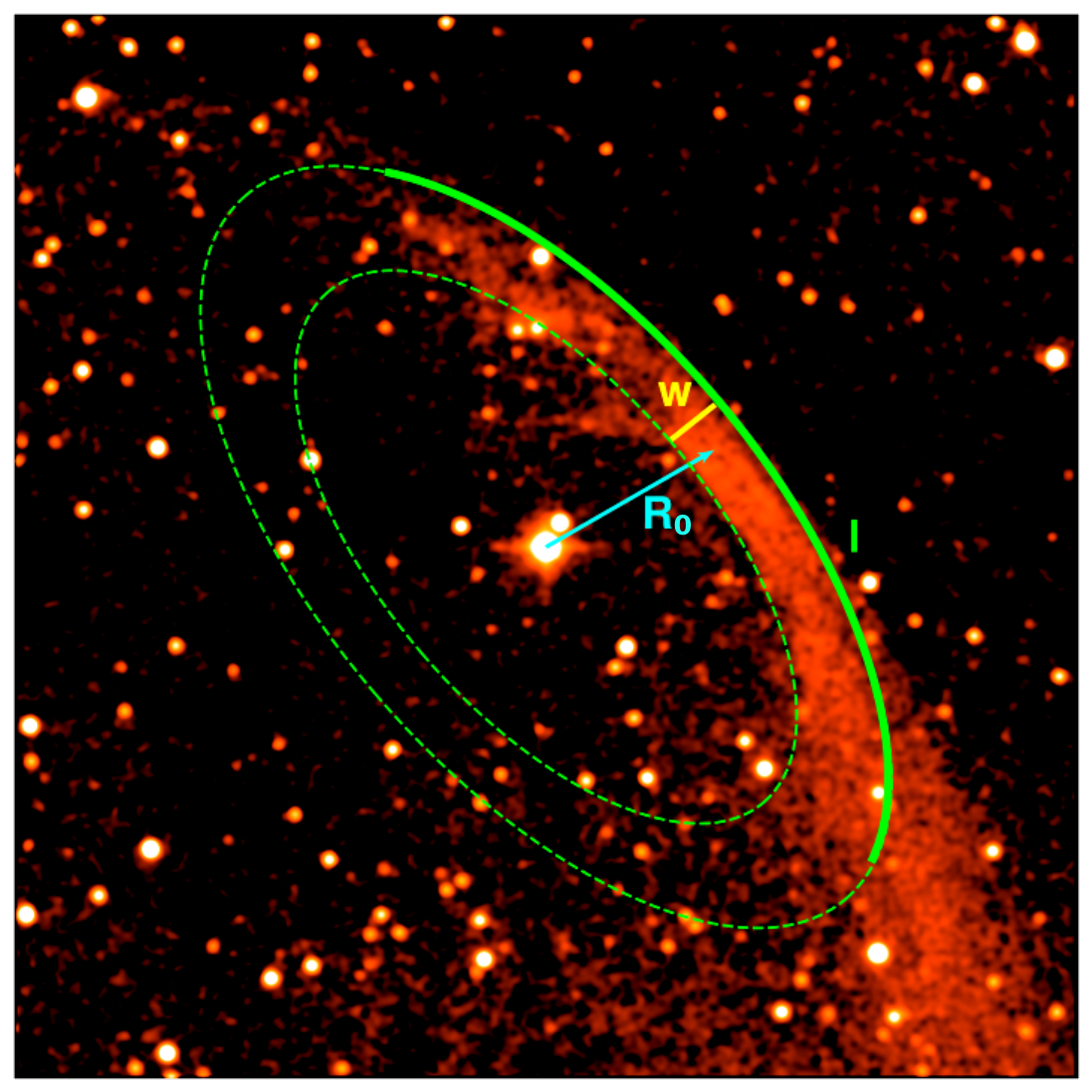}
    \caption{Illustration of the measurements of the bow-shock parameters for the example of K383: CD$-$26\,5136.}
    \label{Fig:BS_Parameters}
\end{figure}

The primary goal of the present survey was the detection and morphological characterization of extended bow-shock nebulae. The observations were carried out using relatively narrow H$\alpha$ filters compared to those employed in large photometric surveys such as IPHAS (9.5\,nm).
No continuum observations were obtained and the H$\alpha$ images are therefore not continuum-subtracted. Possible continuum contribution could become problematic, if a bow shock source is located within a H\,{\sc ii} region that itself has strong H$\alpha$ emission. Furthermore, emission at the wavelength of H$\alpha$ might originate from reflection of starlight from a dust structure, mimicking H$\alpha$ emission. While H\,{\sc ii} regions are easily recognized and bow shocks within them should still be detectable, distinguishing between real H$\alpha$ emission and a reflection nebula is more challenging from our data alone. As a qualitative assessment of the latter, we inspected Digitized Sky Survey 2 (DSS2) blue images for all detected nebulae.

{\tiny
\begin{table*}
\caption{Bow-shock parameters measured from the H$\alpha$ and IR images. For each target, the table lists the stand-off distance, width, length, and eccentricity of the fitted annulus in H$\alpha$ and WISE 22\,\(\mu\)m or {\it Spitzer} 24\,\(\mu\)m images.}
\label{tab:BSParameters}
\centering
\footnotesize
\begin{tabular}{l|ccccccc|ccccccc}
\hline \hline
ID & \multicolumn{7}{c}{H$\alpha$} & \multicolumn{7}{c}{WISE/{\it Spitzer}}  \\
 & \multicolumn{2}{c}{$R_{0}$} & \multicolumn{2}{c}{$w$} & \multicolumn{2}{c}{$l$} & $e$ & \multicolumn{2}{c}{$R_{0}$} & \multicolumn{2}{c}{$w$} & \multicolumn{2}{c}{$l$} & $e$ \\
  & $ \left( \arcsec \right) $ & $ \left( \mathrm{pc} \right) $ & $ \left( \arcsec \right) $ & $ \left( \mathrm{pc} \right) $ & $ \left( \arcsec \right) $ & $ \left( \mathrm{pc} \right) $ &   & $ \left( \arcsec \right) $ & $ \left( \mathrm{pc} \right) $ & $ \left( \arcsec \right) $ & $ \left( \mathrm{pc} \right) $ & $ \left( \arcsec \right) $ & $ \left( \mathrm{pc} \right) $ &   \\
 \hline
EB06 & 790.30 & 6.48 & 272.43 & 2.23 & 3017.20 & 24.73 & 0.75 & 624.35 & 5.12 & 587.82 & 4.82 & 3228.88 & 26.46 & 0.73 \\
EB12 & 126.38 & 1.04 & 35.78 & 0.29 & 383.79 & 3.15 & 0.42 & 65.52 & 0.54 & 100.28 & 0.82 & 473.18 & 3.15 & 0.42 \\
EB15 & 15.92 & 0.39 & 8.95 & 0.22 & 56.08 & 1.36 & 0.64 & 14.66 & 0.36 & 10.34 & 0.25 & 60.35 & 1.46 & 0.56 \\
EB21 & 290.567 & 0.19 & 172.07 & 0.11 & 1488.50 & 0.97 & 0.83 & 295.88 & 0.19 & 129.47 & 0.08 & 1290.76 & 0.84 & 0.82 \\
EB23 & 108.24 & 0.77 & 47.14 & 0.33 & 389.38 & 2.76 & 0.21 & 93.22 & 0.66 & 50.91 & 0.36 & 448.18 & 3.18 & 0.57 \\
BB1 & 605.58 & 7.08 & 288.11 & 3.37 & 2214.76 & 25.89 & 0.65 & 354.11 & 4.14 & 256.32 & 3.00 & 1836.83 & 21.47 & 0.77 \\
BB4 & 807.83 & 3.59 & 329.87 & 1.47 & 1467.33 & 6.53 & 0.28 & 582.58 & 2.59 & 373.07 & 1.66 & 1865.75 & 8.30 & 0.55 \\
BB5 & 71.86 & 0.90 & 35.20 & 0.44 & 418.63 & 5.23 & 0.82 & 46.97 & 0.59 & 40.10 & 0.50 & 385.98 & 4.82 & 0.76 \\
BS11 & 728.78 & 1.36 & 185.54 & 0.35 & 1472.10 & 2.76 & 0.34 & 600.75 & 1.12 & 327.51 & 0.61 & 1868.42 & 3.50 & 0.30  \\
K065 & 24.83 & 0.22 & 8.25 & 0.07 & 64.00 & 0.56 & 0.60 & 25.73 & 0.23 & 12.67 & 0.11 & 91.06 & 0.80 & 0.67 \\
K359 & 48.50 & 0.62 & 27.25 & 0.35 & 278.04 & 3.56 & 0.81 & 44.19 & 0.57 & 38.89 & 0.50 & 216.60 & 2.77 & 0.62 \\ 
K383 & 49.61 & 1.51 & 17.91  & 0.54 & 253.27 & 7.70 & 0.87 & 52.21 & 1.59 & 36.94 & 1.12 & 219.66 & 6.68 & 0.68 \\
K385 & 122.50 & 1.40 & 55.85 & 0.64 & 381.28 & 4.35 & 0.69 & 83.83 & 0.96 & 80.30 & 0.92 & 418.72 & 4.78 & 0.65 \\
K386 & 58.77 & 0.57 & 19.71 & 0.19 & 251.39 & 2.46 & 0.81 & 55.02 & 0.54 & 22.30 & 0.22 & 301.96 & 2.95 & 0.89 \\
K692 & 50.46 & 0.44 & 30.14 & 0.26 & 185.51 & 1.61 & 0.69 & 34.51 & 0.30 & 43.14 & 0.36 & 243.39 & 2.12 & 0.37 \\
N1 & 127.81 & 1.34 & 50.69 & 0.53 & 386.34 & 4.05 & 0.50 & 129.50 & 1.36 & 54.20 & 0.57 & 378.92 & 3.97 & 0.26 \\
\hline
\end{tabular}
\tablefoot{
(1) The uncertainties quoted for \(R_{0}\) correspond to one pixel for the respective H$\alpha$ and IR images. The pixel scales of the instruments are given in Sect.~\ref{sec:observations}. 
(2) To estimate the uncertainty associated with the fitting procedure, the measurements were repeated multiple times for each nebula. The relative uncertainty, quantified as \(100\times\sigma/\mu\), where \(\sigma\) is the standard deviation and \(\mu\) is the mean of the repeated measurements, was found to be \(\sim 2\text{--}15\%\) for \(w\), \(\sim 1\text{--}7\%\) for \(l\), and \(\sim 1\text{--}30\%\) for \(e\) in the H\(\alpha\) images, and \(\sim 3\text{--}14\%\) for \(w\), \(\sim 1\text{--}7\%\) for \(l\), and \(\sim 1\text{--}47\%\) for \(e\) in the IR images.}
\end{table*}
}

\subsection{Archival Infrared Survey Data}

Mid-infrared images were retrieved from the archives of \textit{Spitzer} \citep{werner2004} and WISE \citep{wright2010} through the NASA Infrared Science Archive (IRSA)\footnote{\url{https://irsa.ipac.caltech.edu/applications/Spitzer/SHA}}\footnote{\url{https://irsa.ipac.caltech.edu/applications/wise}}. From \textit{Spitzer}, we used data obtained with the Infrared Array Camera (IRAC; \citealt{fazio2004}) in the 3.6, 4.5, 5.8, and 8.0\,$\mu$m bands, which provide angular resolutions of 1.66, 1.72, 1.88, and 1.98 arcsec, respectively. Additional images at 24\,$\mu$m were taken from MIPS \citep{rieke2004multiband}, which has a beam size of 6.0 arcsec. Complementary data were obtained from the WISE survey in the 3.4, 4.6, 12, and 22\,$\mu$m bands, with angular resolutions of 6.1, 6.4, 6.5, and 12.0 arcsec, respectively.

\section{Results}
\label{sec:results}

From the 78 objects included in Table 1, we identify 15 bow-shock detections (Fig.~\ref{Detections1}), 28 non-detections (not shown), and 35 sources (Figs.~\ref{DE:DK154andPerek}--\ref{DE:IPHAS}) whose classification remains uncertain, mostly because they lie within bright and complex H\,{\sc ii} regions where extended emission masks the underlying structure. These we will refer to in the following as objects with diffuse emission. In addition, we report one previously unrecognized detection, designated N1, which appears in the wide-field SBT frame obtained for BB1: HD\,17505. The arc-like structure of the nebula in H$\alpha$ matches with that seen in the W4 band image. The spectral type of the central star is O9.5\,V, as reported by \citet{garmany1992}. 

\subsection{Bow-Shock Parameters}
\label{sec:BSP}

\begin{figure*}
\centering
\includegraphics[width=0.4\textwidth]{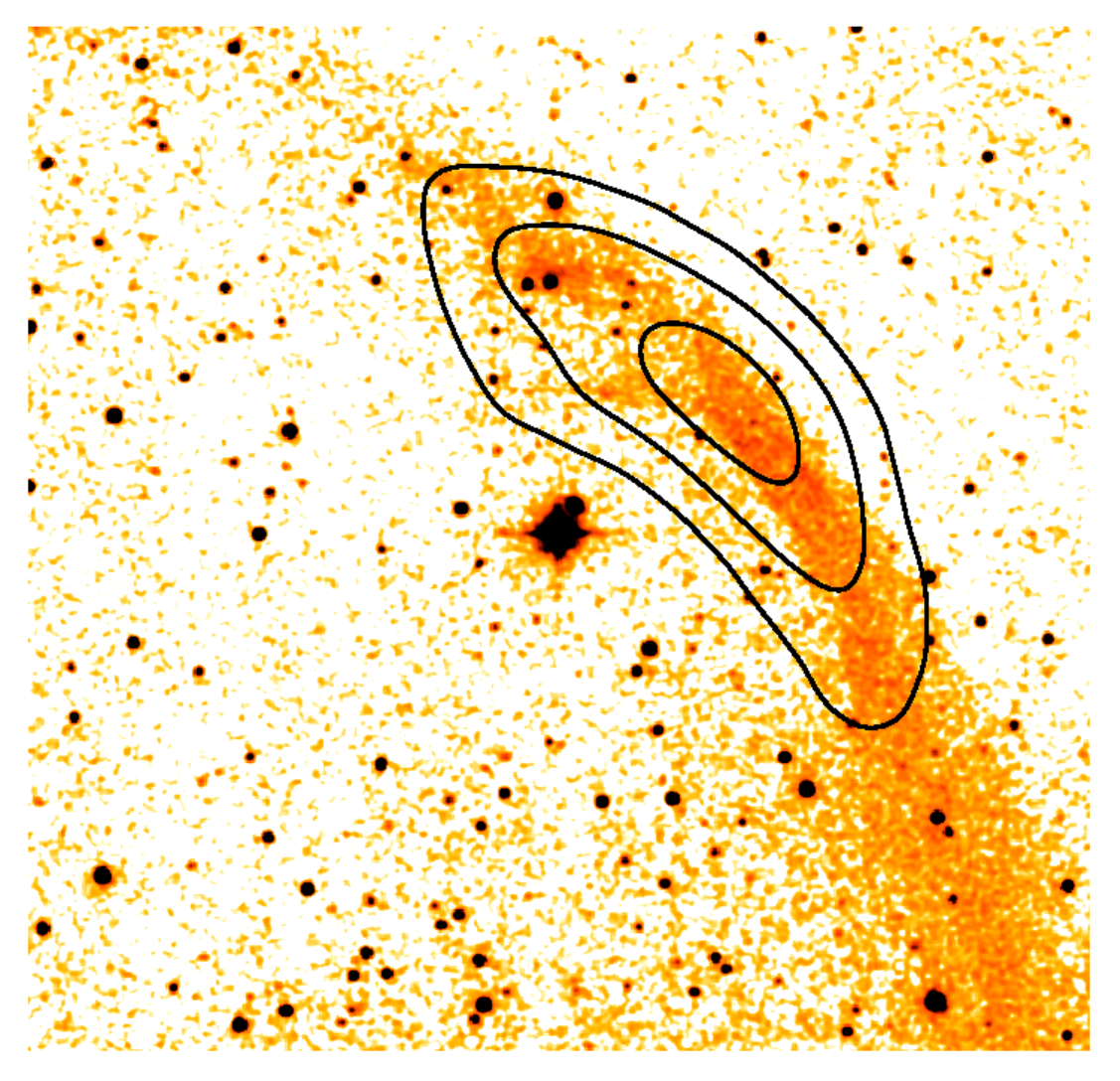}
\includegraphics[width=0.4\textwidth]{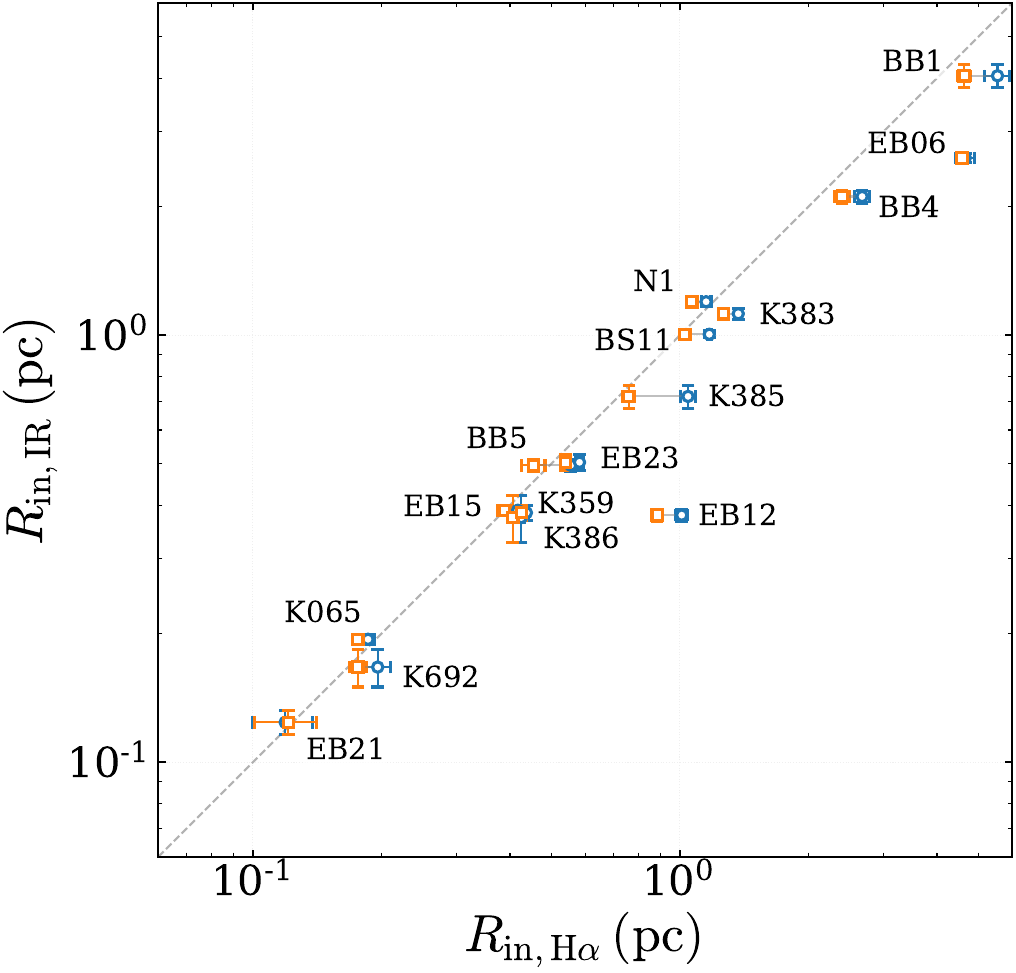}
\caption{Left: H$\alpha$ image of K383 with WISE W4 contours overlaid. The contours correspond to 30\%, 50\%, and 70\% of the arc peak emission above the local background and illustrate that the infrared arc lies slightly closer to the central star than the optical one. Right: Comparison of the distances between the central star and the inner arc
boundary measured along the symmetry axis in H$\alpha$ and infrared emission. Blue circles denote measurements from the original H$\alpha$ images, while orange squares correspond to H$\alpha$ images convolved to the infrared resolution. The dashed line indicates the one-to-one relation.
}
\label{Fig:inner_annulus}
\end{figure*}

\begin{figure}
    \centering
    \includegraphics[width=0.95\linewidth]{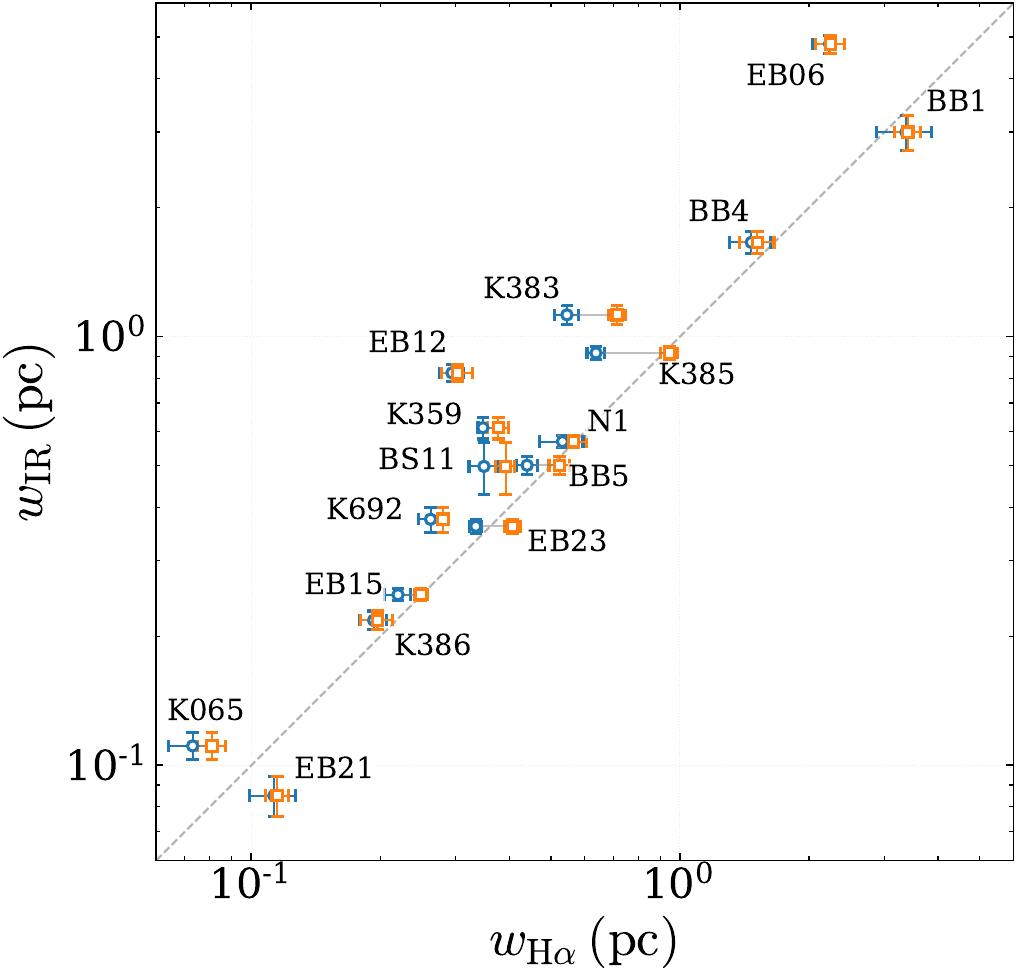}
    \caption{Comparison of the bow-shock widths measured in H$\alpha$ and IR emission for the detected nebulae. Blue circles denote measurements from the original H$\alpha$ images, while orange squares show the corresponding point spread function-matched measurements. The dashed line indicates the one-to-one relation.}
    \label{Fig:scatter_width}
\end{figure}

For each of the 16 detections, we measure the geometrical parameters of the arc-shaped nebula, namely its spatial extent $l$, width $w$, and standoff distance $R_{0}$ using both the H$\alpha$ and mid-infrared images. The measurements are performed by visually fitting each nebula with an elliptical annulus (Fig.~\ref{Fig:BS_Parameters}). The inner and outer ellipses of the annulus were constrained to have the same eccentricity $e$, while the orientation of the ellipse was chosen such that its major/minor axis approximately follows the apparent symmetry axis of the bow shock. In this framework, $l$ corresponds to the partial circumference of the outer ellipse that traces the arc, with the endpoints selected from the extent of the detectable shell emission. These endpoints are not necessarily symmetric with respect to the major or minor axis of the ellipse. The width $w$ is determined from the thickness of the annulus along the symmetry axis, and the stand-off distance $R_{0}$ is measured as the separation between the star and the brightest pixel of the nebula in the W4 band or MIPS 24\,$\mu$m band images, and in the H$\alpha$ images, with the brightest pixel constrained to lie within $\pm 10^{\circ}$ of the symmetry axis of the bow shock. The resulting values are listed in Table~\ref{tab:BSParameters}.  

Although the H$\alpha$ and mid-IR morphologies generally trace the same overall structure,  comparison of the inner boundaries along the symmetry axis indicates that the IR arcs are frequently located slightly closer to the central star. Furthermore, the H$\alpha$ arcs appear systematically thinner than their mid-IR counterparts (left panel of Fig.~\ref{Fig:inner_annulus}). To assess whether these differences arise from resolution effects, the H$\alpha$ images were convolved to the FWHM of the WISE band-4 $(12\arcsec)$ and MIPS 24\,$\mu$m $(6\arcsec)$ images. The measured arc widths in the H$\alpha$ images increase after convolution in all cases, with the effect becoming substantially more pronounced when the images are matched to the lower angular resolution of the WISE data (see Fig.~\ref{Fig:scatter_width}). In several cases, the convolved H$\alpha$ widths become more consistent with the IR measurements, although noticeable discrepancies remain for some targets. Similarly, the distances towards the inner boundary of the arc structures are smaller on the convolved H$\alpha$ images 
(right panel of Fig.~\ref{Fig:inner_annulus}). Nevertheless, a systematic difference of the measured offsets remains for a significant number of objects, suggesting that angular resolution alone cannot fully account for the observed trends.

\subsection{H$\alpha$ Surface Brightness and Emission Measure}
\label{sec:calibration}

For the ten detections laying within the SHASSA survey coverage, we estimated the H$\alpha$ surface brightness by flux-calibrating the observed images against the corresponding SHASSA data. The calibration was obtained by matching the image resolution and spatial sampling to the SHASSA grid, comparing the fluxes measured over identical sky areas, and deriving a linear conversion factor between image counts (ADU) and surface brightness (Rayleighs). A detailed description of the procedure is provided in Appendix~\ref{app:calibration}.

The surface brightness of each bow shock was measured in circular apertures placed along the nebular apex. Aperture radii ranging from 5$\arcsec$ to 40$\arcsec$ were adopted depending on the angular extent of the nebula and the resolution of the observations, to ensure that each aperture fits entirely within the emission region while containing enough pixels for robust statistics. Within each aperture, the mean signal was computed after interquartile-range (IQR) clipping to suppress  contamination from stars, cosmic rays, and other outlying pixels. The local background, including diffuse H\,{\sc ii} region emission where present, was estimated from nearby sky apertures and subtracted from the nebular measurement. For targets embedded in diffuse H{\sc ii} region emission (marked in Table~\ref{tab:SB}), the reported values represent the excess H$\alpha$ surface brightness attributable to the bow shock alone.

The resulting background-subtracted H$\alpha$ surface brightness values are listed in  Table~\ref{tab:SB}. Assuming case-B recombination and an electron temperature of $T_{\mathrm{e}} = 10^4$\,K, the corresponding emission measures were derived using $EM \simeq 2.75\,I_{\rm H\alpha}$, where $EM$ is in $\mathrm{pc}\,\mathrm{cm}^{-6}$ and $I_{\rm H\alpha}$ is the H$\alpha$ surface brightness in Rayleighs \citep[e.g.,][]{osterbrock2006,haffner1999}. The quoted uncertainties include contributions from the scatter among the nebular and background apertures and from the SHASSA-based flux calibration.

\begin{table}
\caption{H$\alpha$ surface brightness of our detections which are within the SHASSA survey coverage, and the corresponding emission measures, ambient densities, and lower limits of expected thermal free--free surface brightness at 6\,GHz.}
\label{tab:SB}
\centering
\footnotesize
\begin{tabular}{lcccc}
\hline \hline
ID & $I_{\rm H\alpha}$ & $EM$ & $n_{\rm ISM}$ & $I_{6\rm\,{GHz}}$ \\
   & (Rayleighs)    & (pc\,cm$^{-6}$) & (cm$^{-3}$) & (kJy sr$^{-1}$) \\
\hline
EB12 & 142 $\pm$ 65 & 391 $\pm$ 179 & $\sim$9 & $\sim$33 \\
EB15$^{*}$  & 4975 $\pm$ 655 & 13681 $\pm$ 1801 & $\sim$62 & $\sim$1152 \\
EB21           & 44 $\pm$ 9     & 121 $\pm$ 25 & $\sim$8 & $\sim$10 \\
EB23           & 171 $\pm$ 6    & 470 $\pm$ 17 & $\sim$9 & $\sim$39 \\
BB4            & 12 $\pm$ 6     & 33 $\pm$ 17 & $\sim$1 & $\sim$3 \\
BB5$^{*}$   & 2775 $\pm$ 404 & 7631 $\pm$ 1111 & $\sim$33 & $\sim$643 \\
K065           & 425 $\pm$ 113  & 1167 $\pm$ 311 & $\sim$32 & $\sim$98 \\
K383           & 49 $\pm$ 9     & 135 $\pm$ 25 & $\sim$4 & $\sim$31 \\
K385           & 95 $\pm$ 14    & 261 $\pm$ 39 & $\sim$5 & $\sim$22 \\
K692$^{*}$ &  230 $\pm$ 46 & 632 $\pm$ 127 & $\sim$12 & $\sim$53 \\
\hline
\end{tabular}
\tablefoot{$^{*}$Excess emission above the local H{\sc ii} region background.}
\end{table}
The emission measure can be used to make order of magnitude estimates for the density of the ambient ISM and the radio free-free emission, for example at 6\,GHz. Assuming a strong shock, the post shock electron density is $n_{\rm e} \approx 4\,n_{\rm ISM}$, where $n_{\rm ISM}$ is the ambient number density of the interstellar medium\footnote{We caution the reader that the compression factor could be less than 4 for weak shocks, and significantly larger for radiative shocks close to the isothermal limit \citep[see, for example,][]{draine1993}}. The emission measure is $EM = n_{\rm e}^2\,l_{\rm eff}$, where $l_{\rm eff}$ is the line of sight path length through the emitting shell. Near the apex, $l_{\rm eff}$ is assumed to be of the same order as the projected H$\alpha$ width $w$ (Table~\ref{tab:BSParameters}). Taking $l_{\rm eff} \sim w$ yields the simple relation, 
\begin{equation}
 n_{\rm ISM} = \frac{1}{4}\sqrt{\frac{EM}{w}}.   
\end{equation}
These values are listed in the fourth column of Table~\ref{tab:SB}.

From the same emission measure, the free-free optical depth follows as \citep{mezger1967,draine2011},
\begin{equation}
    \tau_\nu = 3.28\times10^{-7}\,\left(\frac{T_{\rm e}}{10^{4}}\right)^{-1.35}\,\left(\frac{\nu}{{\rm GHz}}\right)^{-2.1}\, EM.
\end{equation}
For the optically thin case ($\tau_{\nu} \ll 1 $), applying the Rayleigh--Jeans approximation,
\begin{equation}
    I_\nu = \frac{2k\nu^2 T_{\rm e}\tau_\nu}{c^2},
\end{equation}
yields the combined expression \citep{dickinson2003},
\begin{equation}
    I_\nu\,[{\rm Jy\,sr}^{-1}] \approx 11.1\,\left(\frac{T_{\rm e}}{10^{4}}\right)^{0.52}\,\nu_{\rm GHz}^{-0.1}\,I_{\rm H\alpha}[{\rm Rayleigh}], 
\end{equation}
which evaluates to
\(I_\nu \approx 9.3\,I_{\rm H\alpha}\) at \(\nu = 6\)~GHz for \(T_{\rm e} = 10^4\)~K. The resulting radio free-free intensity is given in the last column of Table~\ref{tab:SB}. These values should be considered as lower limits, because the surface brightness values are not extinction corrected.

\subsection{Morphological Classification}

Following the theoretical framework of \citet{henney2019_1}, the interaction of a luminous star with the surrounding medium gives rise to distinct physical scenarios. These scenarios are governed by the stellar luminosity, the wind momentum, and the optical depth of the shocked shell (see their Fig.~1). When stellar radiation effects are negligible ($\tau \ll \eta_{\rm w}$, where $\tau$ is the optical depth in the ultraviolet of the shocked shell and $\eta_{\rm w}$ is the momentum efficiency of the wind), the stellar wind momentum dominates and a classical wind-supported bow shock (WBS) is formed that is controlled by pure (magneto)hydrodynamic interactions. 
With increasing optical depth of the bow shock, the stellar
radiation pressure starts to contribute to the total internal pressure because the radiation is being trapped.  As soon as the shocked layer becomes opaque to the stellar radiation field, the stellar radiation pressure dominates the internal pressure and balances the ISM ram pressure. This optically thick limit 
($\tau > 1$) defines the radiation-supported bow-shock (RBS) regime. For intermediate optical depths, two possibilities exist depending on the strength of the grain-gas coupling. If the coupling is strong, the ISM plasma is decelerated as a whole in response to the radiative force, producing a radiation-supported bow wave (RBW). 
If the coupling is sufficiently weak, the radiative force acts primarily on the dust, allowing the grains to decouple from the gas and producing a dust-wave (DW) scenario. In ionised gas this coupling is maintained mainly by Coulomb drag, and the decoupling occurs at the rip point, where the radiation force exceeds the maximum drag force that the gas can exert. This condition is expressed by the local radiation parameter $\Xi = P_{\mathrm{rad}}/P_{\mathrm{gas}}$, with decoupling occurring when $\Xi$ exceeds the critical value of $ \sim 10^{3}$ \citep{henney2019_2}. For an outer dust wave to form, the rip-point radius must lie outside the wind bow-shock radius; otherwise the dust and gas enter the shocked shell still coupled, where the higher gas pressure further inhibits decoupling.

To estimate the stand-off distance $R_{0}$ in the different interaction regimes, \citet[][see their Sect.~2 for more details]{henney2019_1} introduced a fiducial stand-off distance $R_{1}$ in the optically thick, radiation-only limit, and a corresponding fiducial optical depth $\tau_{1}$ of the shocked shell as,
\begin{equation}
    R_{1} = \left( \frac{L}{4\pi c \rho v^{2}} \right)^{1/2} \quad \text{and} \quad \tau_{1} = \rho \kappa R_{1},
    \label{eq:R1}
\end{equation}
where $L$ is the bolometric luminosity of the star, $\rho$ and $v$ are the ambient ISM mass density and the relative stream velocity, $c$ is the speed of light and $\kappa$ is the total opacity (gas plus dust) per unit mass. The ambient interstellar mass density $\rho$ can be written as $\rho = n_{\rm ISM} \bar{m}$, where $\bar{m}$ is the mean particle mass. Assuming solar abundances, we adopt $\bar{m} \approx 1.3\,m_{\mathrm{p}} \approx 2.17 \times 10^{-24}\,\mathrm{g}$. The combined dust–gas opacity is taken to be $\kappa = 600\,\mathrm{cm^{2}\,g^{-1}}$, corresponding to an effective cross-section of $\sim10^{-21}\,\mathrm{cm^{2}}$ per hydrogen nucleon, in line with typical interstellar dust properties \citep{bertoldi1996}.
The contribution of stellar wind momentum to the support of the bow shock is quantified by the wind-momentum efficiency, defined as
\begin{equation}
\label{eq:windeff}
\eta_{\mathrm{w}} = \frac{\dot{M} v_{\infty}}{L/c},
\end{equation}
where $\dot{M}$ is the stellar mass-loss rate and $v_{\infty}$ is the terminal velocity of the stellar wind. The stand-off distance is then
written as 
\begin{equation}
    R_{0} = xR_{1},
\end{equation}
where $x$ is the solution of 
\begin{equation}
    x^{2} - \left( 1 - e^{-2\tau_{1}x} \right) - \eta_{w} = 0, 
    \label{eq:x}
\end{equation}
and needed to cover all cases from optically thin to optically thick shells. Equation~(\ref{eq:x}) can be solved numerically, and $x$ can be approximated by $x \approx (1 + \eta_{w})^{1/2}$ corresponding to RBS, $x \approx 2\tau_{1}$ for RBW and DW, and $x \approx \eta_{w}^{1/2} $ for WBS. 

We aim to classify our sample according to these theoretical predictions for all objects with stellar and wind parameters available in the literature\footnote{We caution that this approach considers a homogeneous distribution of the ambient dust with a pre-defined dust-to-gas ratio. A different mixing value or fragmented dusty bow-shock structures may alter the optical depth in the ultraviolet and hence the theoretical predictions.}. However, for the computations density values for the ISM embedding our targets are needed.

\begin{landscape}
\begin{table}[t!]
\centering
{\tiny
\caption{Morphological and physical parameters of bow-shock nebulae. See text for detailed explanations.}
\label{tab:morphology}
\begin{tabular}{lccccccccccccccccc}
     \hline \hline
     ID & $\mathrm{log} T_{\rm eff}$ & $\mathrm{log}\left( \frac{L}{L_{\odot}}\right)$ & $\dot{M} \times 10^{7}$ & $v_{\infty}$ & $v_{\mathrm{tg}}$ & $v_{\mathrm{r}}$ & $v$ & Ref. & $n_{\rm ISM}$ & \multicolumn{3}{c}{Expected $R_{0}$} & Obs. $R_{0}$ & Class & BS in & H$\alpha$ & Binary \\
        &  &  &  &  &  &  &  &  &  & RBS & DW & WBS &  &  & radio & &  \\
        & $ \left( \mathrm{K} \right) $ &  & $ \left(M_{\odot} \mathrm{yr}^{-1} \right) $ & $ \left( \mathrm{km}\,\mathrm{s}^{-1} \right) $ & $ \left( \mathrm{km}\,\mathrm{s}^{-1} \right) $ & $ \left( \mathrm{km}\,\mathrm{s}^{-1} \right) $ & $ \left( \mathrm{km}\,\mathrm{s}^{-1} \right) $ &  & $ \left( \mathrm{cm}^{-3} \right) $ & $ \left( \mathrm{pc} \right) $ & $ \left( \mathrm{pc} \right) $ & $ \left( \mathrm{pc} \right) $ & $ \left( \mathrm{pc} \right) $ &  &  &  & \\
    \hline
    EB01 & 4.48 & 5.39 & 0.87   & 3171  & 5.28   & 25.02    & 25.57  & $1,14,27$      & $0.1\,-\,5$ & $13.98\,-\,1.98$ & 0.15   & $3.19\,-\,0.45$ & 0.24    & DW/WBS         &  no   &  no   &   no	\\
    EB02 & 4.38 & 6.24 & 205.56 & 1105  & 16.16  & 0.30     & 16.16  & $1,12,16$    & $1\,-\,100$ & $23.15\,-\,2.32$ & 2.62   & $14.5\,-\,1.45$ & 0.75    & WBS        &  $-$  &  diff &   no	\\
    EB03 & 4.63 & 5.56 & 3.16   & 2800  & 9.21   & $-68.00$ & 68.62  & $7,18$	    & $1\,-\,100$ & $2.05\,-\,0.21$  & 0.03   & $0.67\,-\,0.07$ & 0.43 & RBS/WBS        &  y    &  diff &   $-$  \\
    EB04 & 4.37 & 4.50 & 0.06   & 500   & 17.82  & $28.10$  & 33.27  & $2,16$	    & $0.1\,-\,5$ & $3.74\,-\,0.53$  & 0.01   & $0.08\,-\,0.01$ & 0.08    & WBS    &  no   &  no   &   no	\\ 
    EB05 & 4.16 & 2.90 & 0.01   & 500   & 28.60  & $4.00$   & 28.88  & $3,15,17$    & $0.1\,-\,5$ & $0.70\,-\,0.10$  & 0.0004 & $0.12\,-\,0.02$ & 0.071   & WBS        &  $-$  &  no   &   $-$  \\
    EB06 & 4.47 & 5.63 & 50.00  & 1560  & 57.74  & $10.00$  & 58.60  & $5,18 $      & $0.1\,-\,5$ & $10.75\,-\,1.52$ & 0.05   & $7.41\,-\,1.05$ & 5.12    & RBS/WBS    &  $-$  &  bow  &   $-$  \\
    EB07 & 4.52 & 4.65 & 0.66   & 1500  & 80.71  & $51.00$  & 95.47  & $1,13,18$    & $1\,-\,100$ & $0.51\,-\,0.05$  & 0.002  & $0.16\,-\,0.02$ & 0.03    & WBS        &  no   &  diff &   no	\\
    EB08 & 4.51 & 4.68  & 0.65    & 1500   & 9.52   & 16.00      & 18.62    & $4,\,13 ,\,18 $   & $1\,-\,100$	  & $2.73\,-\,0.27$	     & 0.05 & $0.82\,-\,0.08$ 	&  0.15 &  WBS   &  no   &  no   &   no	\\
    EB09 & 4.44 & 4.21 & 0.14   & 750   & 27.56  & $-15.00$ & 31.38  & $3,15,16$    & $1\,-\,100$ & $0.86\,-\,0.09$  & 0.006  & $0.16\,-\,0.02$ & 0.26    & RBS    &  no   &  diff &   $-$  \\
    EB10 & 4.52 & 5.24 & 4.97   & 2070  & 19.98  & $-16.40$ & 25.85  & $1,12,16$    & $0.1\,-\,5$ & $12.76\,-\,1.81$ & 0.10   & $6.10\,-\,0.86$ & 1.11    & WBS       &  $-$  &  no   &   SB2  \\
    EB11 & 4.32 & 4.09 & 0.01   & 600   & 19.68  & $30.60$  & 36.38  & $3,15,16$    & $0.1\,-\,5$ & $2.12\,-\,0.30$  & 0.004  & $0.10\,-\,0.01$ & 0.24    & WBS &  no   &  diff &   $-$  \\
    EB12 & 4.45 & 5.57 & 15.73    &  1590  & 11.54    & 64.00      &   65.03  & $4,\, 12,\, 18$ & $1\,-\,100$ 	& $2.39\,-\,0.24$	  & 0.03 & $1.19\,-\,0.12$&  0.54 & RBS/WBS   &  no   &  bow   &   $-$  \\
    EB13 & 4.64 & 5.59 & 15.92  & 2960  & 18.29  & $15.00$  & 23.65  & $1 ,12,20$    & $0.1\,-\,5$ & $23.25\,-\,3.29$ & 0.27   & $14.27\,-\,2.02$& 1.12    & WBS        &  no   &  no   &   SB2  \\
    EB14 & 4.60 & 5.65 & 13.86  & 2456  & 26.39  & $42.80$  & 50.28  & $1,12,28$    & $0.1\,-\,5$ & $10.93\,-\,1.55$ & 0.07   & $5.70\,-\,0.81$ & 1.49    & WBS        &  no   &  no   &   yes  \\
    EB15 & 4.60 & 4.46 & 4.41   & 2875  & 98.44  & $50.4$   & 110.59 & $6,13$	    & $1\,-\,100$ & $0.35\,-\,0.04$  & 0.001  & $0.09\,-\,0.01$ & 0.36    & RBS        &  y    &  bow  &   yes  \\
    EB16 & 4.34 & 3.62 & 0.06   & 300   & 20.05  & $46.00$  & 50.18  & $3,15,17$    & $0.1\,-\,5$ & $0.97\,-\,0.14$  & 0.0008 & $0.14\,-\,0.02$ & 0.024   & WBS        &  no   &  no   &   yes  \\
    EB17 & 4.55 & 5.22 & 4.79   & 2545  & 79.96  & $-36.54$ & 87.91  & $8,13$	    & $0.1\,-\,5$ & $3.79\,-\,0.53$  & 0.008  & $1.95\,-\,0.28$ & 1.07    & RBS/WBS    &  no   &  no   &   $-$  \\
    EB18 & 4.35 & 4.16 & 1.4    & 1065  & 60.84  & $-4.00$  & 60.97  & $3,13,15,16$ & $0.1\,-\,5$ & $1.69\,-\,0.24$  & 0.0015 & $0.98\,-\,0.14$ & 0.14    & WBS        &  no   &  no   &   $-$  \\
    EB19 & 4.47 & 5.55 & 2.50   & 1990  & 131.62 & $15.00$  & 132.47 & $2,16 $      & $0.1\,-\,5$ & $3.26\,-\,0.46$  & 0.008  & $0.83\,-\,0.12$ & 1.18    & RBS        &  no   &  no   &   $-$  \\
    EB20 & 4.45 & 4.83 & 1.40   & 1100  & 26.33  & $-7.00$  & 27.24  & $2,21,22$    & $0.1\,-\,5$ & $7.03\,-\,0.99$  & 0.04   & $2.24\,-\,0.32$ & 0.24    & WBS       &  no   &  no   &   $-$  \\
    EB21 & 4.51 & 4.82 & 1.49   & 1505  & 17.2   & $12.20$  & 21.09  & $1,12,24$    & $0.1\,-\,5$ & $9.15\,-\,1.29$  & 0.06   & $4.10\,-\,0.48$ & 0.19    &  WBS      &  no   &  bow  &   no	\\
    EB22 & 4.44 & 4.44 & 0.90   & 1345  & 116.81 & $-53.30$ & 128.40 & $3,12,15,16$ & $0.1\,-\,5$ & $1.00\,-\,0.14$  & 0.0007 & $0.42\,-\,0.06$ & 0.14    & RBS/WBS    &  no   &  no   &   $-$  \\
    EB23 & 4.45 & 5.10 & 2.22   & 1535  & 18.66  & $25.40$  & 31.52  & $1,12,15,16$     & $0.1\,-\,5$ & $8.35\,-\,1.18$  & 0.05   & $2.88\,-\,0.41$ & 0.66    & WBS        &  y    &  bow  &   SB1  \\
    EB24 & 4.55 & 5.14 & 0.4    & 2295  & 5.87   & $-2.00$  & 6.20   & $4,13,15,17$ & $0.1\,-\,5$ & $42.56\,-\,6.02$ & 1.41   & $7.59\,-\,1.07$ & 1.34    & DW/WBS     &  no   &  no   &    ?	\\
    EB25 & 4.50 & 5.65 & 5.00   & 1980  & 92.84  & $19.00$  & 94.76  & $2,18 $      & $0.1\,-\,5$ & $5.18\,-\,0.73$  & 0.02   & $1.63\,-\,0.23$ & 3.29    & RBS        &  no   &  no   &   $-$  \\
    EB26 & 4.37 & 5.57 & 2.3    & 1735  & 23.53  & $-11.00$ & 25.97  & $4,13,15,18$ & $1\,-\,100$ & $5.41\,-\,0.54$  & 0.22   & $1.22\,-\,0.12$ & 1.6	  & RBS        &  $-$  &  diff &   SB2  \\
    EB27 & 4.67 & 6.31 & 254.75 & 2325  & 21.96  & $-53.00$ & 57.37  & $1,13,23$    & $1\,-\,100$ & $8.58\,-\,0.86$  & 0.24   & $6.60\,-\,0.66$ & 1.69    & RBS/WBS    &  y    &  diff &   no	\\
    EB28 & 4.44 & 4.44  & 6.00    & 1535   & 80.71    & $-$125.3      & 149.04   & $3,\,13,\,15,\,16 $   & $1\,-\,100$	  & $0.40\,-\,0.04$	     & 0.0005 & $0.31\,-\,0.03$ 	& 1.06 & $-$    &  $-$  &  diff &   $-$  \\
    BB1  & 4.55 & 5.25 & 1.65   & 2500  & 12.20  & $-27.80$ & 30.36  & $9$	    & $0.1\,-\,5$ & $10.50\,-\,1.49$ & 0.08   & $3.37\,-\,0.48$ & 4.14    & RBS        &  $-$  &  bow  &   yes  \\
    BS07 & 4.51 & 5.80 & 14.13  & 2100  & 52.00  & $-75.10$ & 91.35  & $10,16$      & $1\,-\,100$ & $2.13\,-\,0.21$  & 0.03   & $0.93\,-\,0.09$ & 1.10    & RBS        &  $-$  &  no   &   yes  \\
    BS11 & 4.53 & 5.35 & 7.20  & 2125  & 7.87  & 30.10 & 31.11  & $4,13,16$      & $1\,-\,100$ & $3.89\,-\,0.39$  & 0.09   & $1.95\,-\,0.20$ & 1.12    & RBS/WBS       &  $-$  & bow   &  yes   \\
    K065 & 4.33 & 3.82 & 0.07   & 817   & 13.40  & $-$      & $-$    & $1,14 $      & $1\,-\,100$ & $1.42\,-\,0.14$  & 0.02   & $0.27\,-\,0.03$ & 0.23    & RBS/WBS    &  $-$  &  bow  &   no	\\
    K067 & 4.51 & 4.85 & 0.76   & 1500  & 13.15  & $-$      & $-$    & $1,13 $      & $1\,-\,100$ & $4.61\,-\,0.46$  & 0.16   & $1.26\,-\,0.13$ & 0.067   & DW/WBS        &  $-$  &  diff &   no	\\
    K150 & 4.40 & 4.44 & 0.40   & 1267  & 16.97  & $-$      & $-$    & $1,14 $      & $1\,-\,100$ & $2.24\,-\,0.22$  & 0.04   & $3.36\,-\,0.34$ & 0.04    & DW         &  $-$  &  diff &   ?	\\
    K330 & 4.35 & 5.25 & 10.55  & 1065  & 49.84  & $-22.00$ & 54.48  & $1,13,22$    & $1\,-\,100$ & $1.96\,-\,0.20$  & 0.02   & $0.96\,-\,0.10$ & 0.22    & RBS/WBS    &  $-$  &  diff &   no	\\
    K345 & 4.44 & 4.42 & 0.11   & 1821  & 14.52  & $-45.72$ & 47.97  & $1,14,19$    & $1\,-\,100$ & $0.76\,-\,0.08$  & 0.004  & $0.15\,-\,0.01$ & 0.09    & RBS/WBS    &  $-$  &  diff &   no	\\
    K346 & 4.33 & 5.58 & 19.20  & 908   & 85.36  & $-$      & $-$    & $1,14 $      & $1\,-\,100$ & $1.77\,-\,0.18$  & 0.02   & $0.76\,-\,0.08$ & 0.19    & RBS/WBS    &  $-$  &  diff &   SB1  \\
    K351 & 4.52 & 5.48 & 11.76  & 790   & 151.43 & $-$      & $-$    & $1,13 $      & $0.1\,-\,5$ & $2.73\,-\,0.39$  & 0.005  & $0.99\,-\,0.14$ & 1.28    & RBS        &  $-$  &  no   &   no	\\
    K359 & 4.50 & 5.19 & 3.15   & 2043  & 41.89  & $-50.30$ & 65.46  & $1,14,25$    & $1\,-\,100$ & $1.46\,-\,0.15$  & 0.014  & $0.60\,-\,0.06$ & 0.57    & RBS/WBS    &  diff &  bow  &   no	\\
    K379 & 4.48 & 4.80 & 0.76   & 1505  & 25.29  & $25.00$  & 35.56  & $11,13,21$    & $0.1\,-\,5$ & $5.15\,-\,0.73$  & 0.02   & $1.48\,-\,0.21$ & 0.43    & WBS        &  $-$  &  no   &   $-$  \\
    K383 & 4.54 & 5.95 & 46.99  & 2180  & 173.68 & $57.80$  & 183.05 & $1,13,26$    & $0.1\,-\,5$ & $4.51\,-\,0.64$  & 0.01   & $2.72\,-\,0.38$ & 1.59    & RBS/WBS    &  $-$  &  bow  &   SB2  \\
    K385 & 4.61 & 5.55 & 19.14  & 2300  & 94.75  & $87.79$  & 129.17 & $1,13,19$    & $0.1\,-\,5$ & $4.10\,-\,0.58$  & 0.008  & $2.53\,-\,0.36$ & 0.96    & RBS/WBS    &  $-$  &  bow  &   SB2  \\
    K386 & 4.39 & 5.64 & 51.39  & 1105  & 100.0  & $-3.20$  & 100.05 & $1,12,29$    & $0.1\,-\,5$ & $5.93\,-\,0.84$  & 0.017  & $3.71\,-\,0.52$ & 0.54    & WBS        &  $-$  &  bow  &   EB	\\
    K388 & 4.34 & 5.80 & 75.00  & 750   & 88.86  & $16.80$  & 90.43  & $1,13,16$    & $1\,-\,100$ & $2.33\,-\,0.23$  & 0.03   & $1.29\,-\,0.13$ & 0.67    & RBS/WBS    &  $-$  &  diff &   no	\\
    K389 & 4.39 & 4.85 & 2.97   & 1065  & 63.14  & $-$      & $-$    & $1,13 $      & $1\,-\,100$ & $1.03\,-\,0.10$  & 0.007  & $0.44\,-\,0.04$ & 0.14    & RBS/WBS    &  $-$  &  diff &   no	\\
    K634 & 4.50 & 5.06 & 3.07   & 590   & 22.20  & $-27.00$ & 34.95  & $1,13,22$    & $0.1\,-\,5$ & $7.07\,-\,1.00$  & 0.04   & $1.89\,-\,0.27$ & 0.28    & WBS        &  $-$  &  no   &   no	\\
    K692 & 4.57 & 5.28 & 1.40   & 22.95 & 23.11  & $29.23$  & 37.26  & $1,13,19$    & $1\,-\,100$ & $0.98\,-\,0.10$  & 0.007  & $0.27\,-\,0.03$ & 0.30    & RBS       &  $-$  &  bow  &   SB2  \\
    K705 & 4.43 & 5.04 & 1.22   & 1405  & 31.59  & $-$      & $-$    & $1,13 $      & $1\,-\,100$ & $2.40\,-\,0.24$  & 0.04   & $0.64\,-\,0.06$ & 0.32    & RBS/WBS    &  $-$  &  diff &   no	\\
    \hline
\end{tabular}
\tablefoot{The object K150 is embedded in diffuse emission and it is unclear whether H$\alpha$ emission associated with the infrared detection exists. A classification as RBW is also possible.}
}
\end{table}
\end{landscape}
The ISM is not homogeneous, and measured values for local densities are rare. To use value ranges that appear plausible, 
we adopt ambient ISM densities based on the local H$\alpha$ emission characteristics. 
When no discernible H$\alpha$ background is present, we assume $n_{\rm ISM} \sim 0.1-5\,\mathrm{cm^{-3}}$, characteristic of diffuse phases such as the warm neutral medium and warm ionised medium \citep[][Table~1.3]{draine2011}. 
Such densities are also compatible with values inferred from H$\alpha$ observations of runaway-star bow shocks \citep{brown2005} and with median ambient densities of a few $\mathrm{cm^{-3}}$ derived in infrared surveys \citep{peri2012,kobulnicky2018,carretero2025}. 
Conversely, in regions showing structured H$\alpha$ emission we adopt $n_{\rm ISM} \sim 1-100\,\mathrm{cm^{-3}}$, corresponding to diffuse \ion{H}{ii} gas and the cold neutral medium (\citealt[][Table~1.3]{draine2011}), consistent with densities inferred for prominent infrared bow shocks and shock-compressed gas layers \citep{kobulnicky2018,moutzouri2025}. These adopted ranges are supported by order-of-magnitude density estimates derived from the emission measure and projected H$\alpha$ width (Table.~\ref{tab:SB}). The derived $n_{\rm ISM}$ values separate into two distinct groups. Isolated objects (e.g., K383, K385, EB21) yield $4$--$8$ cm$^{-3}$, consistent within geometric uncertainties with the warm ionised medium, whereas targets embedded in bright H\,{\sc ii} regions (EB15, K065, EB12, EB23, K692) give $9$--$62$ cm$^{-3}$, consistent with the typical H\,{\sc ii}-region range of $1$--$100$ cm$^{-3}$.
With these ranges for the ISM densities, we compute the expected stand-off distances for the different categories (or regimes) and confront them to stand-off distances measured from the mid-infrared images. Since the inclination of the bow shock axis is generally unknown, the observed stand-off distance represents a lower limit to the true value, as projection effects reduce the apparent separation except in the edge-on configuration \citep{yong2018}. In cases where the measured stand-off distance does not fall within any of the predicted ranges, we adopt the nearest higher theoretical value (i.e., the smallest value exceeding the observed $R_{0}$) to assign the corresponding morphological regime.

The results are given in Table~\ref{tab:morphology}, which lists the target identifiers taken from Table~\ref{tab:TargetSelection} in Col.~(1). Col.~(2) gives the effective temperatures. Values are adopted from \citet{patten2025} where available; their spectral classifications agree in most cases with those obtained from SIMBAD within 1--2 subtypes. For the remaining objects, temperatures were estimated from the spectral types using theoretical temperature scales for O stars \citep{martins2005_1}, B supergiants \citep{2008A&A...478..823M}, and B main-sequence stars \citep{2007A&A...471..625T}. Col.~(3) lists stellar luminosities taken from the literature when available or derived from the spectral type using the tables from \citet{martins2005_1} and \citet{hohle2010}. Col.~(4) gives the stellar mass-loss rates derived using the prescription of \citet{vink2001}. Col.~(5) lists the terminal wind velocities taken from, or derived from the 
tabulations of \citet{howrath1997} or \citet{prinja1990} when available; otherwise they are estimated from the escape velocity following \citet{vink2001}. Col.~(6) gives the tangential velocities derived from the proper motions listed in Table~\ref{tab:TargetSelection}. Col.~(7) lists radial velocities adopted from the literature when available, and for binaries we use the systemic $\gamma$-velocity when available in the literature, while Col.~(8) gives the resultant space velocities. Col.~(9) lists the reference numbers corresponding to the literature sources used; all references are provided in Table~\ref{tab:reference}, where references 1--11 correspond to stellar parameters, 12--15 to terminal velocities, and 16--29 to radial velocities. Col.~(10) lists the adopted ISM density range based on the local H$\alpha$ emission characteristics. Cols.~(11)--(13) give the stand-off distances estimated for RBS, RBW/DW, and WBS using the solutions of Equations~(\ref{eq:R1}) - (\ref{eq:x}).
Col.~(14) lists the measured stand-off distances from \textit{WISE} band~4 and \textit{Spitzer} MIPS 24~$\mu$m images. Col.~(15) provides the likely morphological classification. Col.~(16) indicates the detection of bow shocks in radio wavelengths (y --- detection; no --- non-detection; diff --- diffuse emission from a background source). Col.~(17) describes the emission characteristics in the H$\alpha$ images from our survey (bow --- arc-shaped nebula detected; diff --- complex background H$\alpha$ emission; no --- no emission detected). Col.~(18) lists the binary flags: SB1 --- single-lined spectroscopic binary; SB2 --- double-lined spectroscopic binary; EB --- eclipsing binary; no --- no indication \citep[from][]{patten2025}; yes --- binary reported in SIMBAD.

\begin{table}[]
{\tiny
    \centering
    \caption{ Literature references corresponding to the sources used in Table~\ref{tab:morphology}. References 1–11 provide stellar parameters, 12–15 terminal wind velocities, and 16–29 radial velocities.}
    \begin{tabular}{ll|ll}
    \hline \hline
        No. & Ref. & No. & Ref. \\
        \hline
          1  &   \cite{patten2025}      &  16   &   \cite{gontcharov2006}    \\ 
          2  &   \cite{becker2017}      &  17   &   \cite{kharchenko2007}    \\ 
          3  &   \cite{hohle2010}       &  18   &   \cite{holgado2018}       \\ 
          4  &   \cite{martins2005_1}   &  19   &   \cite{gaia2023}          \\ 
          5  &   \cite{crowther2006}    &  20   &   \cite{mahy2010}          \\ 
          6  &   \cite{lorenzo2017}     &  21   &   \cite{hipparcos}         \\ 
          7  &   \cite{martins2005_2}   &  22   &   \cite{duflot1995}        \\ 
          8  &   \cite{mahy2022}        &  23   &   \cite{kobulnicky2022}    \\ 
          9  &   \cite{raucq2018}       &  24   &   \cite{zehe2018}          \\ 
          10  &   \cite{bouret2012}     &  25   &   \cite{abt1963}           \\ 
          11  &   \cite{papics2011}     &  26   &   \cite{williams2011}      \\ 
          12  &   \cite{howrath1997}    &  27   &   \cite{jonsson2020}       \\ 
          13  &   \cite{prinja1990}     &  28   &   \cite{mossoux2018}       \\ 
          14  &   \cite{vink2001}       &  29   &   \cite{pourbaix2004}      \\ 
          15  &   \cite{peri2012}       &       &                            \\ 
        \hline
    \end{tabular}
    \label{tab:reference}
    }
\end{table}

\section{Discussion}
\label{sec:discussion}

This work constitutes the first targeted H$\alpha$ survey of infrared bow shock nebulae, motivated by the detections reported by \cite{brown2005}. Previous searches have relied primarily on all-sky or wide-field surveys such as SHASSA and VTSS, which are constrained by low spatial resolution and short exposure times. The present survey was designed to overcome these limitations.  Within the sample of 78 targets, we obtain clear detections of bow shocks in H$\alpha$ for approximately 19\% of the objects. Another 45\% of the targets are located within regions of complex or diffuse H$\alpha$ emission, where the presence of structures related to bow shocks cannot be unambiguously assessed. The remaining 36\% show no detectable emission associated with the infrared bow shock candidates. For these non-detections, we note that deeper observations may still reveal faint or extended structures.

In the following, we compare our results with those of previous studies performed either in H$\alpha$ or in different wavelength regimes, and we start with the eight objects, for which \cite{brown2005} reported H$\alpha$ detections. We identify clear 
bow-shock structures in H$\alpha$ associated with four of their objects (see Fig.~\ref{Detections1}), namely BB1: HD\,17505, BB4: HD\,57061, BB5: HD\,92206, and EB21: HD\,149757 (=$\zeta$~Oph). In particular, \cite{brown2005} previously classified the nebula around BB5 as small, noting the presence of a bright surrounding H\,{\sc ii} region. Our higher-resolution H$\alpha$ image resolves an arc-like emission structure embedded within this bright H\,{\sc ii} region, whose morphology closely matches the emission observed in the MIPS 24\,$\mu$m band, supporting its interpretation as a bow-shock feature. Two additional targets, EB13: HD\,48099 and BB8: HD\,158186, are located within regions of complex background emission, which hampers the unambiguous identification of bow shock–related H$\alpha$ emission. Finally, in two cases, BB2: HD\,24431 and BB6: HD\,135240, arc-like structures are visible in H$\alpha$ images, but the objects lack a corresponding mid-infrared counterpart. Instead, they were detected on $60\,\mu$m images from the IRAS mission indicating that these objects are associated with cooler ambient dust.
It is also worth mentioning that \cite{peri2012} searched for bow-shock emission in H$\alpha$ using data from the same surveys (SHASSA and VTSS) as \cite{brown2005}. For the 28 targets with mid-infrared bow-shock detections reported in their first E-BOSS release, they did not identify any convincing bow-shock candidates. In contrast, our H$\alpha$ survey reveals four clear detections associated with EB06: HD\,30614, EB15: HD\,64315, EB23: HD\,165319, and the already mentioned object EB21. 
The bow shock in H$\alpha$ connected to the object EB23 was also found by \citet{2008A&A...490.1071G} based on images from the SuperCOSMOS H-alpha Survey \citep{2005MNRAS.362..689P}. The morphology they show in their Fig.\,4 largely agrees with our observations.

Although the H$\alpha$ and mid-infrared morphologies generally trace the same overall structure, we observe the infrared arcs to be located in many cases closer to the central star as mentioned in Sect.~\ref{sec:BSP} (Fig.\,\ref{Fig:inner_annulus}). This offset likely reflects differences in dust–gas coupling, with small grains tightly coupled to the gas and therefore closely following the ionised material traced by H$\alpha$. Larger grains, by contrast, possess greater momentum and are more resistant to drag forces, resulting in longer deceleration timescales and dust distributions that progressively deviate from the gas morphology, producing spatial offsets between infrared and H$\alpha$ emission \citep{vanMarle2011}.
We also observe that the widths of the infrared structures are generally broader than their H$\alpha$ counterparts (see Fig.\,\ref{Fig:scatter_width}).
While the lower spatial resolution and coarser point spread functions of the mid-infrared images lead to an apparent broader width of the nebulae compared to our higher resolution H$\alpha$ images, this effect can only partially account for the diverging widths of the structures.
Therefore, we explore the emission mechanisms of both ionised gas and dust. The surface brightness of recombination lines such as H$\alpha$ scales with the emission measure, $\mathrm{EM} \propto n_{\mathrm e}^2 $, and is therefore highly sensitive to gas compression in shocked regions \citep{osterbrock2006}. In contrast, mid-infrared dust emission depends primarily on the dust column density and grain temperature with the latter following from the balance between  heating of the dust particles by the stellar radiation and cooling by their thermal emission 
\citep{draine2011, 2018MNRAS.473.1576K}.
Consequently, dust emission traces a broader region around the bow shock, producing wider and more diffuse infrared arcs, while the H$\alpha$ emission peaks at the densest post-shock regions and appears correspondingly thinner, as seen in our survey. 

\begin{figure*}
    \centering
    \includegraphics[width=0.8\linewidth]{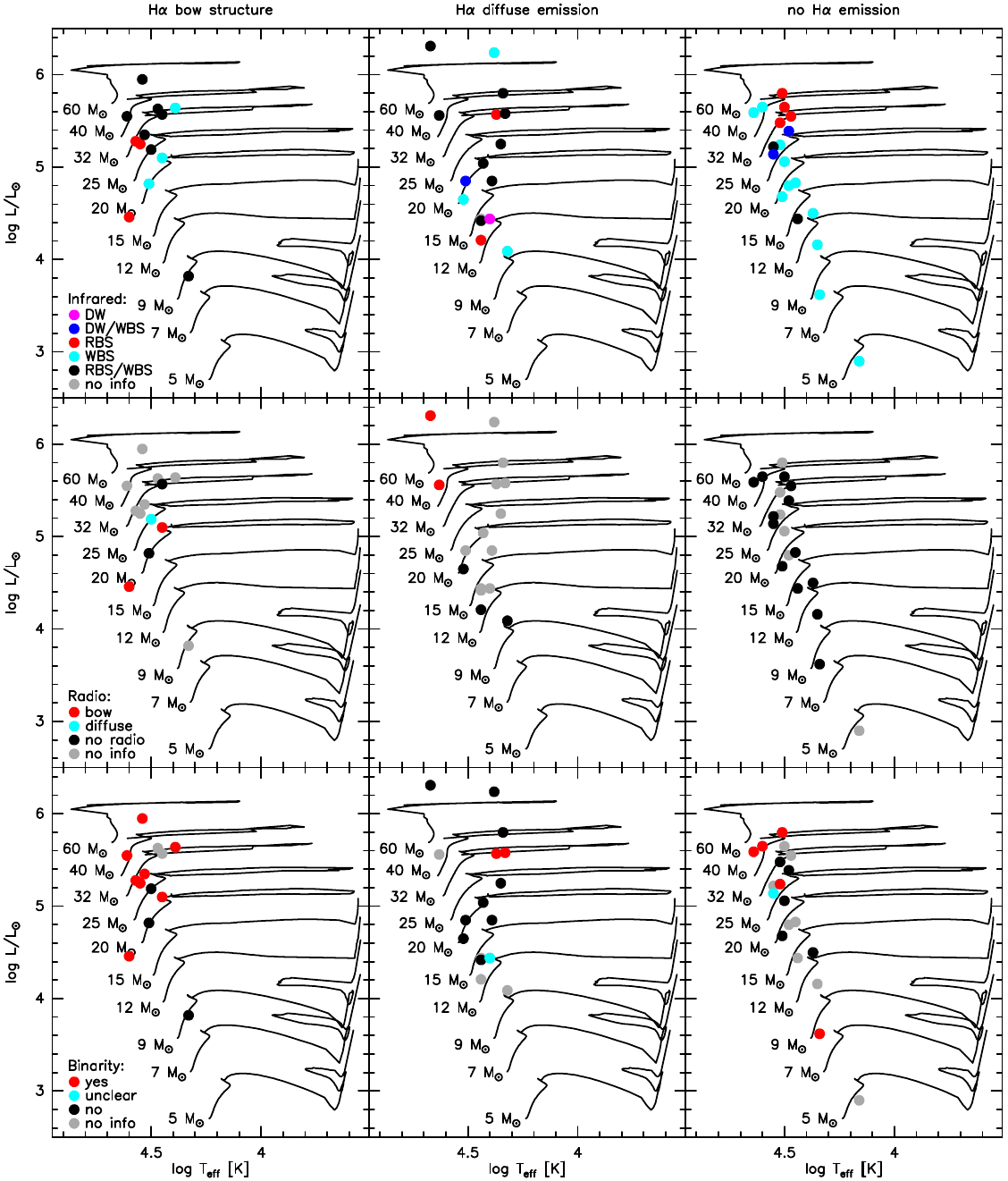}
    \caption{Comparison of the H$\alpha$ detections (bows -- left column, diffuse emission -- middle column, no emission -- right column) with the infrared classification (top row), the presence of radio emission (middle row) and the information on companions (bottom row). The shown evolutionary tracks are for rotating stars with solar metallicity taken from \citet{2012A&A...537A.146E}.}
    \label{Fig:BS_statistics}
\end{figure*}

Bow shock nebulae have also been investigated at radio wavelengths, providing a complementary diagnostic of particle acceleration and non-thermal emission. The first detection of synchrotron radiation from a stellar bow shock was reported for EB27: BD\,+43\,3654 by \cite{benaglia2010,benaglia2021}. In our H$\alpha$ survey, this source is embedded within bright emission from an H {\sc ii} region, which hampers a clear optical identification of the bow shock. In the second E-BOSS release, \cite{peri2015} reported radio features spatially coincident with infrared arcs for EB03: HD\,15629, EB15: HD\,64315, and EB23: HD\,165319. Of these, EB03 is affected by strong background H$\alpha$ emission, whereas EB15 and EB23 are clearly detected in H$\alpha$, reinforcing their interpretation as bow shock nebulae. The study by \citet{eijnden2022} contained 19 of our targets selected from the E-BOSS catalogue. Of these, two were the sources with previously detected radio emission (EB15 and EB23) for which the authors confirmed the emission related to the infrared bow shocks. They also found
radio emission towards EB07: HD\,34078 for which they conclude that it originates from an unrelated background source, and our H$\alpha$ images show that the target is embedded in bright background emission. The remaining 16 
objects were not detected in the radio images used by \citet{eijnden2022}. These are the objects EB21, for which we clearly detect an H$\alpha$ bow, and EB12: HD\,47432, which displays a faint H$\alpha$ arc (Fig.\,\ref{Detections1}), EB09 and EB11 which both display diffuse H$\alpha$ emission (Figs.~\ref{DE:T31}-\ref{DE:IPHAS}), and EB04, EB08, 
EB13, EB14, EB16, EB17, EB18, EB19, EB20, EB22, EB24, and EB25, which also have no H$\alpha$ emission. With the follow-up, deeper radio survey of 24 northern bow shock nebulae conducted by \citet{moutzouri2025} we have nine objects in common. Of these, K072 and EB27 (detected in earlier surveys) were reported to be radio bow-shock detections,  EB23, K159, K359 were marginal detections, and EB01, EB09, EB12, 
and EB21 were not detected.
For K072: G\,018.2660$-$00.2988, our H$\alpha$ images (Fig.~\ref{DE:DK154andPerek}) show that the brightest emission is spatially offset from the infrared bow shock and instead associated with a larger nebular structure
also known as the infrared dust bubble N22 \citep{2012A&A...544A..39J}. Closer inspection of the  different images suggests that the radio emission  better aligns with our H$\alpha$ emission rather than with the dust bow. This calls the radio bow-shock interpretation into question as was also noted by \citet{moutzouri2025}. Of the marginal radio detections, EB23 (as mentioned above) and K359: HD\,215806 are clearly detected in H$\alpha$, whereas K159: G\,031.6770$+$00.1775 remains undetected and likely requires deeper observations, although \citet{moutzouri2025} interpret their radio signal as just diffuse free-free emission from the H\,{\sc ii} region. Among the radio non-detections, EB01: HD\,2083 is also undetected in H$\alpha$ and is likely a dust-wave candidate
whereas EB09: HD\,37032 is dominated by strong background emission.
As mentioned above, the arc seen in H$\alpha$ around EB12 is rather faint and longer radio exposures might be needed to detect the structure.
The last object from this survey with no radio detection is EB21 (= $\zeta$\,Oph), which has already been investigated by \citet{eijnden2022}. The non-detection is surprising given that we see a clear bow structure in H$\alpha$ (Fig.\,\ref{Detections1}) and simulations predict significant  
thermal radio emission at 6\,GHz \citep{green2022}. However, as mentioned by \citet{moutzouri2025}, the predictions of a radio flux of $35$\,kJy\,sr$^{-1}$ by \citet{green2022} were about twice as high as their $3\sigma$ upper limit measurement. Our estimate is with $\sim 10$\,kJy\,sr$^{-1}$ (Table\,\ref{tab:SB}) even lower and in agreement with the non-detection. Information about all radio (non-) detections is included in Table\,\ref{tab:morphology}.

Detailed investigations of possible physical parameters of the bow-shocks in H$\alpha$ are so far rare. In this respect, it is worth mentioning that the environment of EB21 was studied by \citet{gvaramadze2012}, who characterized the physical conditions of the H\,{\sc ii} region surrounding the object as an almost circular photoionised bubble with a diameter of approximately $10^{\circ}$, corresponding to $\sim 9.6$\,pc at a distance of $112$\,pc and derived the gas density to be $\sim 3$\,cm$^{-3}$ and an emission measure of $90$\,cm$^{-6}$\,pc. \citet{green2022}, based on their MHD simulations,  predicted that the bow shock must have a sufficiently high emission measure to be detectable above the background, and explicitly called for new narrow-band H$\alpha$ observations. In their models, the contact discontinuity was found to be located at $0.11-0.13$\,pc and the forward shock at $0.22-0.26$\,pc, with the peak H$\alpha$ emission expected at the contact discontinuity. Our H$\alpha$ images resolve a bow shock around $\zeta$\,Oph (Fig.~\ref{Detections1}) at a stand-off distance of $0.17$\,pc and a width of $\sim 0.1$\,pc, in close agreement with their models.

The mass-loss rates adopted in Table~\ref{tab:morphology} use the \citet{vink2001} prescription, which overpredicts $\dot{M}$ for late O- and B-type dwarfs by up to two orders of magnitude \citep[e.g.][]{puls2008}. For $\zeta$\,Oph (EB21), the UV-derived rate is $\dot{M} \approx 1.6 \times 10^{-9}\,M_\odot\,{\rm yr}^{-1}$ \citep{marcolino2009}, roughly 100 times lower than our adopted value, while the rate from its H\,{\sc ii} region and bow-shock radii is $\sim$10 times lower \citep{gvaramadze2012}. Using the \citet{marcolino2009} mass-loss rate for EB21, the expected WBS stand-off distance drops from $4.10-0.48$\,pc to $0.33-0.05$\,pc, in much better agreement with the observed $0.19$\,pc, though the classification remains WBS. More generally, lowering $\dot{M}$ reduces $\eta_w$ (Eq.~\ref{eq:windeff}) and the WBS stand-off distance ($R_0 \propto \eta_w^{1/2}$), while the RBS and DW stand-off distances are not affected significantly. The values of predicted WBS stand-off distances in Table~\ref{tab:morphology} should then be considered only as upper limits. Hence, a lower mass-loss rate increases the role of radiation pressure in supporting the shell, which can shift borderline objects into the radiation-supported regime.

Another bow-schock object that was studied numerically is $\lambda$~Cep = HD\,210839 (BS07). 
The object has been observed in H$\alpha$ as part of the IPHAS survey \citep{2005MNRAS.362..753D} and a faint arc was detected and interpreted by \citet{2015A&A...576A..97S} as bow shock.
Using 3D MHD simulations and the parameters of the H$\alpha$ arc, \cite{scherer2020} modeled the astrosphere of $\lambda$~Cephei and found the resulting structure to be approximately axially symmetric and with a distance of the bow shock of 1.1\,pc. They also interpret the H$\alpha$ emission to be related to regions where the shocked ISM density is highest.
Despite multiple extensive exposures of the structure, we did not detect any significant H$\alpha$ signal aligned with the infrared bow structure (Fig.\,\ref{DE:SBT}) but mainly diffuse background emission\footnote{A faint arc structure is visible in H$\alpha$ and on the WISE image to the south-east of the object, in agreement with what was reported by \citet{2015A&A...576A..97S}, but based on the full infrared structure, we think its interpretation as bow shock is questionable.}.

Three bow shock candidates were reported in the study of runaway stars from the young star cluster NGC 6611 \citep{2008A&A...490.1071G}. One of them was associated with the object EB23. These authors proposed that the objects might have been ejected from the cluster based on the morphology of the detected structures and proper motion arguments.

From a sample of 23 targets comprising 20 objects from \cite{kobulnicky2018} and three additional sources $\theta^{1}$~Ori~D, LP~Ori, and $\sigma$~Ori, \cite{henney2019_3} identified four strong and three marginal candidates for radiation supported bow waves. While our survey does not include any of the objects classified as dust waves or bow waves in their study, our H$\alpha$ analysis identifies three candidates: EB01, EB24, and K150.
EB01 and EB24 are strong dust wave candidates, as no H$\alpha$ emission associated with the infrared arcs is detected, although deeper H$\alpha$ observations may still reveal faint structures. The classification of K150 remains uncertain due to its location within bright background H$\alpha$ emission. The targets common to both studies EB21, K067, and EB27 are classified as wind supported bow shocks by \cite{henney2019_3}, a conclusion that is also supported by our calculations. Of these, EB21 shows an associated arc like structure in H$\alpha$, whereas K067 and EB27 are dominated by background emission. Notably, the dust wave candidates identified in our work span a wide range of stand off distances from $0.04$ to $1.35$\,pc, in contrast to the small stand off distances below 0.1 pc reported by \cite{henney2019_3}. These candidates are further characterized by low stellar velocities relative to the ISM below $30$\,km\,s$^{-1}$ and weak mass loss rates of $4$ to $8.7 \times 10^{-8}$\,M$_\odot$yr$^{-1}$.

In Fig.~\ref{Fig:BS_statistics} we plot the sample stars with information on both effective temperature and luminosity (Table~\ref{tab:morphology}) along with stellar evolutionary tracks of rotating stars with solar metallicity from \citet{2012A&A...537A.146E}. The panels in the left column refer to objects with a bow structure detected in H$\alpha$, whereas the middle and right columns are for objects with diffuse H$\alpha$ emission and non-detections, respectively. To search for possible correlations with other properties of the objects, we 
color them according to the assigned infrared class of bow shocks (top row), the information on radio emission (middle row) and binarity (bottom row). The information on the radio (non-)detections as described above and on binarity \citep[from][and from SIMBAD]{patten2025} is included in Table~\ref{tab:morphology}.

Only 13 of our objects with H$\alpha$ bow structures have reported stellar parameters. With the exception of K065 and the binary star EB15, H$\alpha$ bow structures are seen around stars more massive than $20$\,M$_{\odot}$. The two exceptions are both embedded in dense \ion{H}{ii} regions. K065 lies in the Eagle Nebula (M16) associated with the young cluster NGC\,6611 \citep{hillenbrand1993,guarcello2007}, whereas EB15 resides in the \ion{H}{ii} region S311 within the NGC\,2467 complex and acts as one of its dominant ionizing sources \citep{yadav2016,lorenzo2017}. For both targets, the proper-motion vectors deviate significantly from the symmetry axis of the bow-shock nebulae, suggesting a weather-vane scenario. Four objects appear to be evolved, of which two are binaries. The binary fraction in this sample is generally high ($\sim 58\%$). So far, only five objects have been observed at radio wavelengths, with two clear detections of bow structures.

From the objects with diffuse H$\alpha$ emission, 16 were found to have stellar parameters. Half of this sample are evolved stars, of which two are in binaries and only one of them was observed in the radio regime with a clear bow structure detected. 
A second target with a radio bow structure seems to be on the main sequence.
The majority of objects in this sample are presumably single stars.

A large portion (19) of objects with known stellar parameters was not detected in our H$\alpha$ survey. While 13 objects of this sample have been observed at radio wavelengths, none was detected. This is the most clear correlation that we observe. 
The sample is composed mostly of main-sequence stars spreading in mass from $5-40$\,M$_{\odot}$, and the majority of them has a classical WBS detected in the infrared. 
Three of the four evolved objects have RBS, and with initial masses in the range $30$ to $40$M$_{\odot}$, it is surprising that these objects have no detectable H$\alpha$ counterparts. The binary fraction in this H$\alpha$ featureless sample is about 
$26\%$ in total, but increases to $45\%$ when excluding objects with unknown status.

Finally, we would like to briefly address the question of whether the arc-like structures detected by the narrow-band H$\alpha$ filter might have been mistakenly interpreted as bow shocks, even though they could also be reflection nebulae. From our comparison of all bow-structure objects with the DSS2 blue images, 
no arc-like counterpart is visible for EB06, EB12, BB4, BS11, K359, K383, K385, and K386.
In addition, we found that two objects, EB15 and BB5 are embedded within extended H\,{\sc ii} regions and show widespread blue emission, but no distinct arc-shaped feature corresponding to the H$\alpha$ morphology. Therefore, we consider these as real detections of bow-shock nebulae with H$\alpha$ emission. For the remaining objects, the situation is less clear. For EB21, the nebular region is saturated by stellar emission in the DSS2 image, preventing a meaningful assessment, whereas arc-like structures are visible for K065, K692, BB1, N1, and EB23. Whether these are due to reflection nebulae or maybe contribution from intense [O\,{\sc iii}] emission from the nebula as in the case of $\zeta$~Oph \citep{gull1979} cannot be finally distinguished. These objects clearly require follow-up observations to either confirm or discard their classification as H$\alpha$-emitting bow shocks.

\section{Conclusions}
\label{sec:conclusions}

We presented the first targeted H$\alpha$ imaging survey of infrared-selected bow shocks around massive OB-type stars. Our sample comprises 78 objects selected from several extensive infrared bow-shock catalogues. Among these targets, we identify clear arc-shaped nebulae in H$\alpha$ for 15 objects, and we additionally report one previously unrecognized bow-shock detection found serendipitously in a wide-field frame. A further 35 objects show diffuse H$\alpha$ emission associated with complex background structures, while 28 objects remain undetected in H$\alpha$. Most of the systems in our sample are associated with runaway stars, although a few objects may instead reflect environmental shaping consistent with the weather-vane scenario, in which large-scale flows in the surrounding ISM determine the orientation of the bow structure.

For the detected systems, we derived geometrical parameters of the bow-shock nebulae from both H$\alpha$ and mid-infrared images. The large-scale morphologies traced by the ionised gas and dust are generally consistent, although systematic differences are observed. In particular, the infrared arcs tend to lie slightly closer to the central star and appear broader than the corresponding H$\alpha$ structures. These differences likely reflect the distinct physical processes governing dust and ionised gas emission in the shocked region.

Using available stellar and wind parameters from the literature, we explored the relative roles of stellar wind momentum and radiation pressure in supporting the bow-shock structures. By comparing the observed stand-off distances with theoretical predictions for plausible ranges of ambient ISM density, several objects appear consistent with radiation-supported bow waves or dust-wave scenarios, while others remain compatible with classical wind-supported bow shocks.

No clear correlation is found between H$\alpha$ detections and the stellar physical parameters or binarity, although a relatively high binary fraction is present among the targets. Where radio observations are available, objects undetected in H$\alpha$ are also generally undetected at radio wavelengths.

The primary aim of this survey was to establish the presence of ionised bow structures associated with infrared bow-shock candidates. Future follow-up optical spectroscopy will be required to determine the physical conditions of the ionised gas, including density, temperature, and kinematics, and to compare these properties with theoretical models and numerical simulations. It would also be instructive to collect information about possible radio counterparts of the objects for which our survey detected H$\alpha$ bow structures. To facilitate radio observations, we provided (where possible) measurements of surface brightness values in H$\alpha$, emission measures, as well as estimates for expected thermal radio emission at $6$\,GHz.

\begin{acknowledgements}
We would like to thank the reviewer for valuable comments and suggestions, which have helped improve the manuscript. Based on observations taken with the Perek 2-m telescope and the Small Binocular Telescope (SBT) at Ond\v{r}ejov Observatory. The Digitized Sky Survey was produced at the Space Telescope Science Institute under U.S. Government grant NAG W–2166. The images of these surveys are based on photographic data obtained using the Oschin Schmidt Telescope on Palomar Mountain and the UK Schmidt Telescope. The plates were processed into the present compressed digital form with the permission of these institutions.
This research made use of the NASA Astrophysics Data System (ADS) and the SIMBAD database, operated at CDS, Strasbourg, France. K.S.S., M.K., D.H.N. and O.V.M. acknowledge
financial support from the Czech Science Foundation (GA \v{C}R, grant number 25-17532S).
The Astronomical Institute of the Czech Academy of Sciences is supported by the project RVO:67985815.
The project is co-funded by the European Union (Project 101183150 - OCEANS) and supported by the Czech Ministry of Education, Youth and Sports (MEYS) (Project No. CZ.02.01.01/00/22\_008/0004632 -- FORTE). 
\end{acknowledgements}

   \bibliographystyle{aa} % style aa.bst
   \bibliography{bow-shocks} % your references Yourfile.bib

@ARTICLE{brown2005,
       author = {{Brown}, D. and {Bomans}, D.~J.},
        title = "{To see or not to see a bow shock. Identifying bow shocks with H{\ensuremath{\alpha}} allsky surveys}",
      journal = {\aap},
         year = 2005,
        month = aug,
       volume = {439},
       number = {1},
        pages = {183-194},
          doi = {10.1051/0004-6361:20041054},
archivePrefix = {arXiv},
       eprint = {astro-ph/0505098},
 primaryClass = {astro-ph},
       adsurl = {https://ui.adsabs.harvard.edu/abs/2005A&A...439..183B}
}

@ARTICLE{2001PASP..113.1326G,
       author = {{Gaustad}, John E. and {McCullough}, Peter R. and {Rosing}, Wayne and {Van Buren}, Dave},
        title = "{A Robotic Wide-Angle H{\ensuremath{\alpha}} Survey of the Southern Sky}",
      journal = {\pasp},
         year = 2001,
        month = nov,
       volume = {113},
       number = {789},
        pages = {1326-1348},
          doi = {10.1086/323969},
archivePrefix = {arXiv},
       eprint = {astro-ph/0108518},
 primaryClass = {astro-ph},
       adsurl = {https://ui.adsabs.harvard.edu/abs/2001PASP..113.1326G}
}

@ARTICLE{2008ApJ...689..242P,
       author = {{Povich}, Matthew S. and {Benjamin}, Robert A. and {Whitney}, Barbara A. and {Babler}, Brian L. and {Indebetouw}, R{\'e}my and {Meade}, Marilyn R. and {Churchwell}, Ed},
        title = "{Interstellar Weather Vanes: GLIMPSE Mid-Infrared Stellar Wind Bow Shocks in M17 and RCW 49}",
      journal = {\apj},
         year = 2008,
        month = dec,
       volume = {689},
       number = {1},
        pages = {242-248},
          doi = {10.1086/592565},
archivePrefix = {arXiv},
       eprint = {0808.2168},
 primaryClass = {astro-ph},
       adsurl = {https://ui.adsabs.harvard.edu/abs/2008ApJ...689..242P}
}

@ARTICLE{kobulnicky2017,
       author = {{Kobulnicky}, Henry A. and {Schurhammer}, Danielle P. and {Baldwin}, Daniel J. and {Chick}, William T. and {Dixon}, Don M. and {Lee}, Daniel and {Povich}, Matthew S.},
        title = "{Infrared Photometric Properties of 709 Candidate Stellar Bowshock Nebulae}",
      journal = {\aj},
         year = 2017,
        month = nov,
       volume = {154},
       number = {5},
          eid = {201},
        pages = {201},
          doi = {10.3847/1538-3881/aa90ba},
archivePrefix = {arXiv},
       eprint = {1710.07892},
 primaryClass = {astro-ph.SR},
       adsurl = {https://ui.adsabs.harvard.edu/abs/2017AJ....154..201K}
}

@ARTICLE{2012A&A...544A..39J,
       author = {{Ji}, W.-G. and {Zhou}, J.-J. and {Esimbek}, J. and {Wu}, Y.-F. and {Wu}, G. and {Tang}, X.-D.},
        title = "{The infrared dust bubble N22: an expanding H ii region and the star formation around it}",
      journal = {\aap},
         year = 2012,
        month = aug,
       volume = {544},
          eid = {A39},
        pages = {A39},
          doi = {10.1051/0004-6361/201218861},
archivePrefix = {arXiv},
       eprint = {1206.2762},
 primaryClass = {astro-ph.GA},
       adsurl = {https://ui.adsabs.harvard.edu/abs/2012A&A...544A..39J}
}

@ARTICLE{kobulnicky2018,
       author = {{Kobulnicky}, Henry A. and {Chick}, William T. and {Povich}, Matthew S.},
        title = "{Demonstration of a Novel Method for Measuring Mass-loss Rates for Massive Stars}",
      journal = {\apj},
         year = 2018,
        month = mar,
       volume = {856},
       number = {1},
          eid = {74},
        pages = {74},
          doi = {10.3847/1538-4357/aab3e0},
archivePrefix = {arXiv},
       eprint = {1803.02794},
 primaryClass = {astro-ph.SR},
       adsurl = {https://ui.adsabs.harvard.edu/abs/2018ApJ...856...74K}
}

@ARTICLE{kobulnicky2016,
       author = {{Kobulnicky}, Henry A. and {Chick}, William T. and {Schurhammer}, Danielle P. and {Andrews}, Julian E. and {Povich}, Matthew S. and {Munari}, Stephan A. and {Olivier}, Grace M. and {Sorber}, Rebecca L. and {Wernke}, Heather N. and {Dale}, Daniel A. and {Dixon}, Don M.},
        title = "{A Comprehensive Search for Stellar Bowshock Nebulae in the Milky Way: A Catalog of 709 Mid-infrared Selected Candidates}",
      journal = {\apjs},
         year = 2016,
        month = dec,
       volume = {227},
       number = {2},
          eid = {18},
        pages = {18},
          doi = {10.3847/0067-0049/227/2/18},
archivePrefix = {arXiv},
       eprint = {1609.02204},
 primaryClass = {astro-ph.SR},
       adsurl = {https://ui.adsabs.harvard.edu/abs/2016ApJS..227...18K}
}

@ARTICLE{peri2015,
       author = {{Peri}, C.~S. and {Benaglia}, P. and {Isequilla}, N.~L.},
        title = "{E-BOSS: An Extensive stellar BOw Shock Survey. II. Catalogue second release}",
      journal = {\aap},
         year = 2015,
        month = jun,
       volume = {578},
          eid = {A45},
        pages = {A45},
          doi = {10.1051/0004-6361/201424676},
archivePrefix = {arXiv},
       eprint = {1504.04264},
 primaryClass = {astro-ph.SR},
       adsurl = {https://ui.adsabs.harvard.edu/abs/2015A&A...578A..45P}
}

@ARTICLE{meyer2014,
       author = {{Meyer}, D.~M. -A. and {Mackey}, J. and {Langer}, N. and {Gvaramadze}, V.~V. and {Mignone}, A. and {Izzard}, R.~G. and {Kaper}, L.},
        title = "{Models of the circumstellar medium of evolving, massive runaway stars moving through the Galactic plane}",
      journal = {\mnras},
         year = 2014,
        month = nov,
       volume = {444},
       number = {3},
        pages = {2754-2775},
          doi = {10.1093/mnras/stu1629},
archivePrefix = {arXiv},
       eprint = {1408.2828},
 primaryClass = {astro-ph.SR},
       adsurl = {https://ui.adsabs.harvard.edu/abs/2014MNRAS.444.2754M}
}

@ARTICLE{van1995,
       author = {{van Buren}, Dave and {Noriega-Crespo}, Alberto and {Dgani}, Ruth},
        title = "{An IRAS/ISSA Survey of Bow Shocks Around Runaway Stars}",
      journal = {\aj},
         year = 1995,
        month = dec,
       volume = {110},
        pages = {2914},
          doi = {10.1086/117739},
       adsurl = {https://ui.adsabs.harvard.edu/abs/1995AJ....110.2914V}
}

@ARTICLE{van1988,
       author = {{van Buren}, Dave and {McCray}, Richard},
        title = "{Bow Shocks and Bubbles Are Seen around Hot Stars by IRAS}",
      journal = {\apjl},
         year = 1988,
        month = jun,
       volume = {329},
        pages = {L93},
          doi = {10.1086/185184},
       adsurl = {https://ui.adsabs.harvard.edu/abs/1988ApJ...329L..93V}
}

@ARTICLE{peri2012,
       author = {{Peri}, C.~S. and {Benaglia}, P. and {Brookes}, D.~P. and {Stevens}, I.~R. and {Isequilla}, N.~L.},
        title = "{E-BOSS: an Extensive stellar BOw Shock Survey. I. Methods and first catalogue}",
      journal = {\aap},
         year = 2012,
        month = feb,
       volume = {538},
          eid = {A108},
        pages = {A108},
          doi = {10.1051/0004-6361/201118116},
archivePrefix = {arXiv},
       eprint = {1109.3689},
 primaryClass = {astro-ph.SR},
       adsurl = {https://ui.adsabs.harvard.edu/abs/2012A&A...538A.108P}
}

@ARTICLE{noriega1997,
       author = {{Noriega-Crespo}, Alberto and {van Buren}, Dave and {Dgani}, Ruth},
        title = "{Bow Shocks Around Runaway Stars.III.The High Resolution Maps}",
      journal = {\aj},
         year = 1997,
        month = feb,
       volume = {113},
        pages = {780-786},
          doi = {10.1086/118298},
       adsurl = {https://ui.adsabs.harvard.edu/abs/1997AJ....113..780N}
}

@ARTICLE{tetzlaff2010,
       author = {{Tetzlaff}, N. and {Neuh{\"a}user}, R. and {Hohle}, M.~M. and {Maciejewski}, G.},
        title = "{Identifying birth places of young isolated neutron stars}",
      journal = {\mnras},
         year = 2010,
        month = mar,
       volume = {402},
       number = {4},
        pages = {2369-2387},
          doi = {10.1111/j.1365-2966.2009.16093.x},
archivePrefix = {arXiv},
       eprint = {0911.4441},
 primaryClass = {astro-ph.GA},
       adsurl = {https://ui.adsabs.harvard.edu/abs/2010MNRAS.402.2369T}
}

@ARTICLE{henney2019_1,
       author = {{Henney}, William J. and {Arthur}, S.~J.},
        title = "{Bow shocks, bow waves, and dust waves - I. Strong coupling limit}",
      journal = {\mnras},
         year = 2019,
        month = jul,
       volume = {486},
       number = {3},
        pages = {3423-3433},
          doi = {10.1093/mnras/stz1043},
archivePrefix = {arXiv},
       eprint = {1903.03737},
 primaryClass = {astro-ph.SR},
       adsurl = {https://ui.adsabs.harvard.edu/abs/2019MNRAS.486.3423H}
}

@ARTICLE{henney2019_2,
       author = {{Henney}, William J. and {Arthur}, S.~J.},
        title = "{Bow shocks, bow waves, and dust waves - II. Beyond the rip point}",
      journal = {\mnras},
         year = 2019,
        month = jul,
       volume = {486},
       number = {3},
        pages = {4423-4442},
          doi = {10.1093/mnras/stz1130},
archivePrefix = {arXiv},
       eprint = {1903.07774},
 primaryClass = {astro-ph.SR},
       adsurl = {https://ui.adsabs.harvard.edu/abs/2019MNRAS.486.4423H}
}

@ARTICLE{bertoldi1996,
       author = {{Bertoldi}, Frank and {Draine}, B.~T.},
        title = "{Nonequilibrium Photodissociation Regions: Ionization-Dissociation Fronts}",
      journal = {\apj},
         year = 1996,
        month = feb,
       volume = {458},
        pages = {222},
          doi = {10.1086/176805},
archivePrefix = {arXiv},
       eprint = {astro-ph/9508067},
 primaryClass = {astro-ph},
       adsurl = {https://ui.adsabs.harvard.edu/abs/1996ApJ...458..222B}
}

@ARTICLE{werner2004,
       author = {{Werner}, M.~W. and {Roellig}, T.~L. and {Low}, F.~J. and {Rieke}, G.~H. and {Rieke}, M. and {Hoffmann}, W.~F. and {Young}, E. and {Houck}, J.~R. and {Brandl}, B. and {Fazio}, G.~G. and {Hora}, J.~L. and {Gehrz}, R.~D. and {Helou}, G. and {Soifer}, B.~T. and {Stauffer}, J. and {Keene}, J. and {Eisenhardt}, P. and {Gallagher}, D. and {Gautier}, T.~N. and {Irace}, W. and {Lawrence}, C.~R. and {Simmons}, L. and {Van Cleve}, J.~E. and {Jura}, M. and {Wright}, E.~L. and {Cruikshank}, D.~P.},
        title = "{The Spitzer Space Telescope Mission}",
      journal = {\apjs},
         year = 2004,
        month = sep,
       volume = {154},
       number = {1},
        pages = {1-9},
          doi = {10.1086/422992},
archivePrefix = {arXiv},
       eprint = {astro-ph/0406223},
 primaryClass = {astro-ph},
       adsurl = {https://ui.adsabs.harvard.edu/abs/2004ApJS..154....1W}
}

@ARTICLE{wright2010,
       author = {{Wright}, Edward L. and {Eisenhardt}, Peter R.~M. and {Mainzer}, Amy K. and {Ressler}, Michael E. and {Cutri}, Roc M. and {Jarrett}, Thomas and {Kirkpatrick}, J. Davy and {Padgett}, Deborah and {McMillan}, Robert S. and {Skrutskie}, Michael and {Stanford}, S.~A. and {Cohen}, Martin and {Walker}, Russell G. and {Mather}, John C. and {Leisawitz}, David and {Gautier}, III, Thomas N. and {McLean}, Ian and {Benford}, Dominic and {Lonsdale}, Carol J. and {Blain}, Andrew and {Mendez}, Bryan and {Irace}, William R. and {Duval}, Valerie and {Liu}, Fengchuan and {Royer}, Don and {Heinrichsen}, Ingolf and {Howard}, Joan and {Shannon}, Mark and {Kendall}, Martha and {Walsh}, Amy L. and {Larsen}, Mark and {Cardon}, Joel G. and {Schick}, Scott and {Schwalm}, Mark and {Abid}, Mohamed and {Fabinsky}, Beth and {Naes}, Larry and {Tsai}, Chao-Wei},
        title = "{The Wide-field Infrared Survey Explorer (WISE): Mission Description and Initial On-orbit Performance}",
      journal = {\aj},
         year = 2010,
        month = dec,
       volume = {140},
       number = {6},
        pages = {1868-1881},
          doi = {10.1088/0004-6256/140/6/1868},
archivePrefix = {arXiv},
       eprint = {1008.0031},
 primaryClass = {astro-ph.IM},
       adsurl = {https://ui.adsabs.harvard.edu/abs/2010AJ....140.1868W}
}

@ARTICLE{fazio2004,
       author = {{Fazio}, G.~G. and {Hora}, J.~L. and {Allen}, L.~E. and {Ashby}, M.~L.~N. and {Barmby}, P. and {Deutsch}, L.~K. and {Huang}, J.-S. and {Kleiner}, S. and {Marengo}, M. and {Megeath}, S.~T. and {Melnick}, G.~J. and {Pahre}, M.~A. and {Patten}, B.~M. and {Polizotti}, J. and {Smith}, H.~A. and {Taylor}, R.~S. and {Wang}, Z. and {Willner}, S.~P. and {Hoffmann}, W.~F. and {Pipher}, J.~L. and {Forrest}, W.~J. and {McMurty}, C.~W. and {McCreight}, C.~R. and {McKelvey}, M.~E. and {McMurray}, R.~E. and {Koch}, D.~G. and {Moseley}, S.~H. and {Arendt}, R.~G. and {Mentzell}, J.~E. and {Marx}, C.~T. and {Losch}, P. and {Mayman}, P. and {Eichhorn}, W. and {Krebs}, D. and {Jhabvala}, M. and {Gezari}, D.~Y. and {Fixsen}, D.~J. and {Flores}, J. and {Shakoorzadeh}, K. and {Jungo}, R. and {Hakun}, C. and {Workman}, L. and {Karpati}, G. and {Kichak}, R. and {Whitley}, R. and {Mann}, S. and {Tollestrup}, E.~V. and {Eisenhardt}, P. and {Stern}, D. and {Gorjian}, V. and {Bhattacharya}, B. and {Carey}, S. and {Nelson}, B.~O. and {Glaccum}, W.~J. and {Lacy}, M. and {Lowrance}, P.~J. and {Laine}, S. and {Reach}, W.~T. and {Stauffer}, J.~A. and {Surace}, J.~A. and {Wilson}, G. and {Wright}, E.~L. and {Hoffman}, A. and {Domingo}, G. and {Cohen}, M.},
        title = "{The Infrared Array Camera (IRAC) for the Spitzer Space Telescope}",
      journal = {\apjs},
         year = 2004,
        month = sep,
       volume = {154},
       number = {1},
        pages = {10-17},
          doi = {10.1086/422843},
archivePrefix = {arXiv},
       eprint = {astro-ph/0405616},
 primaryClass = {astro-ph},
       adsurl = {https://ui.adsabs.harvard.edu/abs/2004ApJS..154...10F}
}

@ARTICLE{rieke2004multiband,
       author = {{Rieke}, G.~H. and {Young}, E.~T. and {Engelbracht}, C.~W. and {Kelly}, D.~M. and {Low}, F.~J. and {Haller}, E.~E. and {Beeman}, J.~W. and {Gordon}, K.~D. and {Stansberry}, J.~A. and {Misselt}, K.~A. and {Cadien}, J. and {Morrison}, J.~E. and {Rivlis}, G. and {Latter}, W.~B. and {Noriega-Crespo}, A. and {Padgett}, D.~L. and {Stapelfeldt}, K.~R. and {Hines}, D.~C. and {Egami}, E. and {Muzerolle}, J. and {Alonso-Herrero}, A. and {Blaylock}, M. and {Dole}, H. and {Hinz}, J.~L. and {Le Floc'h}, E. and {Papovich}, C. and {P{\'e}rez-Gonz{\'a}lez}, P.~G. and {Smith}, P.~S. and {Su}, K.~Y.~L. and {Bennett}, L. and {Frayer}, D.~T. and {Henderson}, D. and {Lu}, N. and {Masci}, F. and {Pesenson}, M. and {Rebull}, L. and {Rho}, J. and {Keene}, J. and {Stolovy}, S. and {Wachter}, S. and {Wheaton}, W. and {Werner}, M.~W. and {Richards}, P.~L.},
        title = "{The Multiband Imaging Photometer for Spitzer (MIPS)}",
      journal = {\apjs},
         year = 2004,
        month = sep,
       volume = {154},
       number = {1},
        pages = {25-29},
          doi = {10.1086/422717},
       adsurl = {https://ui.adsabs.harvard.edu/abs/2004ApJS..154...25R}
}

@article{2007A&A...471..625T,
       author = {{Trundle}, C. and {Dufton}, P.~L. and {Hunter}, I. and {Evans}, C.~J. and {Lennon}, D.~J. and {Smartt}, S.~J. and {Ryans}, R.~S.~I.},
        title = "{The VLT-FLAMES survey of massive stars: evolution of surface N abundances and effective temperature scales in the Galaxy and Magellanic Clouds}",
      journal = {\aap},
         year = 2007,
        month = aug,
       volume = {471},
       number = {2},
        pages = {625-643},
          doi = {10.1051/0004-6361:20077838},
archivePrefix = {arXiv},
       eprint = {0706.1731},
 primaryClass = {astro-ph},
    publisher = {EDP},
       adsurl = {https://ui.adsabs.harvard.edu/abs/2007A&A...471..625T}
}

@ARTICLE{2012A&A...537A.146E,
       author = {{Ekstr{\"o}m}, S. and {Georgy}, C. and {Eggenberger}, P. and {Meynet}, G. and {Mowlavi}, N. and {Wyttenbach}, A. and {Granada}, A. and {Decressin}, T. and {Hirschi}, R. and {Frischknecht}, U. and {Charbonnel}, C. and {Maeder}, A.},
        title = "{Grids of stellar models with rotation. I. Models from 0.8 to 120 M$_{{\ensuremath{\odot}}}$ at solar metallicity (Z = 0.014)}",
      journal = {\aap},
         year = 2012,
        month = jan,
       volume = {537},
          eid = {A146},
        pages = {A146},
          doi = {10.1051/0004-6361/201117751},
archivePrefix = {arXiv},
       eprint = {1110.5049},
 primaryClass = {astro-ph.SR},
       adsurl = {https://ui.adsabs.harvard.edu/abs/2012A&A...537A.146E}
}

@ARTICLE{2008A&A...478..823M,
       author = {{Markova}, N. and {Puls}, J.},
        title = "{Bright OB stars in the Galaxy. IV. Stellar and wind parameters of early to late B supergiants}",
      journal = {\aap},
         year = 2008,
        month = feb,
       volume = {478},
       number = {3},
        pages = {823-842},
          doi = {10.1051/0004-6361:20077919},
archivePrefix = {arXiv},
       eprint = {0711.1110},
 primaryClass = {astro-ph},
       adsurl = {https://ui.adsabs.harvard.edu/abs/2008A&A...478..823M}
}

@ARTICLE{martins2005_1,
       author = {{Martins}, F. and {Schaerer}, D. and {Hillier}, D.~J.},
        title = "{A new calibration of stellar parameters of Galactic O stars}",
      journal = {\aap},
         year = 2005,
        month = jun,
       volume = {436},
       number = {3},
        pages = {1049-1065},
          doi = {10.1051/0004-6361:20042386},
archivePrefix = {arXiv},
       eprint = {astro-ph/0503346},
 primaryClass = {astro-ph},
       adsurl = {https://ui.adsabs.harvard.edu/abs/2005A&A...436.1049M}
}

@article{moutzouri2025,
       author = {{Moutzouri}, M. and {Mackey}, J. and {Castro}, N. and {Gong}, Y. and {Jim{\'e}nez-Hern{\'a}ndez}, P. and {Toal{\'a}}, J.~A. and {Burger-Scheidlin}, C. and {Rugel}, M. and {Carrasco-Gonz{\'a}lez}, C. and {Brose}, R. and {Menten}, K.~M.},
        title = "{A targeted radio survey of infrared-selected bow shock candidates}",
      journal = {\aap},
         year = 2025,
        month = dec,
       volume = {704},
          eid = {A268},
        pages = {A268},
          doi = {10.1051/0004-6361/202557036},
archivePrefix = {arXiv},
       eprint = {2510.23470},
 primaryClass = {astro-ph.GA},
       adsurl = {https://ui.adsabs.harvard.edu/abs/2025A&A...704A.268M}
}

@article{meyer2016,
  title={On the observability of bow shocks of Galactic runaway OB stars},
  author={Meyer, DM-A and Van Marle, A-J and Kuiper, R and Kley, W},
  journal={\mnras},
  volume={459},
  number={2},
  pages={1146--1158},
  year={2016},
  publisher={Oxford University Press}
}

@book{draine2011,
  title={Physics of the interstellar and intergalactic medium},
  author={Draine, Bruce T},
  volume={19},
  year={2011},
  publisher={Princeton University Press}
}

@ARTICLE{maiz2004,
       author = {{Ma{\'\i}z-Apell{\'a}niz}, Jes{\'u}s and {Walborn}, Nolan R. and {Galu{\'e}}, H{\'e}ctor {\'A}. and {Wei}, Lisa H.},
        title = "{A Galactic O Star Catalog}",
      journal = {\apjs},
         year = 2004,
        month = mar,
       volume = {151},
       number = {1},
        pages = {103-148},
          doi = {10.1086/381380},
archivePrefix = {arXiv},
       eprint = {astro-ph/0311196},
 primaryClass = {astro-ph},
       adsurl = {https://ui.adsabs.harvard.edu/abs/2004ApJS..151..103M}
}

@article{hoogerwerf2001,
  title={On the origin of the O and B-type stars with high velocities-II. Runaway stars and pulsars ejected from the nearby young stellar groups},
  author={Hoogerwerf, R and De Bruijne, JHJ and De Zeeuw, PT},
  journal={\aap},
  volume={365},
  number={2},
  pages={49--77},
  year={2001},
  publisher={EDP Sciences}
}

@ARTICLE{garmany1982,
       author = {{Garmany}, C.~D. and {Conti}, P.~S. and {Chiosi}, C.},
        title = "{The initial mass function for massive stars.}",
      journal = {\apj},
         year = 1982,
        month = dec,
       volume = {263},
        pages = {777-790},
          doi = {10.1086/160548},
       adsurl = {https://ui.adsabs.harvard.edu/abs/1982ApJ...263..777G}
}

@ARTICLE{cruz1974,
       author = {{Cruz-Gonz{\'a}lez}, C. and {Recillas-Cruz}, E. and {Costero}, R. and {Peimbert}, M. and {Torres-Peimbert}, S.},
        title = "{A catalogue of galactic O stars and the ionization of the low density interstellar medium by runaway stars.}",
      journal = {\rmxaa},
         year = 1974,
        month = nov,
       volume = {1},
        pages = {211-259},
       adsurl = {https://ui.adsabs.harvard.edu/abs/1974RMxAA...1..211C}
}

@ARTICLE{stone1979,
       author = {{Stone}, R.~C.},
        title = "{Kinematics, close binary evolution, and ages of the O stars.}",
      journal = {\apj},
         year = 1979,
        month = sep,
       volume = {232},
        pages = {520-530},
          doi = {10.1086/157311},
       adsurl = {https://ui.adsabs.harvard.edu/abs/1979ApJ...232..520S}
}

@ARTICLE{patten2025,
       author = {{Patten}, Nikhil and {Kobulnicky}, Henry A. and {Povich}, Matthew S. and {Whisnant}, Angelica S. and {Andrews}, Sydney and {Boone}, Alexandra and {Dandu}, Srujan and {Jones}, Naomi and {Justice}, S. Nick and {Hope}, Dylan and {Larsen}, Alexander and {McCrory}, Ryan and {Meredith}, Julia and {Meza}, Maria Renee and {Rosenthal}, Alexandra C. and {Salazar}, William and {Sterling}, Alexander R. and {Yesmin}, Noshin and {Dale}, Daniel A.},
        title = "{Fundamental Parameters for Central Stars of 103 Infrared Bow Shock Nebulae}",
      journal = {\apj},
         year = 2025,
        month = aug,
       volume = {988},
       number = {2},
          eid = {183},
        pages = {183},
          doi = {10.3847/1538-4357/ade22f},
archivePrefix = {arXiv},
       eprint = {2506.07904},
 primaryClass = {astro-ph.SR},
       adsurl = {https://ui.adsabs.harvard.edu/abs/2025ApJ...988..183P}
}

@ARTICLE{vanMarle2011,
       author = {{van Marle}, A.~J. and {Meliani}, Z. and {Keppens}, R. and {Decin}, L.},
        title = "{Computing the Dust Distribution in the Bow Shock of a Fast-moving, Evolved Star}",
      journal = {\apjl},
         year = 2011,
        month = jun,
       volume = {734},
       number = {2},
          eid = {L26},
        pages = {L26},
          doi = {10.1088/2041-8205/734/2/L26},
archivePrefix = {arXiv},
       eprint = {1105.2387},
 primaryClass = {astro-ph.SR},
       adsurl = {https://ui.adsabs.harvard.edu/abs/2011ApJ...734L..26V}
}

@book{osterbrock2006,
  title={Astrophysics Of Gas Nebulae and Active Galactic Nuclei},
  author={Osterbrock, Donald E and Ferland, Gary J},
  year={2006},
  publisher={University science books}
}

@article{benaglia2010,
  title={Detection of nonthermal emission from the bow shock of a massive runaway star},
  author={Benaglia, Paula and Romero, Gustavo Esteban and Mart{\'\i}, Josep and Peri, Cintia Soledad and Araudo, Anabella Teresa},
  journal={\aap},
  volume={517},
  pages={L10},
  year={2010},
  publisher={EDP Sciences}
}

@article{benaglia2021,
  title={High-sensitivity radio study of the non-thermal stellar bow shock EB27},
  author={Benaglia, Paula and del Palacio, Santiago and Hales, Christopher and Colazo, Marcelo E},
  journal={\mnras},
  volume={503},
  number={2},
  pages={2514--2522},
  year={2021},
  publisher={Oxford University Press}
}

@ARTICLE{eijnden2022,
       author = {{Van den Eijnden}, J. and {Saikia}, P. and {Mohamed}, S.},
        title = "{Radio detections of IR-selected runaway stellar bow shocks}",
      journal = {\mnras},
         year = 2022,
        month = jun,
       volume = {512},
       number = {4},
        pages = {5374-5389},
          doi = {10.1093/mnras/stac823},
archivePrefix = {arXiv},
       eprint = {2203.10842},
 primaryClass = {astro-ph.HE},
       adsurl = {https://ui.adsabs.harvard.edu/abs/2022MNRAS.512.5374V}
}

@ARTICLE{gvaramadze2012,
       author = {{Gvaramadze}, V.~V. and {Langer}, N. and {Mackey}, J.},
        title = "{{\ensuremath{\zeta}} Oph and the weak-wind problem}",
      journal = {\mnras},
         year = 2012,
        month = nov,
       volume = {427},
       number = {1},
        pages = {L50-L54},
          doi = {10.1111/j.1745-3933.2012.01343.x},
archivePrefix = {arXiv},
       eprint = {1209.0455},
 primaryClass = {astro-ph.SR},
       adsurl = {https://ui.adsabs.harvard.edu/abs/2012MNRAS.427L..50G}
}

@ARTICLE{henney2019_3,
       author = {{Henney}, William J. and {Arthur}, S.~J.},
        title = "{Bow shocks, bow waves, and dust waves - III. Diagnostics}",
      journal = {\mnras},
         year = 2019,
        month = oct,
       volume = {489},
       number = {2},
        pages = {2142-2158},
          doi = {10.1093/mnras/stz2283},
archivePrefix = {arXiv},
       eprint = {1904.00343},
 primaryClass = {astro-ph.SR},
       adsurl = {https://ui.adsabs.harvard.edu/abs/2019MNRAS.489.2142H}
}

@ARTICLE{baranov1971,
       author = {{Baranov}, V.~B. and {Krasnobaev}, K.~V. and {Kulikovskii}, A.~G.},
        title = "{A Model of the Interaction of the Solar Wind with the Interstellar Medium}",
      journal = {Soviet Physics Doklady},
         year = 1971,
        month = mar,
       volume = {15},
        pages = {791},
       adsurl = {https://ui.adsabs.harvard.edu/abs/1971SPhD...15..791B}
}

@ARTICLE{wilkin2000,
       author = {{Wilkin}, Francis P.},
        title = "{Modeling Nonaxisymmetric Bow Shocks: Solution Method and Exact Analytic Solutions}",
      journal = {\apj},
         year = 2000,
        month = mar,
       volume = {532},
       number = {1},
        pages = {400-414},
          doi = {10.1086/308576},
archivePrefix = {arXiv},
       eprint = {astro-ph/0003389},
 primaryClass = {astro-ph},
       adsurl = {https://ui.adsabs.harvard.edu/abs/2000ApJ...532..400W}
}

@ARTICLE{wilkin1996,
       author = {{Wilkin}, Francis P.},
        title = "{Exact Analytic Solutions for Stellar Wind Bow Shocks}",
      journal = {\apjl},
         year = 1996,
        month = mar,
       volume = {459},
        pages = {L31},
          doi = {10.1086/309939},
       adsurl = {https://ui.adsabs.harvard.edu/abs/1996ApJ...459L..31W}
}

@ARTICLE{mohamed2012,
       author = {{Mohamed}, S. and {Mackey}, J. and {Langer}, N.},
        title = "{3D simulations of Betelgeuse's bow shock}",
      journal = {\aap},
         year = 2012,
        month = may,
       volume = {541},
          eid = {A1},
        pages = {A1},
          doi = {10.1051/0004-6361/201118002},
archivePrefix = {arXiv},
       eprint = {1109.1555},
 primaryClass = {astro-ph.SR},
       adsurl = {https://ui.adsabs.harvard.edu/abs/2012A&A...541A...1M}
}

@ARTICLE{acreman2016,
       author = {{Acreman}, David M. and {Stevens}, Ian R. and {Harries}, Tim J.},
        title = "{Modelling multiwavelength observational characteristics of bow shocks from runaway early-type stars}",
      journal = {\mnras},
         year = 2016,
        month = feb,
       volume = {456},
       number = {1},
        pages = {136-145},
          doi = {10.1093/mnras/stv2632},
archivePrefix = {arXiv},
       eprint = {1511.03059},
 primaryClass = {astro-ph.SR},
       adsurl = {https://ui.adsabs.harvard.edu/abs/2016MNRAS.456..136A}
}

@ARTICLE{mackey2016,
       author = {{Mackey}, Jonathan and {Haworth}, Thomas J. and {Gvaramadze}, Vasilii V. and {Mohamed}, Shazrene and {Langer}, Norbert and {Harries}, Tim J.},
        title = "{Detecting stellar-wind bubbles through infrared arcs in H II regions}",
      journal = {\aap},
         year = 2016,
        month = feb,
       volume = {586},
          eid = {A114},
        pages = {A114},
          doi = {10.1051/0004-6361/201527569},
archivePrefix = {arXiv},
       eprint = {1512.06857},
 primaryClass = {astro-ph.GA},
       adsurl = {https://ui.adsabs.harvard.edu/abs/2016A&A...586A.114M}
}

@ARTICLE{green2019,
       author = {{Green}, Samuel and {Mackey}, Jonathan and {Haworth}, Thomas J. and {Gvaramadze}, Vasilii V. and {Duffy}, Peter},
        title = "{Thermal emission from bow shocks. I. 2D hydrodynamic models of the Bubble Nebula}",
      journal = {\aap},
         year = 2019,
        month = may,
       volume = {625},
          eid = {A4},
        pages = {A4},
          doi = {10.1051/0004-6361/201834832},
archivePrefix = {arXiv},
       eprint = {1903.05505},
 primaryClass = {astro-ph.GA},
       adsurl = {https://ui.adsabs.harvard.edu/abs/2019A&A...625A...4G}
}

@ARTICLE{meyer2017,
       author = {{Meyer}, D.~M.-A. and {Mignone}, A. and {Kuiper}, R. and {Raga}, A.~C. and {Kley}, W.},
        title = "{Bow shock nebulae of hot massive stars in a magnetized medium}",
      journal = {\mnras},
         year = 2017,
        month = jan,
       volume = {464},
       number = {3},
        pages = {3229-3248},
          doi = {10.1093/mnras/stw2537},
archivePrefix = {arXiv},
       eprint = {1610.00543},
 primaryClass = {astro-ph.SR},
       adsurl = {https://ui.adsabs.harvard.edu/abs/2017MNRAS.464.3229M}
}

@ARTICLE{meyer2021,
       author = {{Meyer}, D.~M.-A. and {Mignone}, A. and {Petrov}, M. and {Scherer}, K. and {Vel{\'a}zquez}, P.~F. and {Boumis}, P.},
        title = "{3D MHD astrospheres: applications to IRC-10414 and Betelgeuse}",
      journal = {\mnras},
         year = 2021,
        month = oct,
       volume = {506},
       number = {4},
        pages = {5170-5189},
          doi = {10.1093/mnras/stab2026},
archivePrefix = {arXiv},
       eprint = {2107.05513},
 primaryClass = {astro-ph.SR},
       adsurl = {https://ui.adsabs.harvard.edu/abs/2021MNRAS.506.5170M}
}

@ARTICLE{mackey2021,
       author = {{Mackey}, Jonathan and {Green}, Samuel and {Moutzouri}, Maria and {Haworth}, Thomas J. and {Kavanagh}, Robert D. and {Zargaryan}, Davit and {Celeste}, Maggie},
        title = "{PION: simulating bow shocks and circumstellar nebulae}",
      journal = {\mnras},
         year = 2021,
        month = jun,
       volume = {504},
       number = {1},
        pages = {983-1008},
          doi = {10.1093/mnras/stab781},
archivePrefix = {arXiv},
       eprint = {2103.07555},
 primaryClass = {astro-ph.GA},
       adsurl = {https://ui.adsabs.harvard.edu/abs/2021MNRAS.504..983M}
}

@ARTICLE{ochsendorf2014_1,
       author = {{Ochsendorf}, B.~B. and {Cox}, N.~L.~J. and {Krijt}, S. and {Salgado}, F. and {Bern{\'e}}, O. and {Bernard}, J.~P. and {Kaper}, L. and {Tielens}, A.~G.~G.~M.},
        title = "{Blowing in the wind: The dust wave around {\ensuremath{\sigma}} Orionis AB}",
      journal = {\aap},
         year = 2014,
        month = mar,
       volume = {563},
          eid = {A65},
        pages = {A65},
          doi = {10.1051/0004-6361/201322873},
archivePrefix = {arXiv},
       eprint = {1401.7185},
 primaryClass = {astro-ph.SR},
       adsurl = {https://ui.adsabs.harvard.edu/abs/2014A&A...563A..65O}
}

@ARTICLE{ochsendorf2014_2,
       author = {{Ochsendorf}, B.~B. and {Verdolini}, S. and {Cox}, N.~L.~J. and {Bern{\'e}}, O. and {Kaper}, L. and {Tielens}, A.~G.~G.~M.},
        title = "{Radiation-pressure-driven dust waves inside bursting interstellar bubbles}",
      journal = {\aap},
         year = 2014,
        month = jun,
       volume = {566},
          eid = {A75},
        pages = {A75},
          doi = {10.1051/0004-6361/201423545},
archivePrefix = {arXiv},
       eprint = {1404.2807},
 primaryClass = {astro-ph.SR},
       adsurl = {https://ui.adsabs.harvard.edu/abs/2014A&A...566A..75O}
}

@ARTICLE{ochsendorf2015,
       author = {{Ochsendorf}, B.~B. and {Tielens}, A.~G.~G.~M.},
        title = "{A bimodal dust grain distribution in the IC 434 H ii region}",
      journal = {\aap},
         year = 2015,
        month = apr,
       volume = {576},
          eid = {A2},
        pages = {A2},
          doi = {10.1051/0004-6361/201424799},
archivePrefix = {arXiv},
       eprint = {1501.02256},
 primaryClass = {astro-ph.GA},
       adsurl = {https://ui.adsabs.harvard.edu/abs/2015A&A...576A...2O}
}

@ARTICLE{green2022,
       author = {{Green}, Samuel and {Mackey}, Jonathan and {Kavanagh}, Patrick and {Haworth}, Thomas J. and {Moutzouri}, Maria and {Gvaramadze}, Vasilii V.},
        title = "{Thermal emission from bow shocks. II. 3D magnetohydrodynamic models of {\ensuremath{\zeta}} Ophiuchi}",
      journal = {\aap},
         year = 2022,
        month = sep,
       volume = {665},
          eid = {A35},
        pages = {A35},
          doi = {10.1051/0004-6361/202243531},
archivePrefix = {arXiv},
       eprint = {2203.06331},
 primaryClass = {astro-ph.GA},
       adsurl = {https://ui.adsabs.harvard.edu/abs/2022A&A...665A..35G}
}

@ARTICLE{gull1979,
       author = {{Gull}, T.~R. and {Sofia}, S.},
        title = "{Discovery of two distorted interstellar bubbles.}",
      journal = {\apj},
         year = 1979,
        month = jun,
       volume = {230},
        pages = {782-785},
          doi = {10.1086/157137},
       adsurl = {https://ui.adsabs.harvard.edu/abs/1979ApJ...230..782G}
}

@ARTICLE{binder2019,
       author = {{Binder}, Breanna A. and {Behr}, Patrick and {Povich}, Matthew S.},
        title = "{Searching for Faint X-Ray Emission from Galactic Stellar Wind Bow Shocks}",
      journal = {\aj},
         year = 2019,
        month = may,
       volume = {157},
       number = {5},
          eid = {176},
        pages = {176},
          doi = {10.3847/1538-3881/ab1073},
archivePrefix = {arXiv},
       eprint = {1903.06279},
 primaryClass = {astro-ph.HE},
       adsurl = {https://ui.adsabs.harvard.edu/abs/2019AJ....157..176B}
}

@ARTICLE{becker2017,
       author = {{De Becker}, M. and {del Valle}, M.~V. and {Romero}, G.~E. and {Peri}, C.~S. and {Benaglia}, P.},
        title = "{X-ray study of bow shocks in runaway stars}",
      journal = {\mnras},
         year = 2017,
        month = nov,
       volume = {471},
       number = {4},
        pages = {4452-4464},
          doi = {10.1093/mnras/stx1826},
archivePrefix = {arXiv},
       eprint = {1708.04082},
 primaryClass = {astro-ph.HE},
       adsurl = {https://ui.adsabs.harvard.edu/abs/2017MNRAS.471.4452D}
}

@ARTICLE{lopez2012,
       author = {{L{\'o}pez-Santiago}, J. and {Miceli}, M. and {del Valle}, M.~V. and {Romero}, G.~E. and {Bonito}, R. and {Albacete-Colombo}, J.~F. and {Pereira}, V. and {de Castro}, E. and {Damiani}, F.},
        title = "{AE Aurigae: First Detection of Non-thermal X-Ray Emission from a Bow Shock Produced by a Runaway Star}",
      journal = {\apjl},
         year = 2012,
        month = sep,
       volume = {757},
       number = {1},
        pages = {L6},
          doi = {10.1088/2041-8205/757/1/L6},
archivePrefix = {arXiv},
       eprint = {1208.6511},
 primaryClass = {astro-ph.HE},
       adsurl = {https://ui.adsabs.harvard.edu/abs/2012ApJ...757L...6L}
}

@INPROCEEDINGS{1986SPIE..627..733T,
       author = {{Tody}, Doug},
        title = "{The IRAF Data Reduction and Analysis System}",
    booktitle = {Instrumentation in astronomy VI},
         year = 1986,
       editor = {{Crawford}, David L.},
       series = {Society of Photo-Optical Instrumentation Engineers (SPIE) Conference Series},
       volume = {627},
        month = jan,
        pages = {733},
          doi = {10.1117/12.968154},
       adsurl = {https://ui.adsabs.harvard.edu/abs/1986SPIE..627..733T}
}

@INPROCEEDINGS{1993ASPC...52..173T,
       author = {{Tody}, Doug},
        title = "{IRAF in the Nineties}",
    booktitle = {Astronomical Data Analysis Software and Systems II},
         year = 1993,
       editor = {{Hanisch}, R.~J. and {Brissenden}, R.~J.~V. and {Barnes}, J.},
       series = {Astronomical Society of the Pacific Conference Series},
       volume = {52},
        month = jan,
        pages = {173},
       adsurl = {https://ui.adsabs.harvard.edu/abs/1993ASPC...52..173T}
}

@ARTICLE{2019AN....340..633S,
       author = {{{\v{S}}trobl}, Jan and {Jel{\'\i}nek}, Martin and {Hudec}, Ren{\'e}},
        title = "{Small Binocular Telescope: The new epoch of Burst Alert Robotic Telescope}",
      journal = {Astronomische Nachrichten},
         year = 2019,
        month = aug,
       volume = {340},
       number = {7},
        pages = {633-637},
          doi = {10.1002/asna.201913668},
       adsurl = {https://ui.adsabs.harvard.edu/abs/2019AN....340..633S}
}

@ARTICLE{2018MNRAS.473.1576K,
       author = {{Katushkina}, O.~A. and {Alexashov}, D.~B. and {Gvaramadze}, V.~V. and {Izmodenov}, V.~V.},
        title = "{An astrosphere around the blue supergiant {\ensuremath{\kappa}} Cas: possible explanation of its filamentary structure}",
      journal = {\mnras},
         year = 2018,
        month = jan,
       volume = {473},
       number = {2},
        pages = {1576-1588},
          doi = {10.1093/mnras/stx2488},
archivePrefix = {arXiv},
       eprint = {1709.09494},
 primaryClass = {astro-ph.SR},
       adsurl = {https://ui.adsabs.harvard.edu/abs/2018MNRAS.473.1576K}
}

@ARTICLE{terada2012,
       author = {{Terada}, Yukikatsu and {Tashiro}, Makoto S. and {Bamba}, Aya and {Yamazaki}, Ryo and {Kouzu}, Tomomi and {Koyama}, Shu and {Seta}, Hiromi},
        title = "{Search for Diffuse X-Rays from the Bow Shock Region of Runaway Star BD +43 3654 with Suzaku}",
      journal = {\pasj},
         year = 2012,
        month = dec,
       volume = {64},
       number = {6},
          eid = {138},
        pages = {138},
          doi = {10.1093/pasj/64.6.138},
archivePrefix = {arXiv},
       eprint = {1207.5577},
 primaryClass = {astro-ph.HE},
       adsurl = {https://ui.adsabs.harvard.edu/abs/2012PASJ...64..138T}
}

@ARTICLE{toala2016,
       author = {{Toal{\'a}}, J.~A. and {Oskinova}, L.~M. and {Gonz{\'a}lez-Gal{\'a}n}, A. and {Guerrero}, M.~A. and {Ignace}, R. and {Pohl}, M.},
        title = "{X-Ray Observations of Bow Shocks around Runaway O Stars. The Case of {\ensuremath{\zeta}} Oph and BD+43{\textdegree}3654}",
      journal = {\apj},
         year = 2016,
        month = apr,
       volume = {821},
       number = {2},
          eid = {79},
        pages = {79},
          doi = {10.3847/0004-637X/821/2/79},
archivePrefix = {arXiv},
       eprint = {1602.07805},
 primaryClass = {astro-ph.HE},
       adsurl = {https://ui.adsabs.harvard.edu/abs/2016ApJ...821...79T}
}

@ARTICLE{eijnden2024,
       author = {{van den Eijnden}, J. and {Mohamed}, S. and {Carotenuto}, F. and {Motta}, S. and {Saikia}, P. and {Williams-Baldwin}, D.~R.~A.},
        title = "{Particle acceleration at the bow shock of runaway star LS 2355: non-thermal radio emission but no {\ensuremath{\gamma}}-ray counterpart}",
      journal = {\mnras},
         year = 2024,
        month = aug,
       volume = {532},
       number = {3},
        pages = {2920-2933},
          doi = {10.1093/mnras/stae1622},
archivePrefix = {arXiv},
       eprint = {2407.00380},
 primaryClass = {astro-ph.HE},
       adsurl = {https://ui.adsabs.harvard.edu/abs/2024MNRAS.532.2920V}
}

@ARTICLE{eijnden2022_2,
       author = {{van den Eijnden}, J. and {Heywood}, I. and {Fender}, R. and {Mohamed}, S. and {Sivakoff}, G.~R. and {Saikia}, P. and {Russell}, T.~D. and {Motta}, S. and {Miller-Jones}, J.~C.~A. and {Woudt}, P.~A.},
        title = "{MeerKAT discovery of radio emission from the Vela X-1 bow shock}",
      journal = {\mnras},
         year = 2022,
        month = feb,
       volume = {510},
       number = {1},
        pages = {515-530},
          doi = {10.1093/mnras/stab3395},
archivePrefix = {arXiv},
       eprint = {2111.10159},
 primaryClass = {astro-ph.HE},
       adsurl = {https://ui.adsabs.harvard.edu/abs/2022MNRAS.510..515V}
}

@ARTICLE{carretero2023,
       author = {{Carretero-Castrillo}, M. and {Rib{\'o}}, M. and {Paredes}, J.~M.},
        title = "{Galactic runaway O and Be stars found using Gaia DR3}",
      journal = {\aap},
         year = 2023,
        month = nov,
       volume = {679},
          eid = {A109},
        pages = {A109},
          doi = {10.1051/0004-6361/202346613},
archivePrefix = {arXiv},
       eprint = {2311.01827},
 primaryClass = {astro-ph.SR},
       adsurl = {https://ui.adsabs.harvard.edu/abs/2023A&A...679A.109C}
}

@ARTICLE{carretero2025,
       author = {{Carretero-Castrillo}, M. and {Benaglia}, P. and {Paredes}, J.~M. and {Rib{\'o}}, M.},
        title = "{New stellar bow shocks and bubbles found around runaway stars}",
      journal = {\aap},
         year = 2025,
        month = feb,
       volume = {694},
          eid = {A250},
        pages = {A250},
          doi = {10.1051/0004-6361/202451336},
archivePrefix = {arXiv},
       eprint = {2502.02658},
 primaryClass = {astro-ph.SR},
       adsurl = {https://ui.adsabs.harvard.edu/abs/2025A&A...694A.250C}
}

@ARTICLE{moutzouri2022,
       author = {{Moutzouri}, M. and {Mackey}, J. and {Carrasco-Gonz{\'a}lez}, C. and {Gong}, Y. and {Brose}, R. and {Zargaryan}, D. and {Toal{\'a}}, J.~A. and {Menten}, K.~M. and {Gvaramadze}, V.~V. and {Rugel}, M.~R.},
        title = "{And then they were two: Detection of non-thermal radio emission from the bow shocks of two runaway stars}",
      journal = {\aap},
         year = 2022,
        month = jul,
       volume = {663},
          eid = {A80},
        pages = {A80},
          doi = {10.1051/0004-6361/202243098},
archivePrefix = {arXiv},
       eprint = {2204.11913},
 primaryClass = {astro-ph.HE},
       adsurl = {https://ui.adsabs.harvard.edu/abs/2022A&A...663A..80M}
}

@ARTICLE{2005MNRAS.362..689P,
       author = {{Parker}, Quentin A. and {Phillipps}, S. and {Pierce}, M.~J. and {Hartley}, M. and {Hambly}, N.~C. and {Read}, M.~A. and {MacGillivray}, H.~T. and {Tritton}, S.~B. and {Cass}, C.~P. and {Cannon}, R.~D. and {Cohen}, M. and {Drew}, J.~E. and {Frew}, D.~J. and {Hopewell}, E. and {Mader}, S. and {Malin}, D.~F. and {Masheder}, M.~R.~W. and {Morgan}, D.~H. and {Morris}, R.~A.~H. and {Russeil}, D. and {Russell}, K.~S. and {Walker}, R.~N.~F.},
        title = "{The AAO/UKST SuperCOSMOS H{\ensuremath{\alpha}} survey}",
      journal = {\mnras},
         year = 2005,
        month = sep,
       volume = {362},
       number = {2},
        pages = {689-710},
          doi = {10.1111/j.1365-2966.2005.09350.x},
archivePrefix = {arXiv},
       eprint = {astro-ph/0506599},
 primaryClass = {astro-ph},
       adsurl = {https://ui.adsabs.harvard.edu/abs/2005MNRAS.362..689P}
}

@ARTICLE{2025ApJ...980..239D,
       author = {{del Valle}, M.~V. and {Santos-Lima}, R. and {Pohl}, M.},
        title = "{Radio Polarization from Runaway Star Bowshocks. I. The General Case}",
      journal = {\apj},
         year = 2025,
        month = feb,
       volume = {980},
       number = {2},
          eid = {239},
        pages = {239},
          doi = {10.3847/1538-4357/adae0a},
archivePrefix = {arXiv},
       eprint = {2505.06777},
 primaryClass = {astro-ph.SR},
       adsurl = {https://ui.adsabs.harvard.edu/abs/2025ApJ...980..239D}
}

@ARTICLE{2025A&A...696A..91M,
       author = {{Mackey}, Jonathan and {Mathew}, Arun and {Ali}, Ahmad A. and {Haworth}, Thomas J. and {Brose}, Robert and {Green}, Sam and {Moutzouri}, Maria and {Walch}, Stefanie},
        title = "{Thermal emission from bow shocks: III. Variable diffuse X-ray emission from stellar-wind bow shocks driven by dynamical instabilities}",
      journal = {\aap},
         year = 2025,
        month = apr,
       volume = {696},
          eid = {A91},
        pages = {A91},
          doi = {10.1051/0004-6361/202553722},
archivePrefix = {arXiv},
       eprint = {2501.06021},
 primaryClass = {astro-ph.HE},
       adsurl = {https://ui.adsabs.harvard.edu/abs/2025A&A...696A..91M}
}

@ARTICLE{2008A&A...490.1071G,
       author = {{Gvaramadze}, V.~V. and {Bomans}, D.~J.},
        title = "{Search for OB stars running away from young star clusters. I. NGC 6611}",
      journal = {\aap},
         year = 2008,
        month = nov,
       volume = {490},
       number = {3},
        pages = {1071-1077},
          doi = {10.1051/0004-6361:200810411},
archivePrefix = {arXiv},
       eprint = {0809.0650},
 primaryClass = {astro-ph},
       adsurl = {https://ui.adsabs.harvard.edu/abs/2008A&A...490.1071G}
}

@ARTICLE{gvaramadze2014,
       author = {{Gvaramadze}, V.~V. and {Menten}, K.~M. and {Kniazev}, A.~Y. and {Langer}, N. and {Mackey}, J. and {Kraus}, A. and {Meyer}, D.~M.-A. and {Kami{\'n}ski}, T.},
        title = "{IRC -10414: a bow-shock-producing red supergiant star}",
      journal = {\mnras},
         year = 2014,
        month = jan,
       volume = {437},
       number = {1},
        pages = {843-856},
          doi = {10.1093/mnras/stt1943},
archivePrefix = {arXiv},
       eprint = {1310.2245},
 primaryClass = {astro-ph.SR},
       adsurl = {https://ui.adsabs.harvard.edu/abs/2014MNRAS.437..843G}
}

@ARTICLE{hipparcos,
       author = {{van Leeuwen}, F.},
        title = "{Validation of the new Hipparcos reduction}",
      journal = {\aap},
         year = 2007,
        month = nov,
       volume = {474},
       number = {2},
        pages = {653-664},
          doi = {10.1051/0004-6361:20078357},
archivePrefix = {arXiv},
       eprint = {0708.1752},
 primaryClass = {astro-ph},
       adsurl = {https://ui.adsabs.harvard.edu/abs/2007A&A...474..653V}
}

@ARTICLE{2005MNRAS.362..753D,
       author = {{Drew}, Janet E. and {Greimel}, R. and {Irwin}, M.~J. and {Aungwerojwit}, A. and {Barlow}, M.~J. and {Corradi}, R.~L.~M. and {Drake}, J.~J. and {G{\"a}nsicke}, B.~T. and {Groot}, P. and {Hales}, A. and {Hopewell}, E.~C. and {Irwin}, J. and {Knigge}, C. and {Leisy}, P. and {Lennon}, D.~J. and {Mampaso}, A. and {Masheder}, M.~R.~W. and {Matsuura}, M. and {Morales-Rueda}, L. and {Morris}, R.~A.~H. and {Parker}, Q.~A. and {Phillipps}, S. and {Rodriguez-Gil}, P. and {Roelofs}, G. and {Skillen}, I. and {Sokoloski}, J.~L. and {Steeghs}, D. and {Unruh}, Y.~C. and {Viironen}, K. and {Vink}, J.~S. and {Walton}, N.~A. and {Witham}, A. and {Wright}, N. and {Zijlstra}, A.~A. and {Zurita}, A.},
        title = "{The INT Photometric H{\ensuremath{\alpha}} Survey of the Northern Galactic Plane (IPHAS)}",
      journal = {\mnras},
         year = 2005,
        month = sep,
       volume = {362},
       number = {3},
        pages = {753-776},
          doi = {10.1111/j.1365-2966.2005.09330.x},
archivePrefix = {arXiv},
       eprint = {astro-ph/0506726},
 primaryClass = {astro-ph},
       adsurl = {https://ui.adsabs.harvard.edu/abs/2005MNRAS.362..753D}
}

@ARTICLE{2015A&A...576A..97S,
       author = {{Scherer}, K. and {van der Schyff}, A. and {Bomans}, D.~J. and {Ferreira}, S.~E.~S. and {Fichtner}, H. and {Kleimann}, J. and {Strauss}, R.~D. and {Weis}, K. and {Wiengarten}, T. and {Wodzinski}, T.},
        title = "{Cosmic rays in astrospheres}",
      journal = {\aap},
         year = 2015,
        month = apr,
       volume = {576},
          eid = {A97},
        pages = {A97},
          doi = {10.1051/0004-6361/201425091},
archivePrefix = {arXiv},
       eprint = {1502.04277},
 primaryClass = {astro-ph.HE},
       adsurl = {https://ui.adsabs.harvard.edu/abs/2015A&A...576A..97S}
}

@ARTICLE{poveda1967,
       author = {{Poveda}, A. and {Ruiz}, J. and {Allen}, C.},
        title = "{Run-away Stars as the Result of the Gravitational Collapse of Proto-stellar Clusters}",
      journal = {Boletin de los Observatorios Tonantzintla y Tacubaya},
         year = 1967,
        month = apr,
       volume = {4},
        pages = {86-90},
       adsurl = {https://ui.adsabs.harvard.edu/abs/1967BOTT....4...86P}
}

@ARTICLE{fujii2011,
       author = {{Fujii}, Michiko S. and {Portegies Zwart}, Simon},
        title = "{The Origin of OB Runaway Stars}",
      journal = {Science},
         year = 2011,
        month = dec,
       volume = {334},
       number = {6061},
        pages = {1380},
          doi = {10.1126/science.1211927},
archivePrefix = {arXiv},
       eprint = {1111.3644},
 primaryClass = {astro-ph.GA},
       adsurl = {https://ui.adsabs.harvard.edu/abs/2011Sci...334.1380F}
}

@ARTICLE{pflamm2010,
       author = {{Pflamm-Altenburg}, Jan and {Kroupa}, Pavel},
        title = "{The two-step ejection of massive stars and the issue of their formation in isolation}",
      journal = {\mnras},
         year = 2010,
        month = may,
       volume = {404},
       number = {3},
        pages = {1564-1568},
          doi = {10.1111/j.1365-2966.2010.16376.x},
archivePrefix = {arXiv},
       eprint = {1001.3671},
 primaryClass = {astro-ph.GA},
       adsurl = {https://ui.adsabs.harvard.edu/abs/2010MNRAS.404.1564P}
}

@ARTICLE{blaauw1961,
       author = {{Blaauw}, A.},
        title = "{On the origin of the O- and B-type stars with high velocities (the ``run-away'' stars), and some related problems}",
      journal = {\bain},
         year = 1961,
        month = may,
       volume = {15},
        pages = {265},
       adsurl = {https://ui.adsabs.harvard.edu/abs/1961BAN....15..265B}
}

@ARTICLE{garmany1992,
       author = {{Garmany}, C.~D. and {Stencel}, R.~E.},
        title = "{Galactic OB associations in the Northern Milky Way Galaxy. I. Longitudes 55 to 150.}",
      journal = {\aaps},
         year = 1992,
        month = aug,
       volume = {94},
        pages = {211-244},
       adsurl = {https://ui.adsabs.harvard.edu/abs/1992A&AS...94..211G}
}

@ARTICLE{hohle2010,
       author = {{Hohle}, M.~M. and {Neuh{\"a}user}, R. and {Schutz}, B.~F.},
        title = "{Masses and luminosities of O- and B-type stars and red supergiants}",
      journal = {Astronomische Nachrichten},
         year = 2010,
        month = apr,
       volume = {331},
       number = {4},
        pages = {349},
          doi = {10.1002/asna.200911355},
archivePrefix = {arXiv},
       eprint = {1003.2335},
 primaryClass = {astro-ph.SR},
       adsurl = {https://ui.adsabs.harvard.edu/abs/2010AN....331..349H}
}

@ARTICLE{crowther2006,
       author = {{Crowther}, P.~A. and {Lennon}, D.~J. and {Walborn}, N.~R.},
        title = "{Physical parameters and wind properties of galactic early B supergiants}",
      journal = {\aap},
         year = 2006,
        month = jan,
       volume = {446},
       number = {1},
        pages = {279-293},
          doi = {10.1051/0004-6361:20053685},
archivePrefix = {arXiv},
       eprint = {astro-ph/0509436},
 primaryClass = {astro-ph},
       adsurl = {https://ui.adsabs.harvard.edu/abs/2006A&A...446..279C}
}

@ARTICLE{lorenzo2017,
       author = {{Lorenzo}, J. and {Sim{\'o}n-D{\'\i}az}, S. and {Negueruela}, I. and {Vilardell}, F. and {Garcia}, M. and {Evans}, C.~J. and {Montes}, D.},
        title = "{The massive multiple system HD 64315}",
      journal = {\aap},
         year = 2017,
        month = oct,
       volume = {606},
          eid = {A54},
        pages = {A54},
          doi = {10.1051/0004-6361/201731352},
archivePrefix = {arXiv},
       eprint = {1708.00849},
 primaryClass = {astro-ph.SR},
       adsurl = {https://ui.adsabs.harvard.edu/abs/2017A&A...606A..54L}
}

@ARTICLE{martins2005_2,
       author = {{Martins}, F. and {Schaerer}, D. and {Hillier}, D.~J. and {Meynadier}, F. and {Heydari-Malayeri}, M. and {Walborn}, N.~R.},
        title = "{On stars with weak winds: the Galactic case}",
      journal = {\aap},
         year = 2005,
        month = oct,
       volume = {441},
       number = {2},
        pages = {735-762},
          doi = {10.1051/0004-6361:20052927},
archivePrefix = {arXiv},
       eprint = {astro-ph/0507278},
 primaryClass = {astro-ph},
       adsurl = {https://ui.adsabs.harvard.edu/abs/2005A&A...441..735M}
}

@ARTICLE{mahy2022,
       author = {{Mahy}, L. and {Sana}, H. and {Shenar}, T. and {Sen}, K. and {Langer}, N. and {Marchant}, P. and {Abdul-Masih}, M. and {Banyard}, G. and {Bodensteiner}, J. and {Bowman}, D.~M. and {Dsilva}, K. and {Fabry}, M. and {Hawcroft}, C. and {Janssens}, S. and {Van Reeth}, T. and {Eldridge}, C.},
        title = "{Identifying quiescent compact objects in massive Galactic single-lined spectroscopic binaries}",
      journal = {\aap},
         year = 2022,
        month = aug,
       volume = {664},
          eid = {A159},
        pages = {A159},
          doi = {10.1051/0004-6361/202243147},
archivePrefix = {arXiv},
       eprint = {2207.07752},
 primaryClass = {astro-ph.SR},
       adsurl = {https://ui.adsabs.harvard.edu/abs/2022A&A...664A.159M}
}

@ARTICLE{raucq2018,
       author = {{Raucq}, F. and {Rauw}, G. and {Mahy}, L. and {Sim{\'o}n-D{\'\i}az}, S.},
        title = "{Fundamental parameters of massive stars in multiple systems: The cases of HD 17505A and HD 206267A}",
      journal = {\aap},
         year = 2018,
        month = jun,
       volume = {614},
          eid = {A60},
        pages = {A60},
          doi = {10.1051/0004-6361/201732376},
archivePrefix = {arXiv},
       eprint = {1803.00243},
 primaryClass = {astro-ph.SR},
       adsurl = {https://ui.adsabs.harvard.edu/abs/2018A&A...614A..60R}
}

@ARTICLE{bouret2012,
       author = {{Bouret}, J.-C. and {Hillier}, D.~J. and {Lanz}, T. and {Fullerton}, A.~W.},
        title = "{Properties of Galactic early-type O-supergiants. A combined FUV-UV and optical analysis}",
      journal = {\aap},
         year = 2012,
        month = aug,
       volume = {544},
          eid = {A67},
        pages = {A67},
          doi = {10.1051/0004-6361/201118594},
archivePrefix = {arXiv},
       eprint = {1205.3075},
 primaryClass = {astro-ph.SR},
       adsurl = {https://ui.adsabs.harvard.edu/abs/2012A&A...544A..67B}
}

@ARTICLE{papics2011,
       author = {{P{\'a}pics}, P.~I. and {Briquet}, M. and {Auvergne}, M. and {Aerts}, C. and {Degroote}, P. and {Niemczura}, E. and {Vu{\v{c}}kovi{\'c}}, M. and {Smolders}, K. and {Poretti}, E. and {Rainer}, M. and {Hareter}, M. and {Baglin}, A. and {Baudin}, F. and {Catala}, C. and {Michel}, E. and {Samadi}, R.},
        title = "{CoRoT high-precision photometry of the B0.5 IV star HD 51756}",
      journal = {\aap},
         year = 2011,
        month = apr,
       volume = {528},
          eid = {A123},
        pages = {A123},
          doi = {10.1051/0004-6361/201016131},
archivePrefix = {arXiv},
       eprint = {1101.3428},
 primaryClass = {astro-ph.SR},
       adsurl = {https://ui.adsabs.harvard.edu/abs/2011A&A...528A.123P}
}

@ARTICLE{howrath1997,
       author = {{Howarth}, Ian D. and {Siebert}, Kaj W. and {Hussain}, Gaitee A.~J. and {Prinja}, Raman K.},
        title = "{Cross-correlation characteristics of OB stars from IUE spectroscopy}",
      journal = {\mnras},
         year = 1997,
        month = jan,
       volume = {284},
       number = {2},
        pages = {265-285},
          doi = {10.1093/mnras/284.2.265},
       adsurl = {https://ui.adsabs.harvard.edu/abs/1997MNRAS.284..265H}
}

@ARTICLE{prinja1990,
       author = {{Prinja}, Raman K. and {Barlow}, M.~J. and {Howarth}, Ian D.},
        title = "{Terminal Velocities for a Large Sample of O Stars, B Supergiants, and Wolf-Rayet Stars}",
      journal = {\apj},
         year = 1990,
        month = oct,
       volume = {361},
        pages = {607},
          doi = {10.1086/169224},
       adsurl = {https://ui.adsabs.harvard.edu/abs/1990ApJ...361..607P}
}

@ARTICLE{vink2001,
       author = {{Vink}, Jorick S. and {de Koter}, A. and {Lamers}, H.~J.~G.~L.~M.},
        title = "{Mass-loss predictions for O and B stars as a function of metallicity}",
      journal = {\aap},
         year = 2001,
        month = apr,
       volume = {369},
        pages = {574-588},
          doi = {10.1051/0004-6361:20010127},
archivePrefix = {arXiv},
       eprint = {astro-ph/0101509},
 primaryClass = {astro-ph},
       adsurl = {https://ui.adsabs.harvard.edu/abs/2001A&A...369..574V}
}

@ARTICLE{gontcharov2006,
       author = {{Gontcharov}, G.~A.},
        title = "{Radial velocities of 35495 Hipparcos stars in a common system}",
      journal = {Astronomical and Astrophysical Transactions},
         year = 2006,
        month = apr,
       volume = {25},
       number = {2},
        pages = {145-148},
          doi = {10.1080/10556790600916780},
       adsurl = {https://ui.adsabs.harvard.edu/abs/2006A&AT...25..145G}
}

@ARTICLE{kharchenko2007,
       author = {{Kharchenko}, N.~V. and {Scholz}, R.-D. and {Piskunov}, A.~E. and {R{\"o}ser}, S. and {Schilbach}, E.},
        title = "{Astrophysical supplements to the ASCC-2.5: Ia. Radial velocities of {\ensuremath{\sim}}55000 stars and mean radial velocities of 516 Galactic open clusters and associations}",
      journal = {Astronomische Nachrichten},
         year = 2007,
        month = nov,
       volume = {328},
       number = {9},
        pages = {889},
          doi = {10.1002/asna.200710776},
archivePrefix = {arXiv},
       eprint = {0705.0878},
 primaryClass = {astro-ph},
       adsurl = {https://ui.adsabs.harvard.edu/abs/2007AN....328..889K}
}

@ARTICLE{holgado2018,
       author = {{Holgado}, G. and {Sim{\'o}n-D{\'\i}az}, S. and {Barb{\'a}}, R.~H. and {Puls}, J. and {Herrero}, A. and {Castro}, N. and {Garcia}, M. and {Ma{\'\i}z Apell{\'a}niz}, J. and {Negueruela}, I. and {Sab{\'\i}n-Sanjuli{\'a}n}, C.},
        title = "{The IACOB project. V. Spectroscopic parameters of the O-type stars in the modern grid of standards for spectral classification}",
      journal = {\aap},
         year = 2018,
        month = jun,
       volume = {613},
          eid = {A65},
        pages = {A65},
          doi = {10.1051/0004-6361/201731543},
archivePrefix = {arXiv},
       eprint = {1711.10043},
 primaryClass = {astro-ph.SR},
       adsurl = {https://ui.adsabs.harvard.edu/abs/2018A&A...613A..65H}
}

@ARTICLE{gaia2023,
       author = {{Gaia Collaboration} and {Vallenari}, A. and {Brown}, A.~G.~A. and {Prusti}, T. and {de Bruijne}, J.~H.~J. and {Arenou}, F. and {Babusiaux}, C. and {Biermann}, M. and {Creevey}, O.~L. and {Ducourant}, C. and {Evans}, D.~W. and {Eyer}, L. and {Guerra}, R. and {Hutton}, A. and {Jordi}, C. and {Klioner}, S.~A. and {Lammers}, U.~L. and {Lindegren}, L. and {Luri}, X. and {Mignard}, F. and {Panem}, C. and {Pourbaix}, D. and {Randich}, S. and {Sartoretti}, P. and {Soubiran}, C. and {Tanga}, P. and {Walton}, N.~A. and {Bailer-Jones}, C.~A.~L. and {Bastian}, U. and {Drimmel}, R. and {Jansen}, F. and {Katz}, D. and {Lattanzi}, M.~G. and {van Leeuwen}, F. and {Bakker}, J. and {Cacciari}, C. and {Casta{\~n}eda}, J. and {De Angeli}, F. and {Fabricius}, C. and {Fouesneau}, M. and {Fr{\'e}mat}, Y. and {Galluccio}, L. and {Guerrier}, A. and {Heiter}, U. and {Masana}, E. and {Messineo}, R. and {Mowlavi}, N. and {Nicolas}, C. and {Nienartowicz}, K. and {Pailler}, F. and {Panuzzo}, P. and {Riclet}, F. and {Roux}, W. and {Seabroke}, G.~M. and {Sordo}, R. and {Th{\'e}venin}, F. and {Gracia-Abril}, G. and {Portell}, J. and {Teyssier}, D. and {Altmann}, M. and {Andrae}, R. and {Audard}, M. and {Bellas-Velidis}, I. and {Benson}, K. and {Berthier}, J. and {Blomme}, R. and {Burgess}, P.~W. and {Busonero}, D. and {Busso}, G. and {C{\'a}novas}, H. and {Carry}, B. and {Cellino}, A. and {Cheek}, N. and {Clementini}, G. and {Damerdji}, Y. and {Davidson}, M. and {de Teodoro}, P. and {Nu{\~n}ez Campos}, M. and {Delchambre}, L. and {Dell'Oro}, A. and {Esquej}, P. and {Fern{\'a}ndez-Hern{\'a}ndez}, J. and {Fraile}, E. and {Garabato}, D. and {Garc{\'\i}a-Lario}, P. and {Gosset}, E. and {Haigron}, R. and {Halbwachs}, J.-L. and {Hambly}, N.~C. and {Harrison}, D.~L. and {Hern{\'a}ndez}, J. and {Hestroffer}, D. and {Hodgkin}, S.~T. and {Holl}, B. and {Jan{\ss}en}, K. and {Jevardat de Fombelle}, G. and {Jordan}, S. and {Krone-Martins}, A. and {Lanzafame}, A.~C. and {L{\"o}ffler}, W. and {Marchal}, O. and {Marrese}, P.~M. and {Moitinho}, A. and {Muinonen}, K. and {Osborne}, P. and {Pancino}, E. and {Pauwels}, T. and {Recio-Blanco}, A. and {Reyl{\'e}}, C. and {Riello}, M. and {Rimoldini}, L. and {Roegiers}, T. and {Rybizki}, J. and {Sarro}, L.~M. and {Siopis}, C. and {Smith}, M. and {Sozzetti}, A. and {Utrilla}, E. and {van Leeuwen}, M. and {Abbas}, U. and {{\'A}brah{\'a}m}, P. and {Abreu Aramburu}, A. and {Aerts}, C. and {Aguado}, J.~J. and {Ajaj}, M. and {Aldea-Montero}, F. and {Altavilla}, G. and {{\'A}lvarez}, M.~A. and {Alves}, J. and {Anders}, F. and {Anderson}, R.~I. and {Anglada Varela}, E. and {Antoja}, T. and {Baines}, D. and {Baker}, S.~G. and {Balaguer-N{\'u}{\~n}ez}, L. and {Balbinot}, E. and {Balog}, Z. and {Barache}, C. and {Barbato}, D. and {Barros}, M. and {Barstow}, M.~A. and {Bartolom{\'e}}, S. and {Bassilana}, J.-L. and {Bauchet}, N. and {Becciani}, U. and {Bellazzini}, M. and {Berihuete}, A. and {Bernet}, M. and {Bertone}, S. and {Bianchi}, L. and {Binnenfeld}, A. and {Blanco-Cuaresma}, S. and {Blazere}, A. and {Boch}, T. and {Bombrun}, A. and {Bossini}, D. and {Bouquillon}, S. and {Bragaglia}, A. and {Bramante}, L. and {Breedt}, E. and {Bressan}, A. and {Brouillet}, N. and {Brugaletta}, E. and {Bucciarelli}, B. and {Burlacu}, A. and {Butkevich}, A.~G. and {Buzzi}, R. and {Caffau}, E. and {Cancelliere}, R. and {Cantat-Gaudin}, T. and {Carballo}, R. and {Carlucci}, T. and {Carnerero}, M.~I. and {Carrasco}, J.~M. and {Casamiquela}, L. and {Castellani}, M. and {Castro-Ginard}, A. and {Chaoul}, L. and {Charlot}, P. and {Chemin}, L. and {Chiaramida}, V. and {Chiavassa}, A. and {Chornay}, N. and {Comoretto}, G. and {Contursi}, G. and {Cooper}, W.~J. and {Cornez}, T. and {Cowell}, S. and {Crifo}, F. and {Cropper}, M. and {Crosta}, M. and {Crowley}, C. and {Dafonte}, C. and {Dapergolas}, A. and {David}, M. and {David}, P. and {de Laverny}, P. and {De Luise}, F. and {De March}, R.},
        title = "{Gaia Data Release 3. Summary of the content and survey properties}",
      journal = {\aap},
         year = 2023,
        month = jun,
       volume = {674},
          eid = {A1},
        pages = {A1},
          doi = {10.1051/0004-6361/202243940},
archivePrefix = {arXiv},
       eprint = {2208.00211},
 primaryClass = {astro-ph.GA},
       adsurl = {https://ui.adsabs.harvard.edu/abs/2023A&A...674A...1G}
}

@ARTICLE{mahy2010,
       author = {{Mahy}, L. and {Rauw}, G. and {Martins}, F. and {Naz{\'e}}, Y. and {Gosset}, E. and {De Becker}, M. and {Sana}, H. and {Eenens}, P.},
        title = "{A New Investigation of the Binary HD 48099}",
      journal = {\apj},
         year = 2010,
        month = jan,
       volume = {708},
       number = {2},
        pages = {1537-1544},
          doi = {10.1088/0004-637X/708/2/1537},
archivePrefix = {arXiv},
       eprint = {0912.0605},
 primaryClass = {astro-ph.SR},
       adsurl = {https://ui.adsabs.harvard.edu/abs/2010ApJ...708.1537M}
}

@ARTICLE{duflot1995,
       author = {{Duflot}, M. and {Figon}, P. and {Meyssonnier}, N.},
        title = "{Vitesses radiales. Catalogue WEB: Wilson Evans Batten. Subtittle: Radial velocities: The Wilson-Evans-Batten catalogue.}",
      journal = {\aaps},
         year = 1995,
        month = dec,
       volume = {114},
        pages = {269},
       adsurl = {https://ui.adsabs.harvard.edu/abs/1995A&AS..114..269D}
}

@ARTICLE{kobulnicky2022,
       author = {{Kobulnicky}, Henry A. and {Chick}, William T.},
        title = "{Kinematics of the Central Stars Powering Bowshock Nebulae and the Large Multiplicity Fraction of Runaway OB Stars}",
      journal = {\aj},
         year = 2022,
        month = sep,
       volume = {164},
       number = {3},
          eid = {86},
        pages = {86},
          doi = {10.3847/1538-3881/ac7f2b},
archivePrefix = {arXiv},
       eprint = {2206.09281},
 primaryClass = {astro-ph.SR},
       adsurl = {https://ui.adsabs.harvard.edu/abs/2022AJ....164...86K}
}

@ARTICLE{zehe2018,
       author = {{Zehe}, T. and {Mugrauer}, M. and {Neuh{\"a}user}, R. and {Pannicke}, A. and {Lux}, O. and {Bischoff}, R. and {W{\"o}ckel}, D. and {Wagner}, D.},
        title = "{The radial and rotational velocity of {\ensuremath{\zeta}} Ophiuchi}",
      journal = {Astronomische Nachrichten},
         year = 2018,
        month = jan,
       volume = {339},
       number = {1},
        pages = {46-52},
          doi = {10.1002/asna.201713383},
archivePrefix = {arXiv},
       eprint = {1801.06409},
 primaryClass = {astro-ph.SR},
       adsurl = {https://ui.adsabs.harvard.edu/abs/2018AN....339...46Z}
}

@ARTICLE{abt1963,
       author = {{Abt}, Helmut A.},
        title = "{Stellar Radial Velocities in the Perseus Arm.}",
      journal = {\aj},
         year = 1963,
        month = sep,
       volume = {68},
        pages = {271},
          doi = {10.1086/109086},
       adsurl = {https://ui.adsabs.harvard.edu/abs/1963AJ.....68R.271A}
}

@ARTICLE{williams2011,
       author = {{Williams}, S.~J. and {Gies}, D.~R. and {Hillwig}, T.~C. and {McSwain}, M.~V. and {Huang}, W.},
        title = "{Radial Velocities of Galactic O-type Stars. I. Short-term Constant Velocity Stars}",
      journal = {\aj},
         year = 2011,
        month = nov,
       volume = {142},
       number = {5},
          eid = {146},
        pages = {146},
          doi = {10.1088/0004-6256/142/5/146},
       adsurl = {https://ui.adsabs.harvard.edu/abs/2011AJ....142..146W}
}

@ARTICLE{jonsson2020,
       author = {{J{\"o}nsson}, Henrik and {Holtzman}, Jon A. and {Allende Prieto}, Carlos and {Cunha}, Katia and {Garc{\'\i}a-Hern{\'a}ndez}, D.~A. and {Hasselquist}, Sten and {Masseron}, Thomas and {Osorio}, Yeisson and {Shetrone}, Matthew and {Smith}, Verne and {Stringfellow}, Guy S. and {Bizyaev}, Dmitry and {Edvardsson}, Bengt and {Majewski}, Steven R. and {M{\'e}sz{\'a}ros}, Szabolcs and {Souto}, Diogo and {Zamora}, Olga and {Beaton}, Rachael L. and {Bovy}, Jo and {Donor}, John and {Pinsonneault}, Marc H. and {Poovelil}, Vijith Jacob and {Sobeck}, Jennifer},
        title = "{APOGEE Data and Spectral Analysis from SDSS Data Release 16: Seven Years of Observations Including First Results from APOGEE-South}",
      journal = {\aj},
         year = 2020,
        month = sep,
       volume = {160},
       number = {3},
          eid = {120},
        pages = {120},
          doi = {10.3847/1538-3881/aba592},
archivePrefix = {arXiv},
       eprint = {2007.05537},
 primaryClass = {astro-ph.GA},
       adsurl = {https://ui.adsabs.harvard.edu/abs/2020AJ....160..120J}
}

@ARTICLE{mossoux2018,
       author = {{Mossoux}, E. and {Mahy}, L. and {Rauw}, G.},
        title = "{The long-period massive binary HD 54662 revisited}",
      journal = {\aap},
         year = 2018,
        month = jul,
       volume = {615},
          eid = {A19},
        pages = {A19},
          doi = {10.1051/0004-6361/201732095},
archivePrefix = {arXiv},
       eprint = {1802.06535},
 primaryClass = {astro-ph.SR},
       adsurl = {https://ui.adsabs.harvard.edu/abs/2018A&A...615A..19M}
}

@ARTICLE{pourbaix2004,
       author = {{Pourbaix}, D. and {Tokovinin}, A.~A. and {Batten}, A.~H. and {Fekel}, F.~C. and {Hartkopf}, W.~I. and {Levato}, H. and {Morrell}, N.~I. and {Torres}, G. and {Udry}, S.},
        title = "{S$_{B$^{9}$}$: The ninth catalogue of spectroscopic binary orbits}",
      journal = {\aap},
         year = 2004,
        month = sep,
       volume = {424},
        pages = {727-732},
          doi = {10.1051/0004-6361:20041213},
archivePrefix = {arXiv},
       eprint = {astro-ph/0406573},
 primaryClass = {astro-ph},
       adsurl = {https://ui.adsabs.harvard.edu/abs/2004A&A...424..727P}
}

@ARTICLE{scherer2020,
       author = {{Scherer}, K. and {Baalmann}, L.~R. and {Fichtner}, H. and {Kleimann}, J. and {Bomans}, D.~J. and {Weis}, K. and {Ferreira}, S.~E.~S. and {Herbst}, K.},
        title = "{MHD-shock structures of astrospheres: {\ensuremath{\lambda}} Cephei -like astrospheres}",
      journal = {\mnras},
         year = 2020,
        month = apr,
       volume = {493},
       number = {3},
        pages = {4172-4185},
          doi = {10.1093/mnras/staa497},
archivePrefix = {arXiv},
       eprint = {2002.06966},
 primaryClass = {astro-ph.SR},
       adsurl = {https://ui.adsabs.harvard.edu/abs/2020MNRAS.493.4172S}
}

@ARTICLE{hillenbrand1993,
       author = {{Hillenbrand}, Lynne A. and {Massey}, Philip and {Strom}, Stephen E. and {Merrill}, K. Michael},
        title = "{NGC 6611: A Cluster Caught in the Act}",
      journal = {\aj},
         year = 1993,
        month = nov,
       volume = {106},
        pages = {1906},
          doi = {10.1086/116774},
       adsurl = {https://ui.adsabs.harvard.edu/abs/1993AJ....106.1906H}
}

@ARTICLE{guarcello2007,
       author = {{Guarcello}, M.~G. and {Prisinzano}, L. and {Micela}, G. and {Damiani}, F. and {Peres}, G. and {Sciortino}, S.},
        title = "{Correlation between the spatial distribution of circumstellar disks and massive stars in the open cluster NGC 6611. Compiled catalog and cluster parameters}",
      journal = {\aap},
         year = 2007,
        month = jan,
       volume = {462},
       number = {1},
        pages = {245-255},
          doi = {10.1051/0004-6361:20066124},
archivePrefix = {arXiv},
       eprint = {astro-ph/0610401},
 primaryClass = {astro-ph},
       adsurl = {https://ui.adsabs.harvard.edu/abs/2007A&A...462..245G}
}

@ARTICLE{yadav2016,
       author = {{Yadav}, Ram Kesh and {Pandey}, A.~K. and {Sharma}, Saurabh and {Ojha}, D.~K. and {Samal}, M.~R. and {Mallick}, K.~K. and {Jose}, J. and {Ogura}, K. and {Richichi}, Andrea and {Irawati}, Puji and {Kobayashi}, N. and {Eswaraiah}, C.},
        title = "{A multiwavelength investigation of the H II region S311: young stellar population and star formation}",
      journal = {\mnras},
         year = 2016,
        month = sep,
       volume = {461},
       number = {3},
        pages = {2502-2518},
          doi = {10.1093/mnras/stw1356},
       adsurl = {https://ui.adsabs.harvard.edu/abs/2016MNRAS.461.2502Y}
}

@ARTICLE{yong2018,
       author = {{Tarango-Yong}, Jorge A. and {Henney}, William J.},
        title = "{True versus apparent shapes of bow shocks}",
      journal = {\mnras},
         year = 2018,
        month = jun,
       volume = {477},
       number = {2},
        pages = {2431-2454},
          doi = {10.1093/mnras/sty669},
archivePrefix = {arXiv},
       eprint = {1712.02300},
 primaryClass = {astro-ph.SR},
       adsurl = {https://ui.adsabs.harvard.edu/abs/2018MNRAS.477.2431T}
}

@ARTICLE{marcolino2009,
       author = {{Marcolino}, W.~L.~F. and {Bouret}, J.-C. and {Martins}, F. and {Hillier}, D.~J. and {Lanz}, T. and {Escolano}, C.},
        title = "{Analysis of Galactic late-type O dwarfs: more constraints on the weak wind problem}",
      journal = {\aap},
         year = 2009,
        month = may,
       volume = {498},
       number = {3},
        pages = {837-852},
          doi = {10.1051/0004-6361/200811289},
archivePrefix = {arXiv},
       eprint = {0902.1833},
 primaryClass = {astro-ph.SR},
       adsurl = {https://ui.adsabs.harvard.edu/abs/2009A&A...498..837M}
}

@ARTICLE{puls2008,
       author = {{Puls}, Joachim and {Vink}, Jorick S. and {Najarro}, Francisco},
        title = "{Mass loss from hot massive stars}",
      journal = {\aapr},
         year = 2008,
        month = dec,
       volume = {16},
       number = {3-4},
        pages = {209-325},
          doi = {10.1007/s00159-008-0015-8},
archivePrefix = {arXiv},
       eprint = {0811.0487},
 primaryClass = {astro-ph},
       adsurl = {https://ui.adsabs.harvard.edu/abs/2008A&ARv..16..209P}
}

@ARTICLE{haffner1999,
       author = {{Haffner}, L.~M. and {Reynolds}, R.~J. and {Tufte}, S.~L.},
        title = "{WHAM Observations of H{\ensuremath{\alpha}}, [S II], and [N II] toward the Orion and Perseus Arms: Probing the Physical Conditions of the Warm Ionized Medium}",
      journal = {\apj},
         year = 1999,
        month = sep,
       volume = {523},
       number = {1},
        pages = {223-233},
          doi = {10.1086/307734},
archivePrefix = {arXiv},
       eprint = {astro-ph/9904143},
 primaryClass = {astro-ph},
       adsurl = {https://ui.adsabs.harvard.edu/abs/1999ApJ...523..223H}
}

@ARTICLE{mezger1967,
       author = {{Mezger}, P.~G. and {Henderson}, A.~P.},
        title = "{Galactic H II Regions. I. Observations of Their Continuum Radiation at the Frequency 5 GHz}",
      journal = {\apj},
         year = 1967,
        month = feb,
       volume = {147},
        pages = {471},
          doi = {10.1086/149030},
       adsurl = {https://ui.adsabs.harvard.edu/abs/1967ApJ...147..471M}
}

@ARTICLE{dickinson2003,
       author = {{Dickinson}, C. and {Davies}, R.~D. and {Davis}, R.~J.},
        title = "{Towards a free-free template for CMB foregrounds}",
      journal = {\mnras},
         year = 2003,
        month = may,
       volume = {341},
       number = {2},
        pages = {369-384},
          doi = {10.1046/j.1365-8711.2003.06439.x},
archivePrefix = {arXiv},
       eprint = {astro-ph/0302024},
 primaryClass = {astro-ph},
       adsurl = {https://ui.adsabs.harvard.edu/abs/2003MNRAS.341..369D}
}

@ARTICLE{dennison1998,
       author = {{Dennison}, B. and {Simonetti}, J.~H. and {Topasna}, G.~A.},
        title = "{An imaging survey of northern galactic H{\ensuremath{\alpha}} emmission with arcminute resolution}",
      journal = {\pasa},
         year = 1998,
        month = apr,
       volume = {15},
       number = {1},
        pages = {147-48},
          doi = {10.1071/AS98147},
       adsurl = {https://ui.adsabs.harvard.edu/abs/1998PASA...15..147D}
}

@ARTICLE{draine1993,
       author = {{Draine}, Bruce T. and {McKee}, Christopher F.},
        title = "{Theory of interstellar shocks.}",
      journal = {\araa},
         year = 1993,
        month = jan,
       volume = {31},
        pages = {373-432},
          doi = {10.1146/annurev.aa.31.090193.002105},
       adsurl = {https://ui.adsabs.harvard.edu/abs/1993ARA&A..31..373D}
}

\onecolumn

\begin{appendix}
\nolinenumbers
\section{Observation Log}
Table~\ref{Tab:observation_log} presents the complete observation log for all 78 targets in our survey. The columns list the target identifier, equatorial coordinates (J2000), observing date, telescope, total exposure time, number of individual frames, seeing (where available), and V-band magnitude. Multiple entries for the same target correspond to observations obtained on different nights. For objects lacking a V-band magnitude in SIMBAD, the Gaia DR3 G-band magnitude is provided instead and marked with an asterisk.

{\small 
\begin{longtable}{lcccccccc}
\caption{\label{Tab:observation_log} Observation Log. Multiple observations of the same object are listed in separate rows.}\\
\hline\hline
Name & RA & DEC & Date &
Telescope & Total $T_{\rm exp}$ & No. of  & Seeing & V  \\
     & (J2000)  &  (J2000)  &   &   &   (s) & Frames & ($\arcsec$) & (mag) \\                  
\hline
\endfirsthead
\caption{continued.}\\
\hline\hline
Name & RA & DEC  & Date &
Telescope & Total $T_{\rm exp}$ & No. of  & Seeing & V  \\
     &  (J2000)  &  (J2000)  &   &   &   (s) & Frames & ($\arcsec$)  &  (mag) \\ 
\hline
\endhead
\hline
\endfoot

\object{HD 2083} & 00:25:51.23 & $+$71:48:25.65 & 2020-07-09 & SBT & 8460 & 47 $\times$ 180\,s & 6.97 & 6.91 \\
\object{HD 2905} & 00:32:59.99 & $+$62:55:54.41 & 2020-02-08 & SBT & 5400 & 30 $\times$ 180\,s & 7.06 & 4.16 \\
\object{HD 17505} & 02:51:07.97 & $+$60:25:03.87 & 2020-06-12 & SBT & 2340 & 13 $\times$ 180\,s & 7.54 & 7.07 \\
\object{HD 21856} & 03:32:40.01 & $+$35:27:42.21 & 2020-07-12 & SBT & 3780 & 21 $\times$ 180\,s & 9.86 & 5.9 \\
\object{HD 22928} & 03:42:55.50 & $+$47:47:15.17 & 2020-07-09 & SBT & 6120 & 34 $\times$ 180\,s & 7.63 & 3.01 \\
\object{HD 24431} & 03:55:38.42 & $+$52:38:28.73 & 2020-03-16 & SBT & 4320 & 24 $\times$ 180\,s & 7.13 & 6.74 \\
\object{HD 30614} & 04:54:03.01 & $+$66:20:33.63 & 2020-07-05 & SBT & 3240 & 18 $\times$ 180\,s & 6.88 & 4.29 \\
  &  &  &  2020-07-06 & SBT & 3420 & 19 $\times$ 180\,s & 6.92 & 4.29 \\
\object{HD 36512} & 05:31:55.86 & $-$07:18:05.53 & 2026-02-20/03-03 & SBT & 35460 & 197 $\times$ 180\,s & 8.20 & 4.63 \\
 \object{BD+09 879} & 05:35:08.27 & $+$09:56:02.96 & 2026-02-20/03-03 & SBT & 41400 & 230 $\times$ 180\,s & 8.18 & 3.66 \\
\object{HD 41161} & 06:05:52.45 & $+$48:14:57.42 & 2020-11-30 & SBT & 7020 & 39 $\times$ 180\,s & 7.54 & 6.76 \\
\object{HD 42933} & 06:10:17.91 & $-$54:58:07.11 & 2021-03-09 & T31 & 3600 & 12 $\times$ 300\,s & 4.23 & 4.72 \\
\object{HD 47432} & 06:38:38.18 & $+$01:36:48.67 & 2026-02-20/03-02 & SBT & 28080 & 156 $\times$ 180\,s  & 8.91 & 6.31 \\
\object{HD 48099} & 06:41:59.23 & $+$06:20:43.53 & 2026-01-20 & SBT & 9540 & 53 $\times$ 180\,s & 8.94 & 6.37 \\
\object{HD 50896} & 06:54:13.04 & $-$23:55:42.02 & 2021-03-05 & T31 & 3600 & 12 $\times$ 300\,s & 3.42 & 6.91 \\
\object{HD 51756} & 06:58:28.16 & $-$03:01:25.40 & 2022-04-03 & DK154 & 900 & 5 $\times$ 180\,s & 1.48 & 7.83$^{*}$ \\
\object{HD 54662} & 07:09:20.24 & $-$10:20:47.62 & 2021-03-09 & T17 & 3600 & 12 $\times$ 300\,s & 1.91 & 6.21 \\
\object{HD 57061} & 07:18:42.48 & $-$24:57:15.70 & 2021-04-06 & T31 & 3600 & 12 $\times$ 300\,s & 3.81 & 4.4 \\
\object{HD 64315}  & 07:52:20.28 & $-$26:25:46.68 & 2020-12-12 & DK154 & 3600 & 20 $\times$ 180\,s & 1.01 & 9.24 \\
\object{CD-26 5136} & 07:53:01.01 & $-$27:06:57.70 & 2020-12-13 & DK154 & 4500 & 25 $\times$ 180\,s & 0.95 & 9.67 \\
\object{HD 75860} & 08:50:53.24 & $-$43:45:05.40 & 2022-04-02 & DK154 & 3600 & 20 $\times$ 180\,s & 1.38 & 7.58 \\
\object{HD 76031} & 08:52:04.13 & $-$44:00:35.00 & 2020-12-13 & DK154 & 3600 & 20 $\times$ 180\,s & 0.85 & 9.5 \\
\object{CD-41 4637} & 08:55:27.66 & $-$41:35:22.20 & 2020-03-22 & DK154 & 2700 & 3 $\times$ 900\,s & 2.00 & 9.87 \\
\object{HD 77207} & 08:59:14.91 & $-$48:49:49.20 & 2020-12-13 & DK154 & 1260 & 7 $\times$ 180\,s & 0.73 & 9.91 \\
\object{HD 77581} & 09:02:06.86 & $-$40:33:16.90 & 2021-03-14 & DK154 & 3600 & 30 $\times$ 120\,s & 1.09 & 6.87 \\
G\,284.0765\,$-$00.4323$^{\mathrm{a}}$ & 10:22:23.08 & $-$57:44:28.00 & 2020-02-23 & DK154 & 900 & 1 $\times$ 900\,s & 0.81 & 16.54$^{*}$ \\
G\,286.0498\,$-$01.6583$^{\mathrm{b}}$ & 10:30:23.28 & $-$59:49:53.20 & 2020-03-22 & DK154 & 900 & 1 $\times$ 900\,s & 2.49 & 12.44 \\
\object{HD 92206} & 10:37:22.24 & $-$58:37:22.86 & 2020-12-12 & DK154 & 3960 & 22 $\times$ 180\,s & 1.07 & 8.11$^{*}$ \\
\object{HD 110879} & 12:46:16.80 & $-$68:06:29.21 & 2021-03-14 & DK154 & 915 & 61 $\times$ 15\,s & 0.97 & 3.98$^{*}$ \\
G\,311.1657\,$+$00.0448$^{\mathrm{c}}$  & 14:01:45.74 & $-$61:41:51.50 & 2020-03-22 & DK154 & 900 & 1 $\times$ 900\,s & 2.79 & 13.44$^{*}$ \\
\object{HD 126593} & 14:28:50.87 & $-$60:32:25.10 & 2022-04-02 & DK154 & 1800 & 5 $\times$ 180\,s & 1.33 & 8.65 \\
 &  &  &  2022-04-03 & DK154 & 1800 & 10 $\times$ 180\,s & 1.44 & 8.65 \\
G\,315.3719\,$+$00.6043$^{\mathrm{d}}$ & 14:32:50.57 & $-$59:47:44.10 & 2020-03-22 & DK154 & 900 & 1 $\times$ 900\,s & 2.01 & 12.73$^{*}$ \\
\object{HD 130298} & 14:49:33.76 & $-$56:25:38.47 & 2022-04-02 & DK154 & 1800 & 10 $\times$ 180\,s & 1.26 & 9.83 \\
G\,319.7182\,$-$00.7290$^{\mathrm{e}}$ & 15:08:10.24 & $-$59:04:12.60 & 2020-03-22 & DK154 & 2700 & 3 $\times$ 900\,s & 1.80 & 13.43$^{*}$ \\
\object{HD 135240} & 15:16:56.89 & $-$60:32:25.10 & 2020-07-24 & T31  & 3300 & 11 $\times$ 300\,s & 3.79 & 5.09 \\
G\,322.1986\,$+$00.5183$^{\mathrm{f}}$ & 15:19:16.98 & $-$56:43:06.30 & 2020-03-22 & DK154 & 900 & 1 $\times$ 900\,s & 1.85 & 13.31$^{*}$ \\
\object{HD 136003} & 15:20:42.81 & $-$56:07:57.44 & 2022-04-02 & DK154 & 2760 & 23 $\times$ 120\,s & 1.23 & 6.79 \\
\object{HD 329905} & 15:47:54.90 & $-$48:37:48.48 & 2022-04-03 & DK154 & 2100 & 7 $\times$ 300\,s & 1.25 & 10.43 \\
\object{HD 143275} & 16:00:20.01 & $-$22:37:18.14 & 2021-07-06 & T31 & 1200 & 10 $\times$ 120\,s & 3.52 & 2.32 \\
\object{HD 149757} & 16:37:09.53 & $-$10:34:01.52 & 2020-06-12 & SBT & 5580 & 31 $\times$ 180\,s & 9.12 & 2.56 \\
\object{HD 150898} & 16:47:19.65 & $-$58:20:29.19 & 2022-04-02 & DK154 & 2580 & 43 $\times$ 60\,s & 1.63 & 5.61 \\
\object{HD 152756} & 16:57:15.03 & $-$43:43:15.80 & 2022-04-02 & DK154 & 2640 & 44 $\times$ 60\,s & 1.48 & 9.41 \\
G\,350.6032\,$+$01.0332$^{\mathrm{g}}$ & 17:17:00.53 & $-$36:13:36.15 & 2022-04-02 & DK154 & 1800 & 6 $\times$ 300\,s & 1.58 & 14.89 \\
G\,353.4162\,$+$00.4482$^{\mathrm{h}}$ & 17:27:11.24 & $-$34:14:35.09 & 2021-03-13 & DK154 & 3300 & 11 $\times$ 300\,s & 1.24 & 10.79$^{*}$ \\
\object{HD 158186} & 17:29:12.93 & $-$31:32:03.42 & 2020-07-24 & T31 & 2700 & 9 $\times$ 300\,s & 3.78 & 7.04 \\
\object{CD-29 13925} & 17:45:08.21 & $-$29:56:45.30 & 2022-04-02 & DK154 & 3600 & 12 $\times$ 300\,s & 1.54 & 10.69 \\
G\,003.8417\,$-$01.0440$^{\mathrm{i}}$ & 17:58:30.65 & $-$26:09:49.12 & 2021-08-31 & T31 & 3600 & 12 $\times$ 300\,s & 3.41 & 12.74$^{*}$ \\
G\,006.2977\,$-$00.2012$^{\mathrm{j}}$ & 18:00:40.59 & $-$23:36:50.70 & 2022-04-03 & DK154 & 1800 & 6 $\times$ 300\,s & 1.45 & 16.75$^{*}$ \\
\object{HD 165319} & 18:05:58.83 & $-$14:11:52.99 & 2021-03-13 & DK154 & 3600 & 12 $\times$ 300\,s & 1.25 & 8.04 \\
\object{HD 166996} & 18:13:49.35 & $-$15:48:32.50 & 2022-04-03 & DK154 & 1800 & 10 $\times$ 180\,s & 1.36 & 9.52 \\
\object{NGC 6611 584} & 18:18:23.65 & $-$13:36:27.90 & 2021-03-14 & DK154 & 2700 & 9 $\times$ 300\,s & 1.28 & 12.02 \\
\object{BD-13 4934} & 18:19:05.58 & $-$13:54:50.60 & 2022-04-02 & DK154 & 3600 & 12 $\times$ 300\,s & 1.28 & 10.78 \\
\object{BD-13 4937} & 18:19:20.04 & $-$13:54:21.30 & 2022-04-02 & DK154 & 3600 & 12 $\times$ 300\,s & 1.28 & 10.78 \\
\object{NGC 6618 258} & 18:20:22.75 & $-$16:08:34.20 & 2022-04-03 & DK154 & 1620 & 9 $\times$ 180\,s & 1.28 & 13.69 \\
\object{NGC 6618 326} & 18:20:25.91 & $-$16:08:32.40 & 2022-04-03 & DK154 & 1620 & 9 $\times$ 180\,s & 1.28 & 14.2 \\
G\,015.1032\,$-$00.6489$^{\mathrm{k}}$ & 18:20:26.71 & $-$16:07:09.10 & 2022-04-03 & DK154 & 1620 & 9 $\times$ 180\,s & 1.28 & 17.5 \\
G\,018.2660\,$-$00.2988$^{\mathrm{l}}$ & 18:25:18.13 & $-$13:09:43.00 & 2021-03-13 & DK154 & 2700 & 9 $\times$ 300\,s & 1.23 & 16.75$^{*}$ \\
G\,023.1100\,$+$00.5458$^{\mathrm{m}}$ & 18:31:25.47 & $-$08:28:49.17 & 2021-09-01 & T31 & 3600 & 12 $\times$ 300\,s & 3.94 & 12.06$^{*}$ \\
\object{HD 171491} & 18:35:08.21 & $+$00:02:34.71 & 2020-07-06 & SBT & 3600 & 20 $\times$ 180\,s & 8.18 & 7.99 \\
G\,030.3745\,$+$00.0252$^{\mathrm{n}}$ & 18:46:40.09 & $-$02:15:51.90 & 2022-04-02 & DK154 & 900 & 3 $\times$ 300\,s & 1.92 & - \\
G\,031.6770\,$+$00.1775$^{\mathrm{o}}$ & 18:48:31.03 & $-$01:02:09.20 & 2022-04-02 & DK154 & 900 & 3 $\times$ 300\,s & 1.69 & - \\
\object{TYC 5118-279-1} & 18:49:25.06 & $-$02:21:09.70 &  2021-09-02 & T31 & 3600 & 12 $\times$ 300\,s & 3.25 & 10.48 \\
\object{HD 175514} & 18:55:23.12 & $+$09:20:48.08 & 2020-07-12 & SBT & 5400 & 30 $\times$ 180\,s & 7.37 & 8.64 \\
G\,049.4683\,$-$00.2527$^{\mathrm{p}}$ & 19:23:11.97 & $+$14:33:15.70 & 2023-08-09 & Perek & 7200 & 24 $\times$ 300\,s & 1.76 & - \\
\object{HD 188001} & 19:52:21.76 & $+$18:40:18.75 & 2020-06-12 & SBT & 5400 & 30 $\times$ 180\,s & 8.83 & 6.23 \\
\object{HD 229159} & 20:22:54.07 & $+$39:12:29.40 & 2020-09-16 & Perek & 4260 & 142 $\times$ 30\,s & 1.69 & 8.62 \\
\object{HD 195592} & 20:30:34.96 & $+$44:18:54.85 & 2020-06-11 & SBT & 6480 & 36 $\times$ 180\,s & 8.72 & 7.08 \\
\object{BD+43 3654} & 20:33:36.07 & $+$43:59:07.36 & 2020-06-11 & SBT &  6480 & 36 $\times$ 180\,s & 10.49 & 10.06 \\
G\,083.9246\,$-$00.6778$^{\mathrm{q}}$ & 20:51:50.20 & $+$43:19:29.80 & 2020-09-16 & Perek & 3600  & 12 $\times$ 300\,s & 1.05 & 12.53$^{*}$ \\
\object{HD 199021} & 20:52:53.20 & $+$42:36:27.80 & 2023-05-21 & Perek & 3780 & 126 $\times$ 30\,s  & 1.19 & 8.49 \\
\object{BD+42 3914} & 20:56:24.08 & $+$43:07:46.50 & 2023-05-21 & Perek & 1980 & 100 $\times$ 20\,s & 1.18 & 8.49 \\
  &  &  & 2023-06-25 & Perek & 1700 & 85 $\times$ 20\,s & 1.18 & 8.49 \\
\object{HD 203467} & 21:19:22.22 & $+$64:52:18.68 & 2020-07-04 & SBT & 7020 & 39 $\times$ 180\,s & 6.93 & 5.18 \\
\object{HD 210839} & 22:11:30.57 & $+$59:24:52.14 & 2020-07-04 & SBT & 1080 & 6 $\times$ 180\,s & 7.67 & 5.05 \\
  &  &  & 2020-07-05 & SBT & 9540 & 53 $\times$ 180\,s & 7.41 & 5.05 \\
\object{LS III +55 26} & 22:12:29.95 & $+$55:32:05.40 & 2020-09-16 & Perek & 3720 & 31 $\times$ 120\,s & 1.79 & 11.42 \\
\object{HD 215806} & 22:46:40.45 & $+$58:17:43.80 & 2020-09-06 & Perek & 3720 & 62 $\times$ 60\,s & 1.57 & 9.23 \\
  
\end{longtable}
\tablefoot{Some targets were observed on multiple nights. In these cases, we report the start and end dates of the observations. For objects for which no V-band magnitude is available in SIMBAD, the Gaia DR3 G-band magnitude is listed instead. These entries are indicated by an asterisk.\\
The SIMBAD identifiers of the marked infrared nebulae are as follows: (a)~\object{2MASS J10222305-5744280}, (b)~\object{2MASS J10302331-5949542}, (c)~\object{2MASS J14014576-6141513}, (d)~\object{2MASS J14325057-5947441}, (e)~\object{2MASS J15081025-5904126}, (f)~\object{2MASS J15191694-5643060}, (g)~\object{2MASS J17170051-3613362}, (h)~\object{2MASS J17271123-3414349}, (i)~\object{2MASS J17583064-2609490}, (j)~\object{2MASS J18004059-2336505}, (k)~\object{2MASS J18202666-1607088}, (l)~\object{2MASS J18251808-1309427}, (m)~\object{2MASS J18312547-0828490}, (n)~\object{[KCS2016] J184640.09-021551.9}, (o)~\object{2MASS J18483100-0102094}, (p)~\object{[ABB2014] WISE G049.465-00.254}, (q)~\object{2MASS J20515020+4319298}.
}
}

\section{Objects with diffuse H$\alpha$ emission}

Figs.~\ref{DE:DK154andPerek}-~\ref{DE:IPHAS} present the H$\alpha$ images of 35 sources classified as having diffuse background emission. These targets are located within bright and complex H\,{\sc ii} regions, which prevents the unambiguous identification of bow-shock-related H$\alpha$ structures. The image quality varies across the sample due to the range of telescopes and instrumental setups used (see Sect.~\ref{sec:observations}). Below, we group the objects by instrument and provide brief notes on individual sources.

\subsection{DK154 and Perek Observations} 
DK154 and Perek provide high-resolution images with pixel scales of $\sim$0\farcs4\,px$^{-1}$ (Sect.~\ref{sec:observations}) and seeing ranging from 
$\sim$0\farcs7 to $\sim$1\farcs9 (Table~\ref{Tab:observation_log}), over fields of view smaller than $10' \times 10'$. These instruments resolve fine H$\alpha$ structures, as detailed below.\\
\\
\textbf{K051, K052, K053:} 
All three targets lie within a single $7'\times 7'$ DK154 frame covering the bright M17 (Omega Nebula) H\,{\sc ii} region. Structured diffuse H$\alpha$ emission is seen throughout the field, but no clear arc-like structure is discernible. The corresponding WISE band~3 and band~4 images are corrupted.
\begin{figure*}[b!]
    \centering
\includegraphics[width=0.49\linewidth]{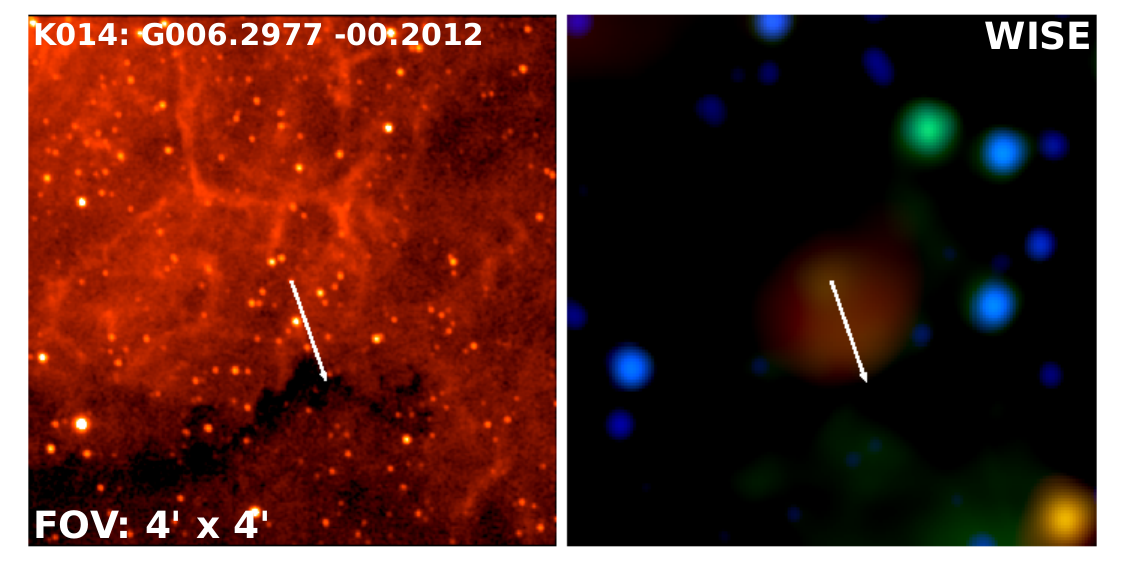}
\includegraphics[width=0.49\linewidth]{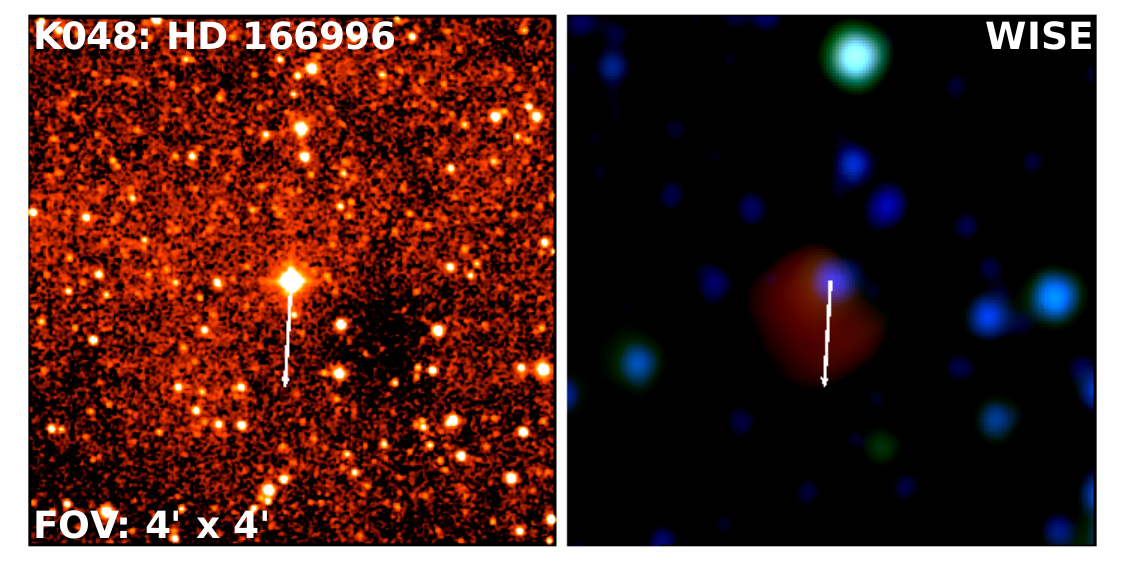}
\caption{Objects with diffuse background H$\alpha$ emission, as observed with the DK154 and Perek telescopes. The colour scheme is identical to that in Fig.~\ref{Detections1}. North is up and east is to the left.}
\label{DE:DK154andPerek}
\end{figure*}

\begin{figure*}[ht!]
\ContinuedFloat
    \centering
\includegraphics[width=0.49\linewidth]{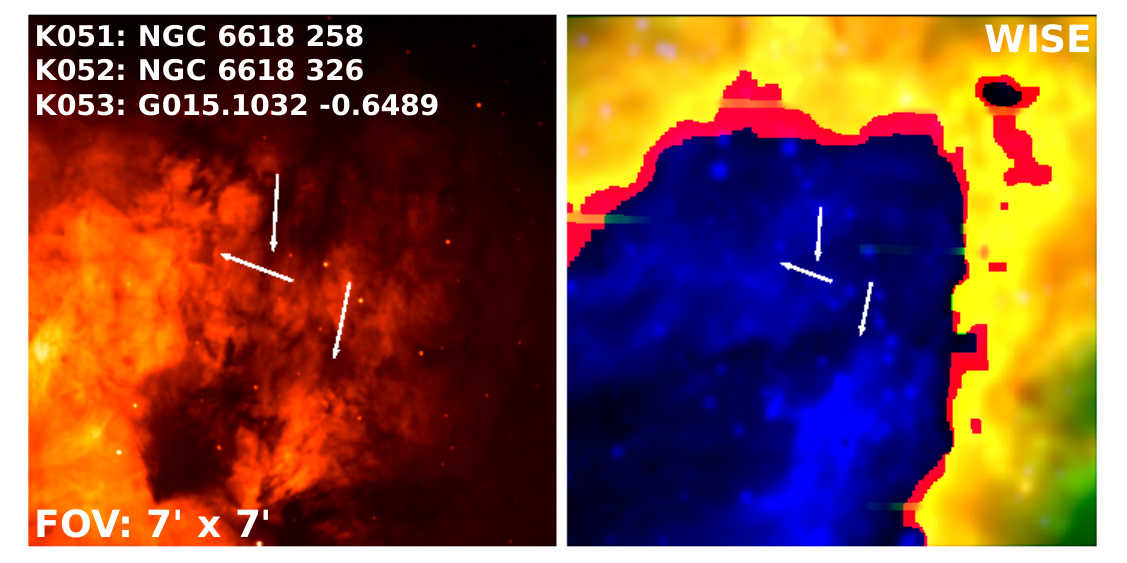} 
\includegraphics[width=0.49\linewidth]{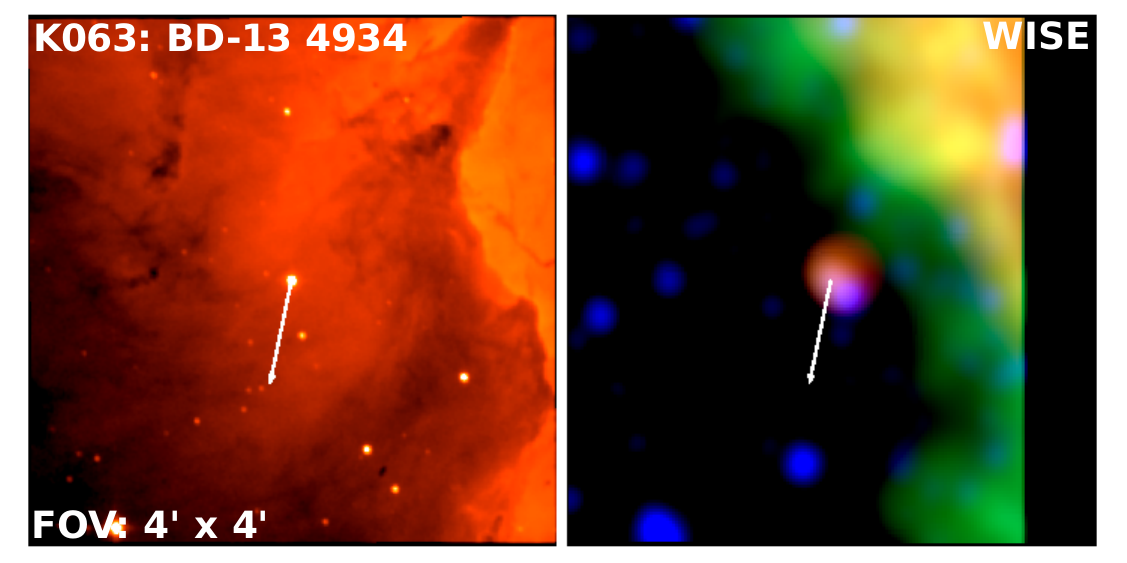}
\includegraphics[width=0.49\linewidth]{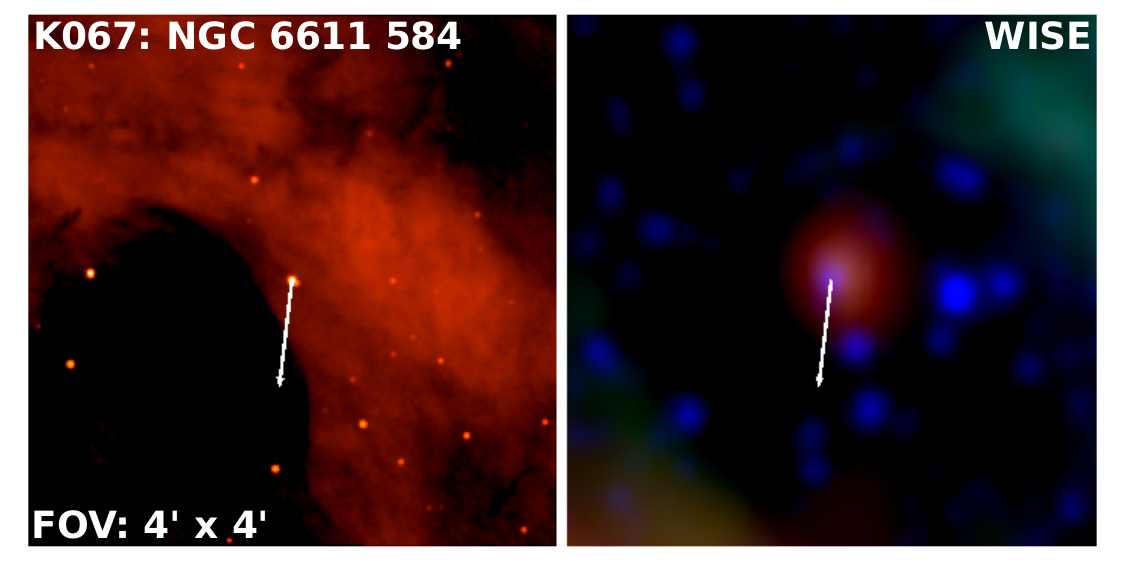}
\includegraphics[width=0.49\linewidth]{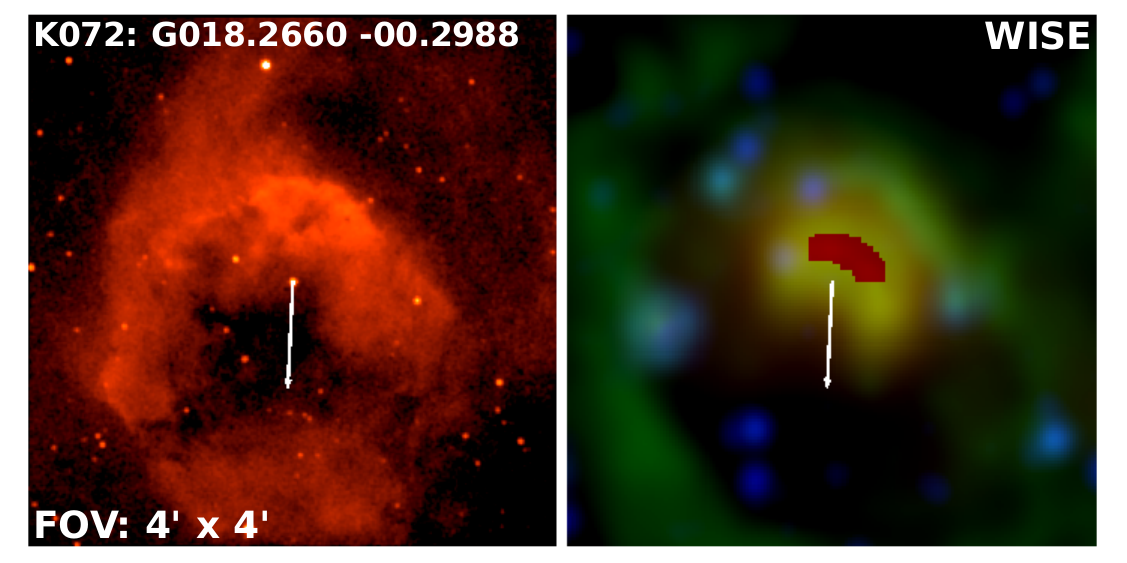}
\includegraphics[width=0.49\linewidth]{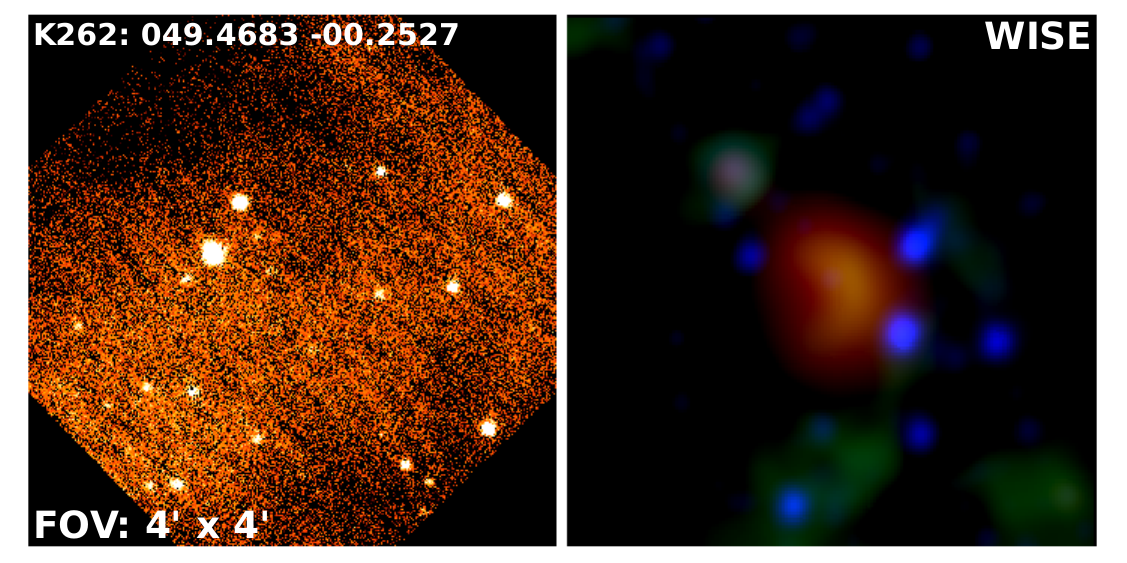}
\includegraphics[width=0.49\linewidth]{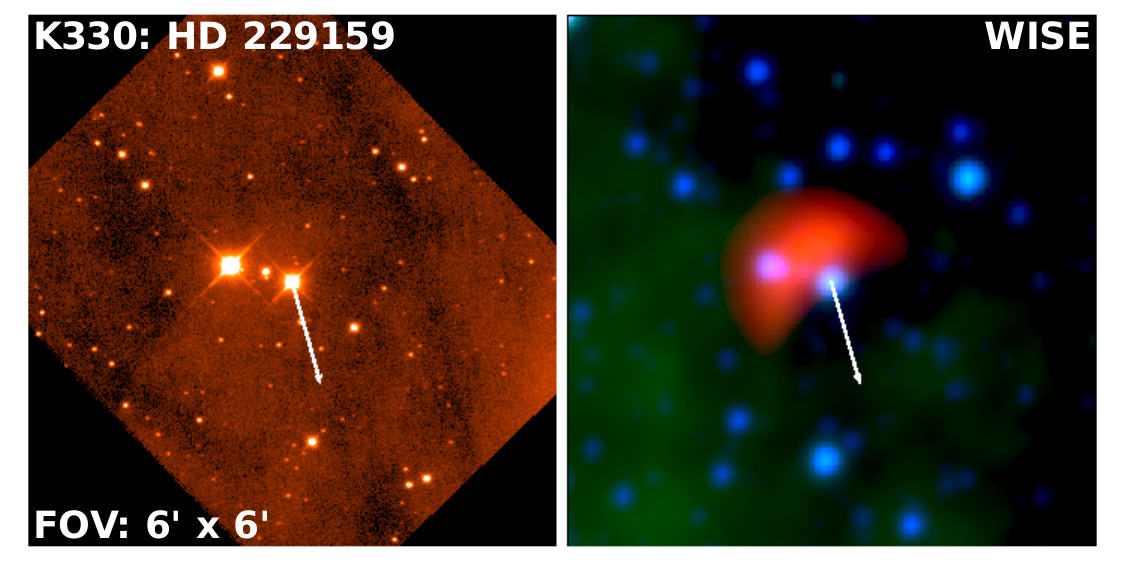}
\includegraphics[width=0.49\linewidth]{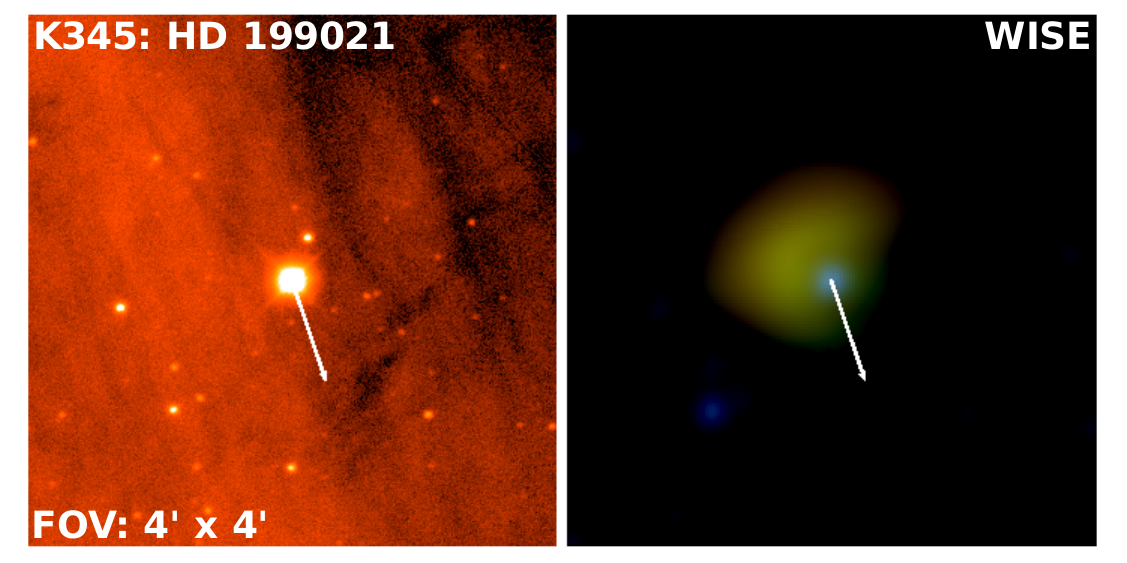} 
\includegraphics[width=0.49\linewidth]{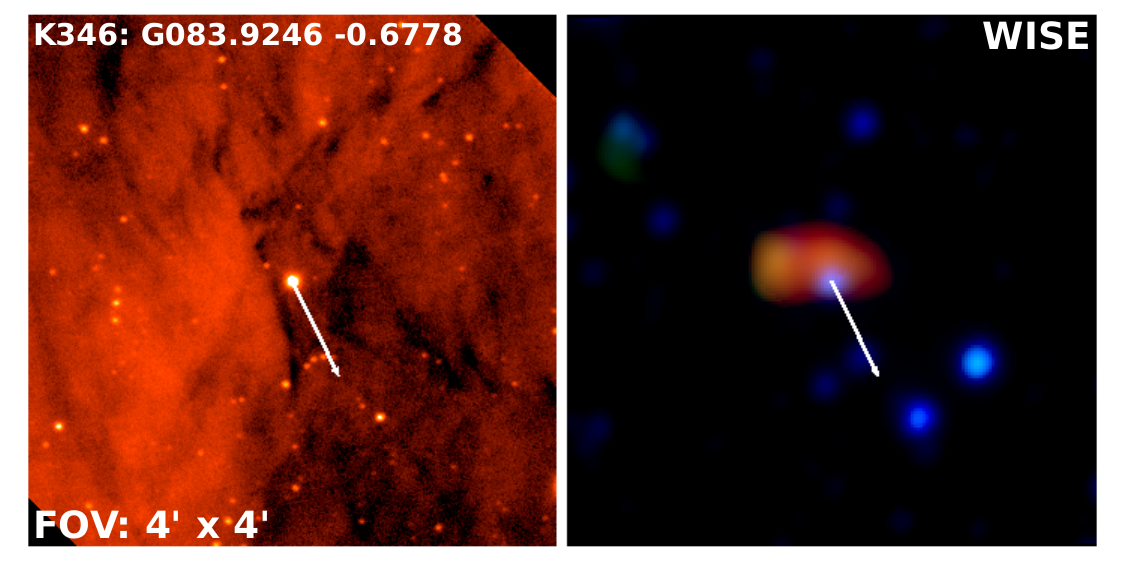}
\includegraphics[width=0.49\linewidth]{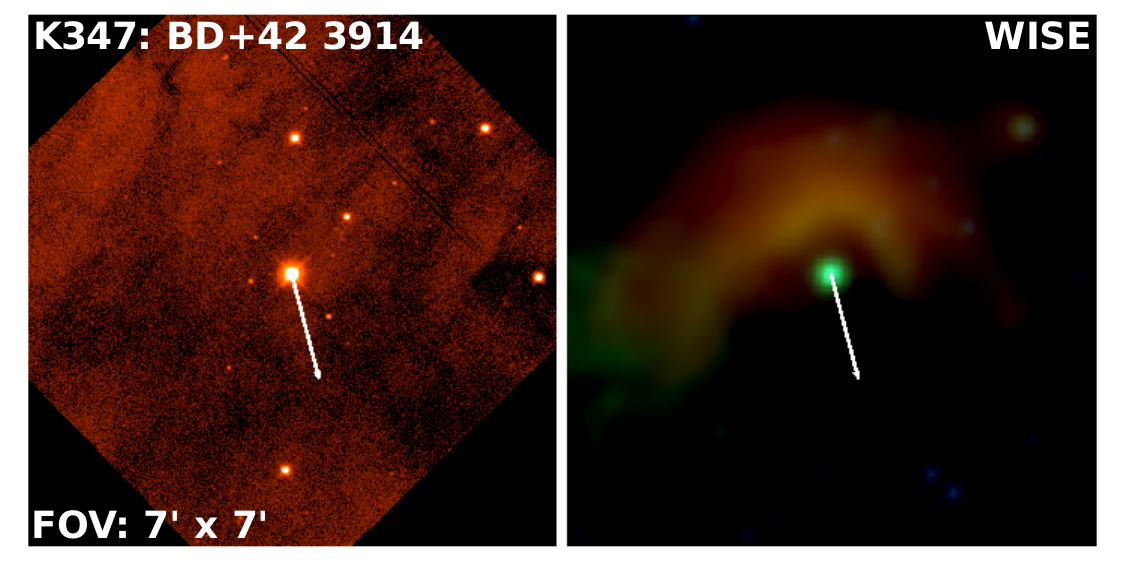}
\includegraphics[width=0.49\linewidth]{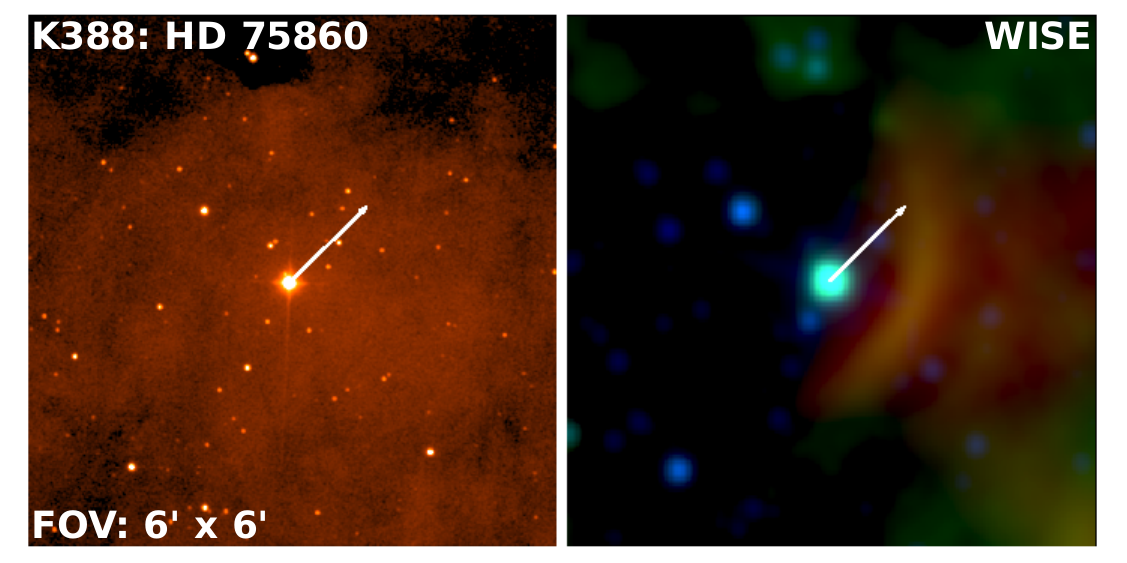}
   \caption{continued.}
\end{figure*}

\begin{figure*}[ht!]
\ContinuedFloat
    \centering
\includegraphics[width=0.49\linewidth]{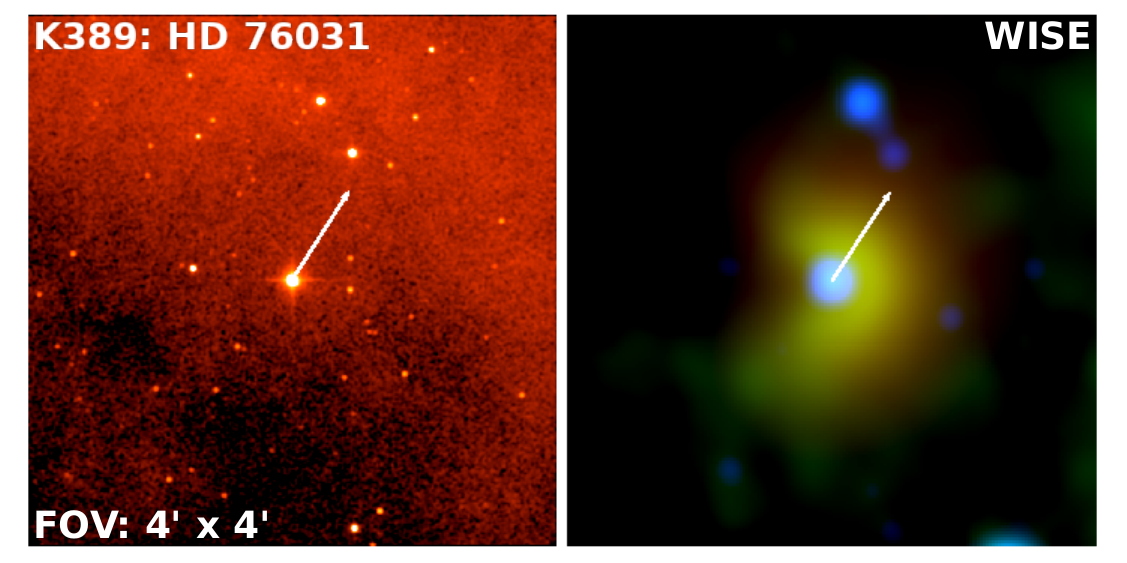}
\includegraphics[width=0.49\linewidth]{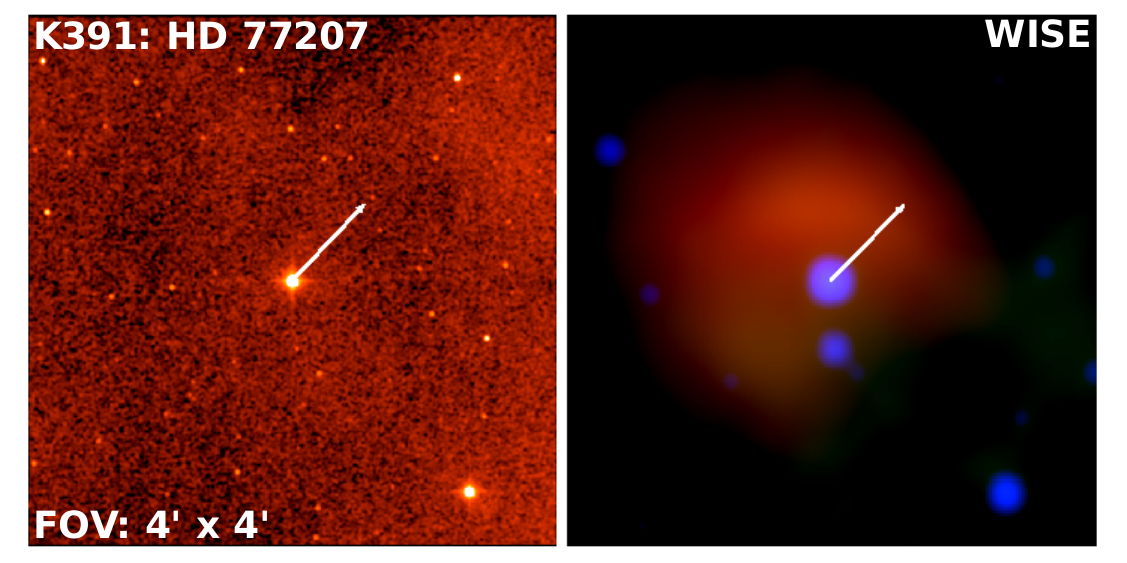}
\includegraphics[width=0.49\linewidth]{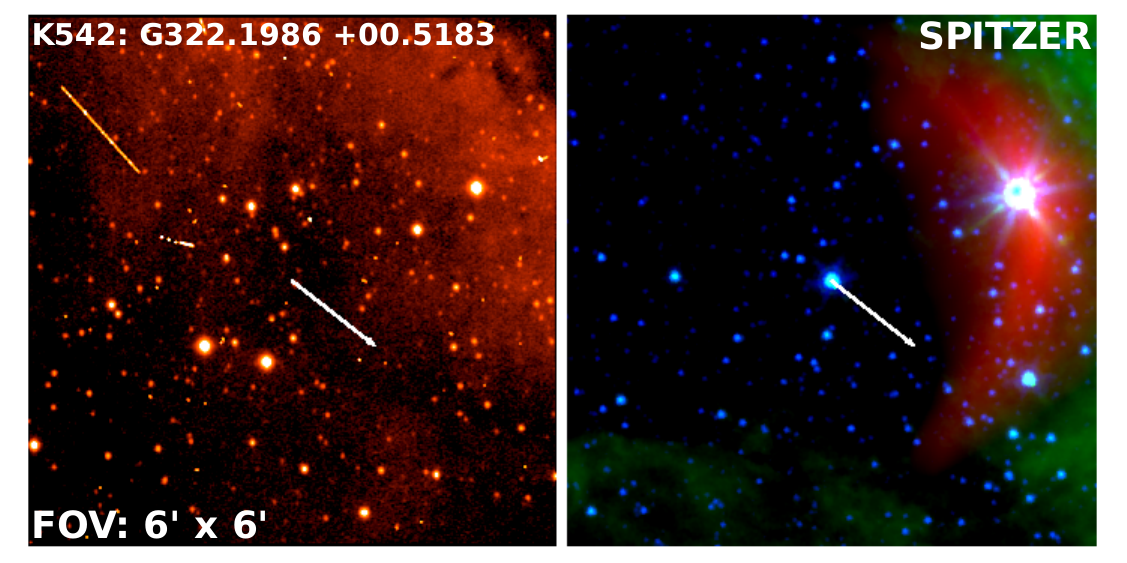}
\includegraphics[width=0.49\linewidth]{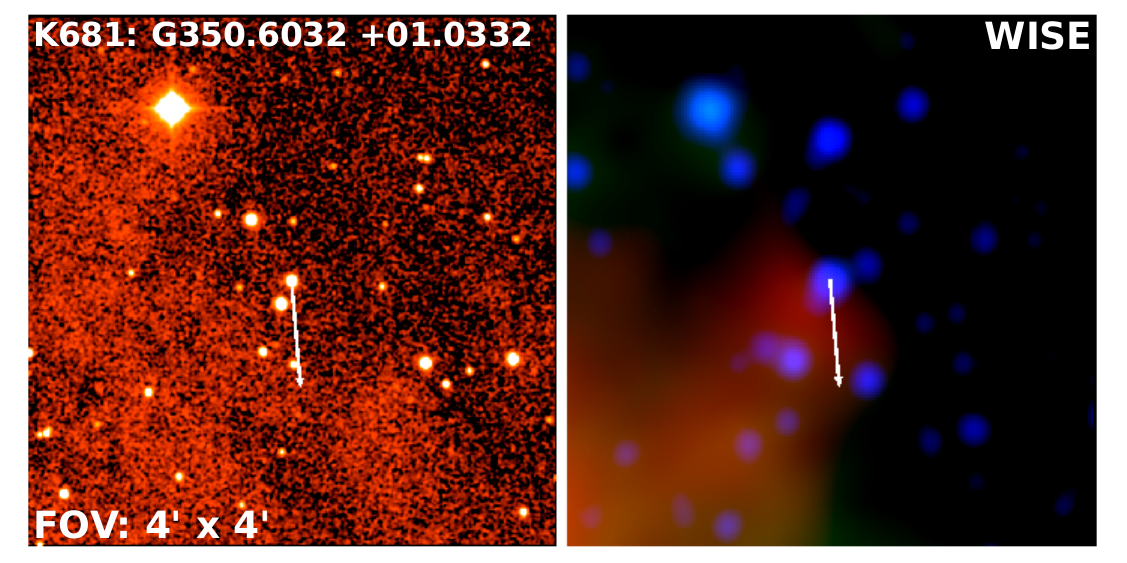}
\includegraphics[width=0.49\linewidth]{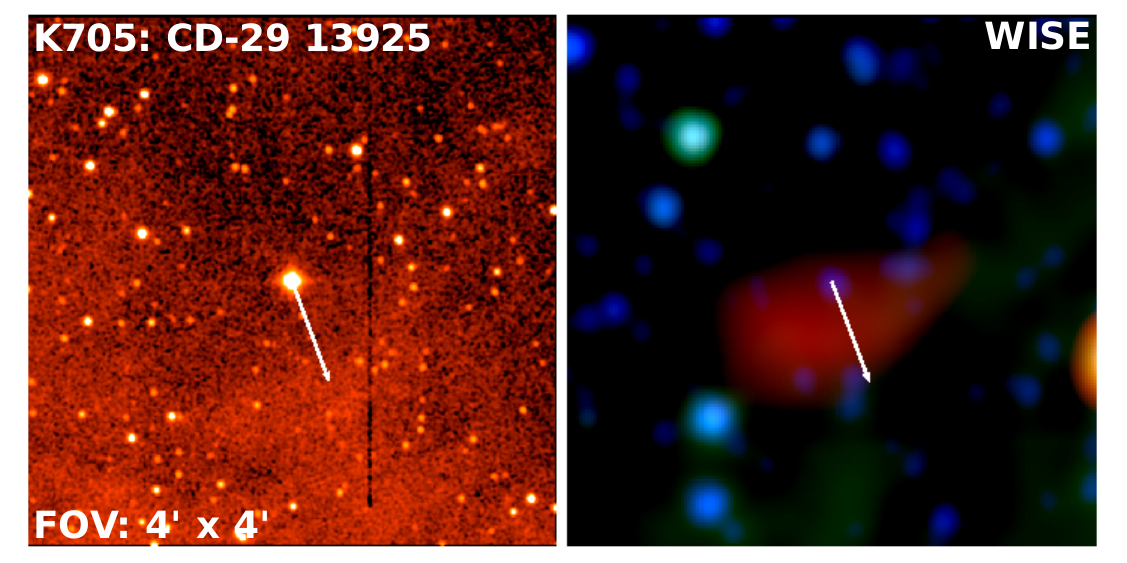}
   \caption{continued.}
\end{figure*}

\noindent\textbf{K063, K067:} Both targets lie in the M16 (Eagle Nebula) H\,{\sc ii} region. The bright, structured nebular emission prevents an unambiguous H$\alpha$ detection.\\
\textbf{K072:} The H$\alpha$ emission is dominated by the infrared dust bubble N22. As noted in Sect.~\ref{sec:discussion}, the radio emission aligns better with the H$\alpha$ nebula than with the infrared bow shock.\\
\\
The remaining targets are not associated with any prominent 
H\,{\sc ii} region. Their DK154 and Perek images reveal 
detailed nebular emission on small scales, but no arc-shaped 
structure is discernible.

\subsection{T31 Observations}
The T31 telescope offers a pixel scale of $1\farcs1$\,px$^{-1}$ (Sect.~\ref{sec:observations}) over a $55\farcm9 \times 55\farcm9$ field, with seeing typically $3\arcsec-5\arcsec$  (Table~\ref{Tab:observation_log}). The coarser resolution limits the detection of compact structures compared to DK154 and Perek, but the wide field was used to observe extended emission from southern-hemisphere targets. Details of individual objects are given below.\\
\begin{figure*}[h!]
    \centering
\includegraphics[width=0.49\linewidth]{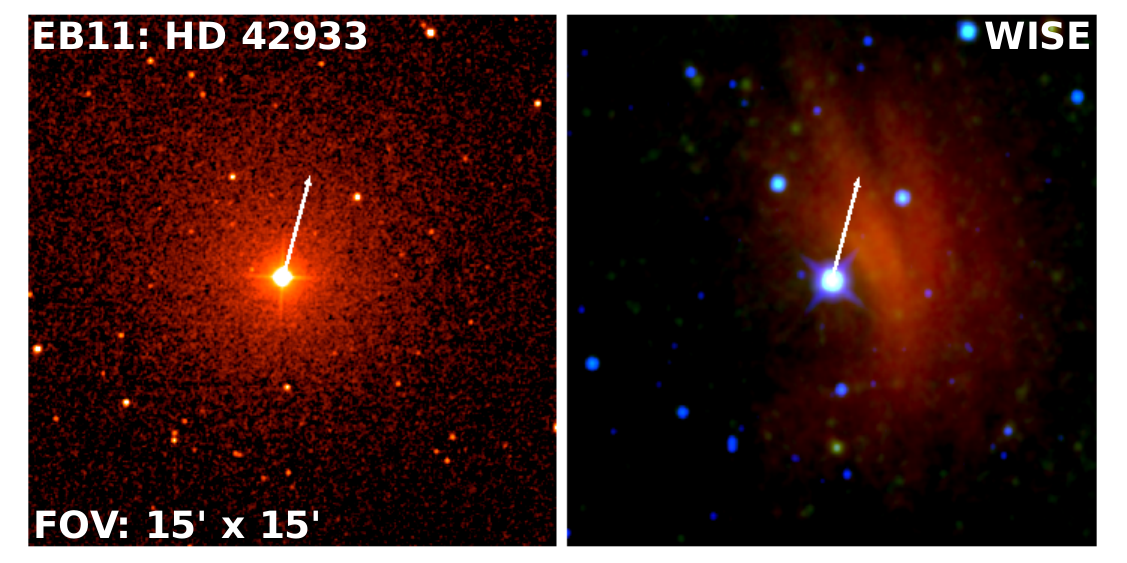} 
\includegraphics[width=0.49\linewidth]{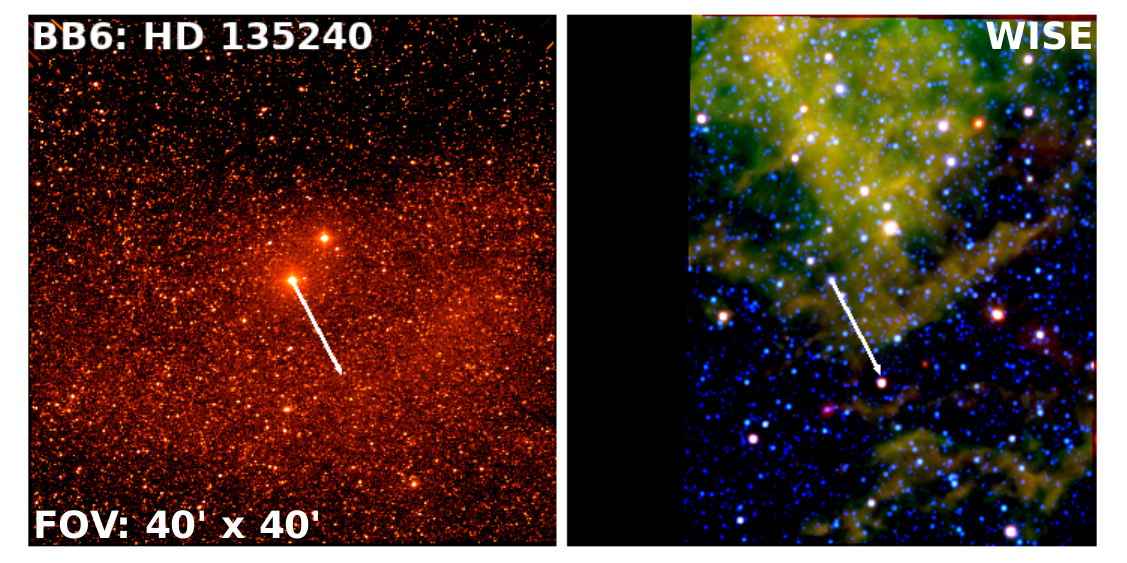} 
\caption{As Fig.\,\ref{DE:DK154andPerek}, but showing the observations with the T31 telescope.}
\label{DE:T31}
\end{figure*}
\\
\textbf{BB6:} An arc-like H$\alpha$ structure is visible in the T31 image, but no corresponding mid-infrared counterpart is detected; the bow shock was originally identified on IRAS 60\,$\mu$m images \citep{van1988}, indicating cooler ambient dust.\\
\textbf{BB8:} \citet{brown2005} reported arc-like H$\alpha$ emission on scales of $\sim1\degr \times 1\degr$ using SHASSA data. The WISE W4 band reveals a compact infrared arc within a $5\arcmin \times 5\arcmin$ field. Our deeper T31 image shows the object embedded in diffuse H$\alpha$ emission, but no arc-like structure corresponding to the IR arc is discernible.\\
\textbf{BS12:} A WN4 Wolf-Rayet star embedded within the wind-blown bubble S308 (Sh~2-308). The T31 image is dominated by diffuse H$\alpha$ emission from the bubble, and no compact arc-like structure is resolved.\\
\textbf{EB11, K101:} No published follow-up exists. No distinct arc is resolved.\\
\textbf{K150:} Embedded in bright diffuse H$\alpha$ emission. The object is a candidate dust wave / 
radiation-supported bow wave (Table.~\ref{tab:morphology}), but the surrounding background prevents an unambiguous classification.
\begin{figure*}[t!]
\ContinuedFloat
    \centering 
\includegraphics[width=0.49\linewidth]{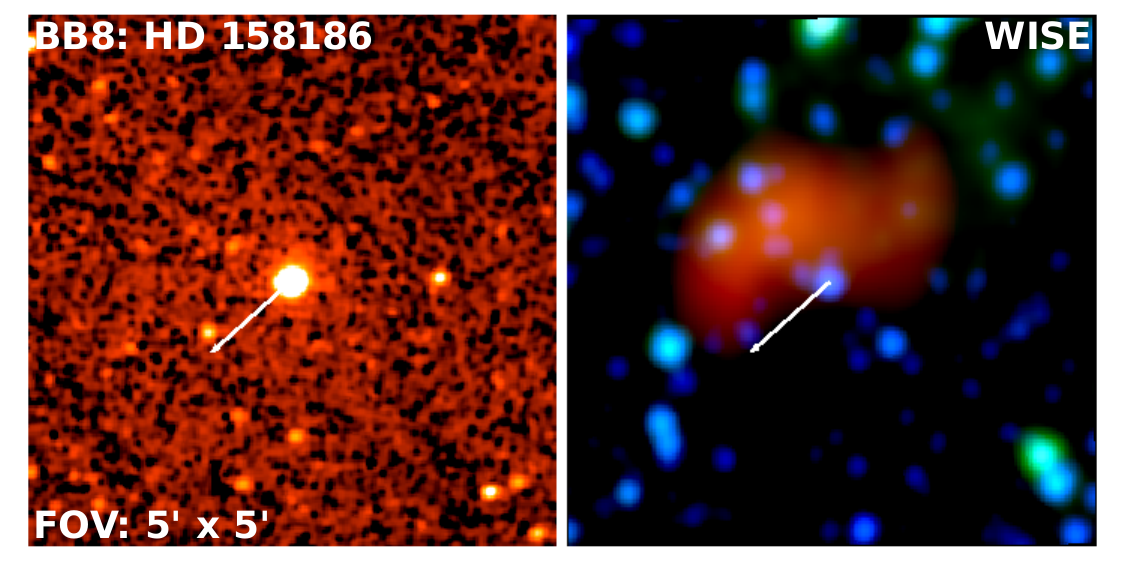} 
\includegraphics[width=0.49\linewidth]{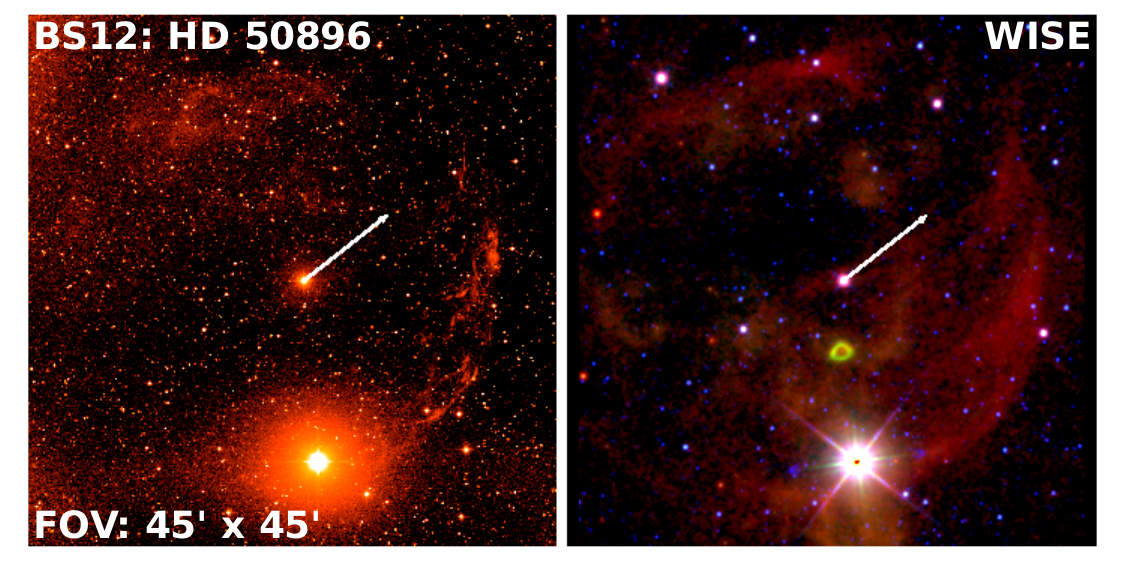} 
\includegraphics[width=0.49\linewidth]{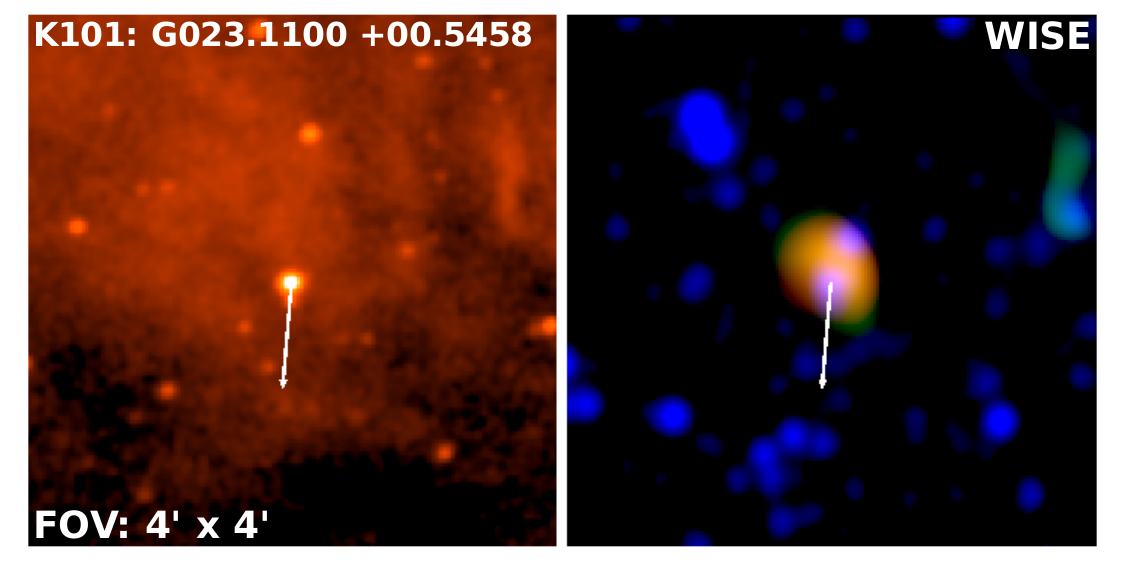}
\includegraphics[width=0.49\linewidth]{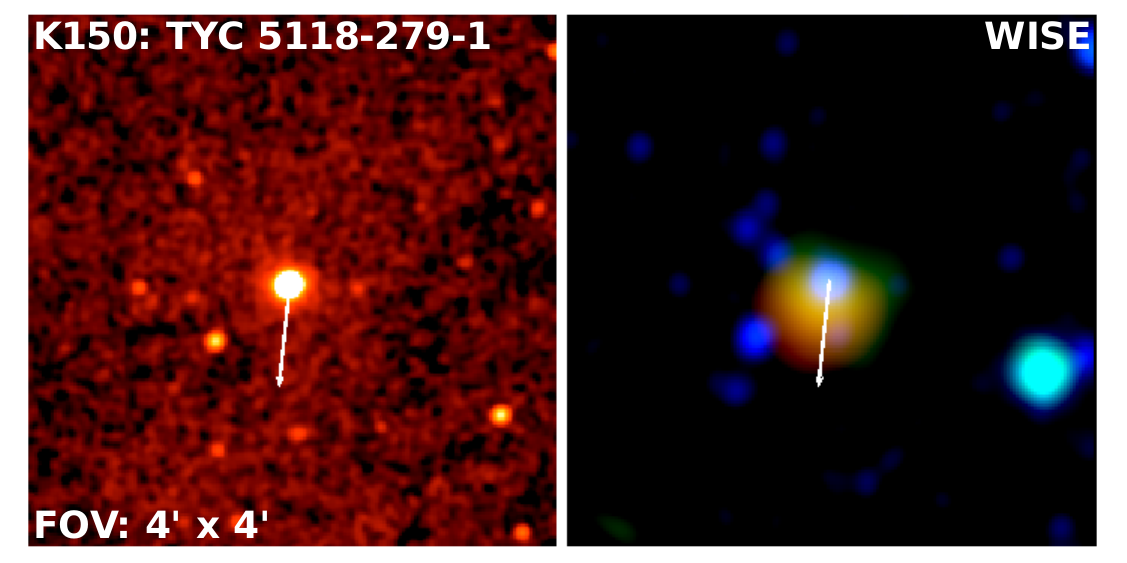}
\caption{Continued.}
\label{DE:T31}
\end{figure*}

\subsection{SBT observations}
The SBT provides a wide $3\fdg5 \times 3\fdg5$ field of view with a pixel scale of $3\farcs13$\,px$^{-1}$ (Sect.~\ref{sec:observations}), making it well suited for mapping the full extent of large-scale nebulae. The SBT was primarily used to observe extended emission from northern-hemisphere targets, as well as sources that could not be scheduled on DK154 or Perek due to observational and telescope constraints. The coarse angular resolution limits the detection of compact structures compared to the 
higher-resolution instruments, and the seeing is typically $\sim 6\arcsec-10\arcsec$ (Table~\ref{Tab:observation_log}). Details of individual targets are given below.\\

\begin{figure*}[b!]
\ContinuedFloat
\centering
\includegraphics[width=0.49\linewidth]{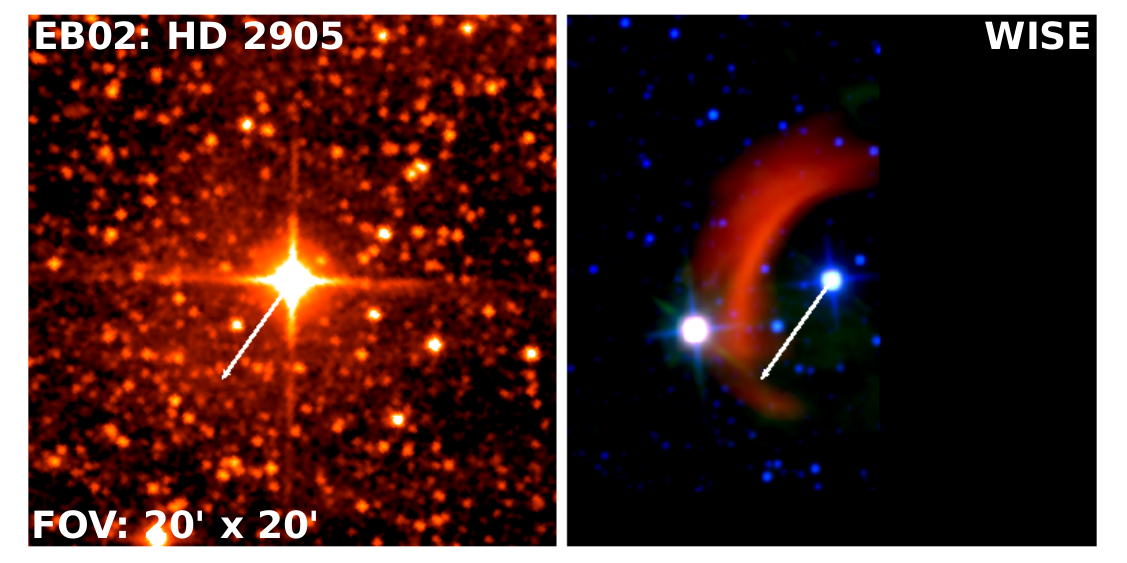}
\includegraphics[width=0.49\linewidth]{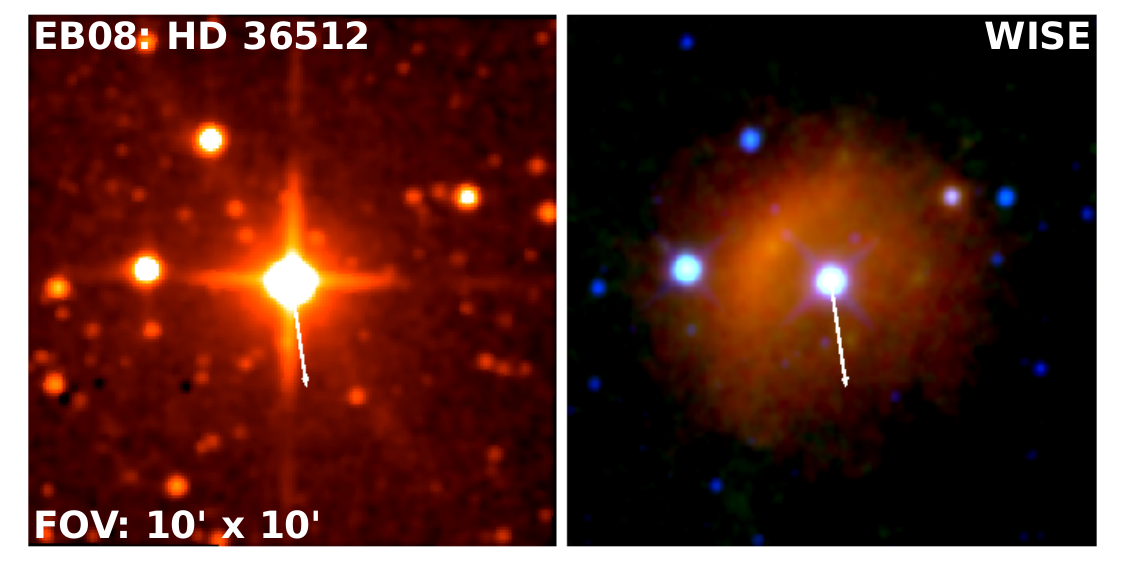}
\caption{As Fig.\,\ref{DE:DK154andPerek}, but for the observations with the SBT telescope.}
\end{figure*}

\begin{figure*}[ht!]
\ContinuedFloat
    \centering
\includegraphics[width=0.49\linewidth]{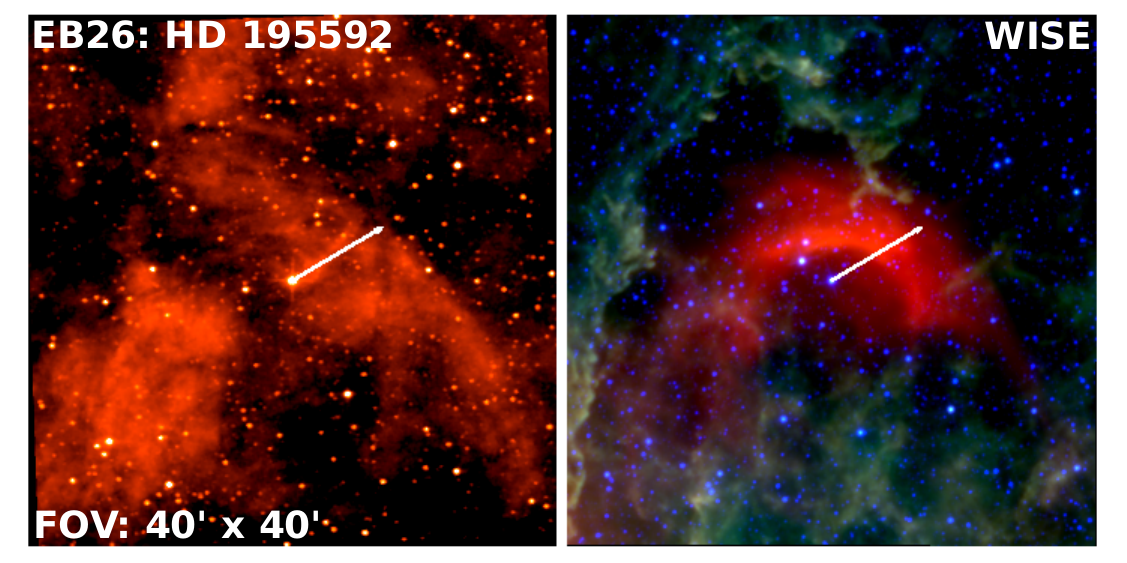}
\includegraphics[width=0.49\linewidth]{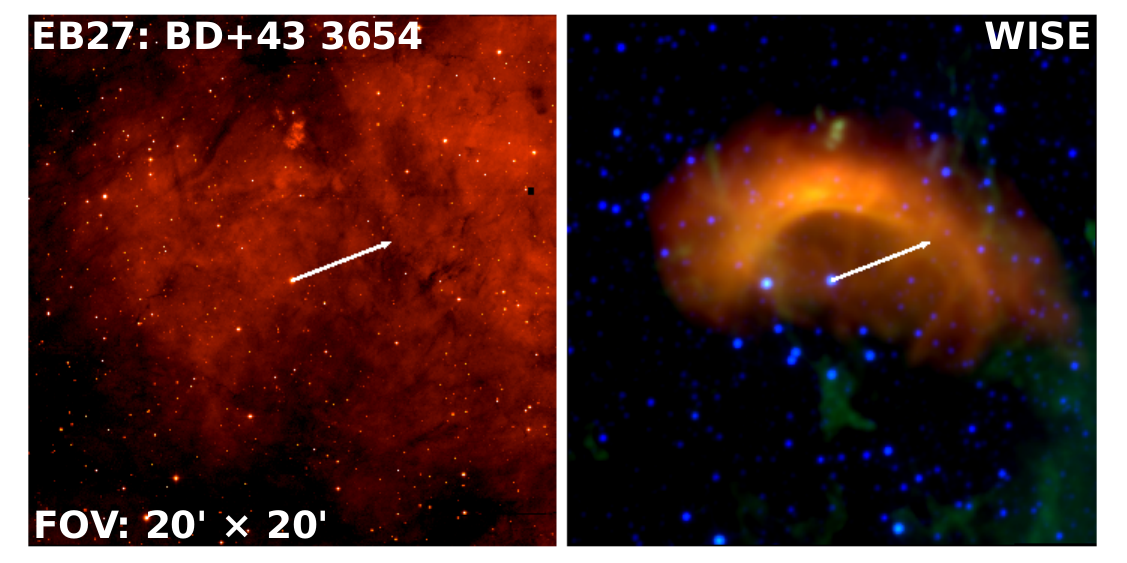} 
\includegraphics[width=0.49\linewidth]{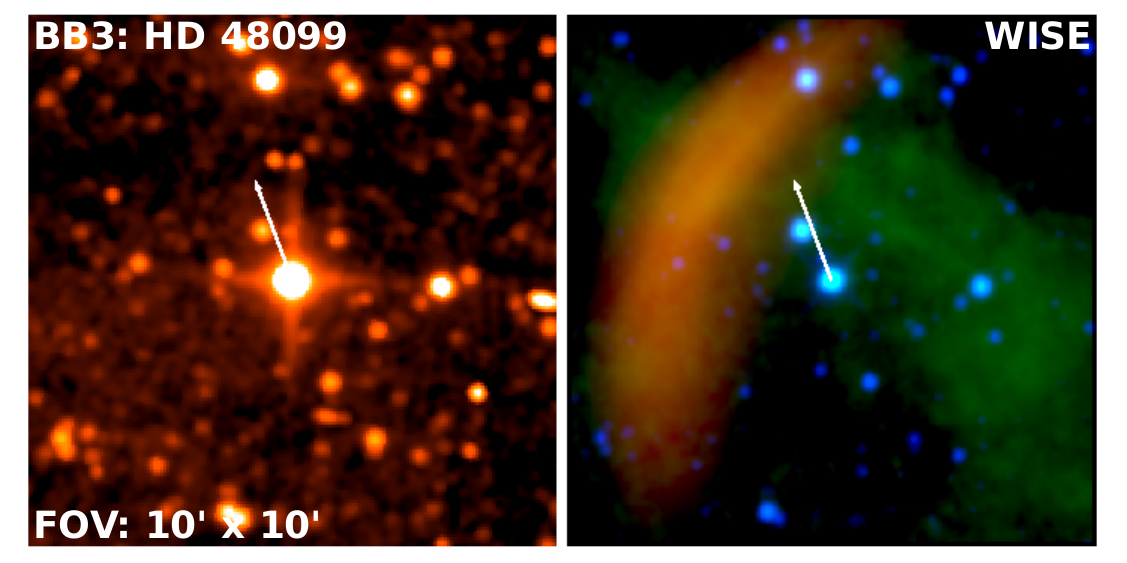}
\includegraphics[width=0.49\linewidth]{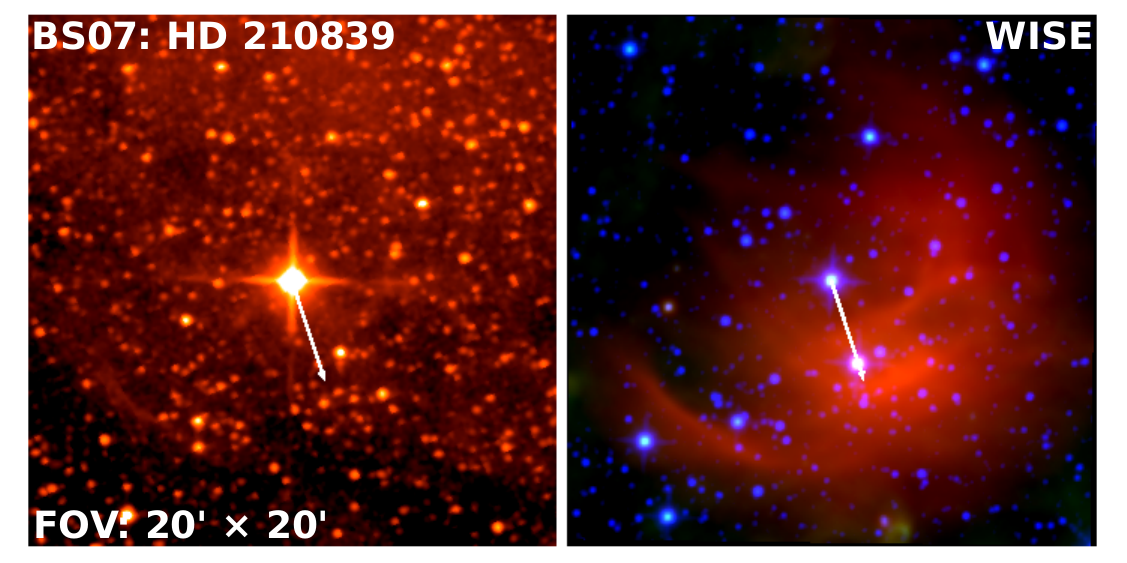}
 \caption{continued.}
\label{DE:SBT}%
\end{figure*}
\noindent\textbf{EB27:} Known synchrotron bow shock that has been studied extensively at radio wavelengths (Sect.~\ref{sec:discussion}). Embedded in a bright H\,{\sc ii} region, the SBT image is dominated by the surrounding emission, preventing an unambiguous optical identification.\\
\noindent\textbf{BS07:} A faint H$\alpha$ arc was reported from IPHAS \citep{2015A&A...576A..97S}, but despite our deep SBT exposures, no significant H$\alpha$ counterpart to the infrared bow shock is detected.\\
\\
Rest of the objects show diffuse H$\alpha$ emission in the SBT images with no distinct arc resolved.

\subsection{IPHAS Archival Images}

For EB03, EB07, EB09, and EB28, which could not be scheduled during our dedicated runs, we used archival H$\alpha$ images from the IPHAS survey to complete the sample. All four images show diffuse H$\alpha$ emission, and no distinct arc is discernible in any field. Deeper follow-up observations may be required to reveal faint bow-shock structures.
\begin{figure*}[b!]
    \centering
\includegraphics[width=0.489\linewidth]{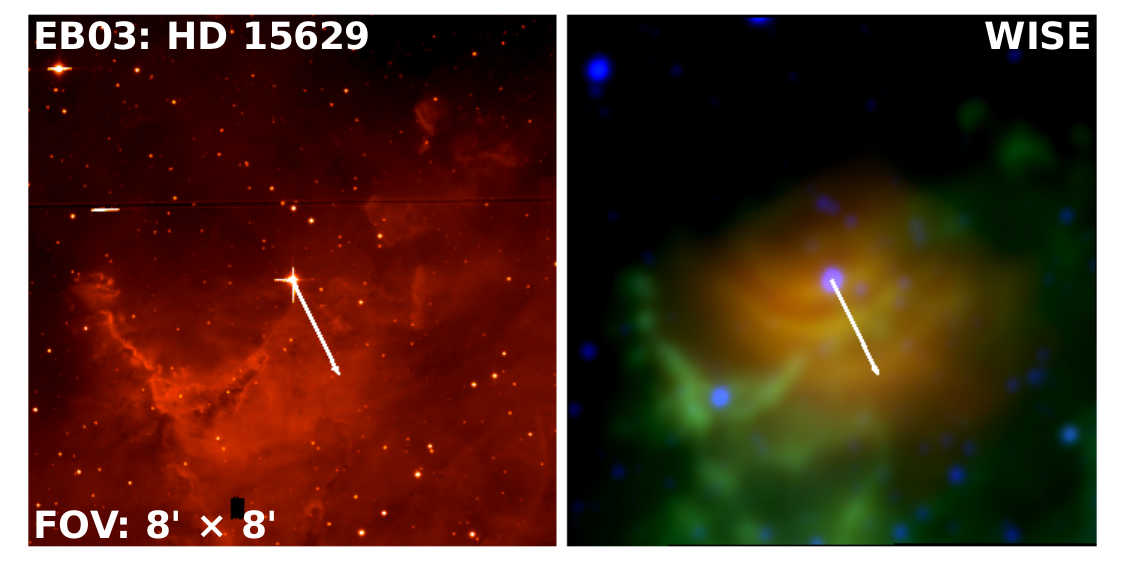} 
\includegraphics[width=0.489\linewidth]{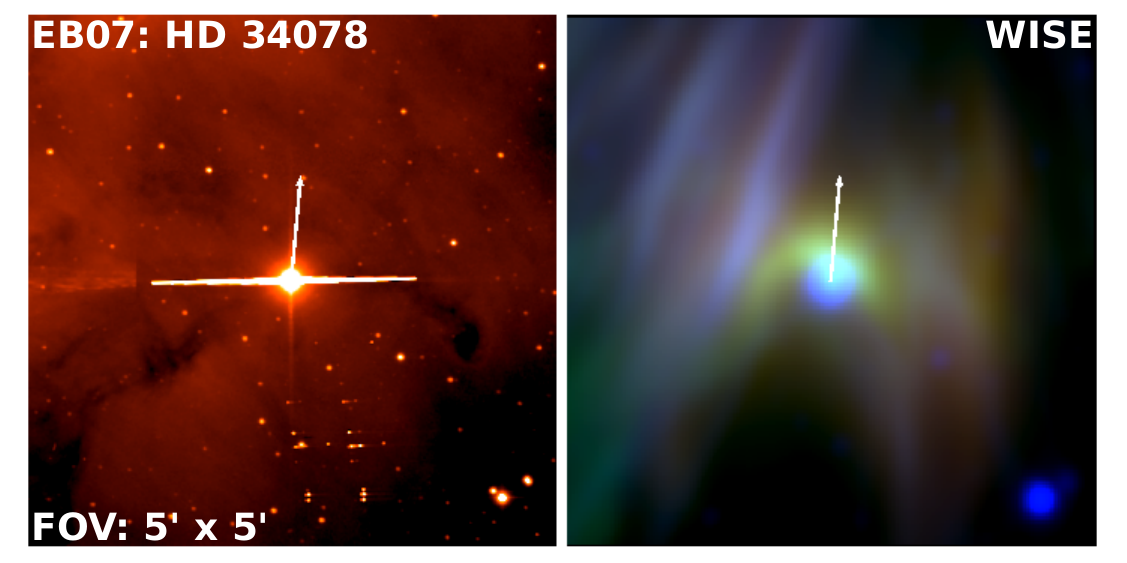} 
\includegraphics[width=0.489\linewidth]{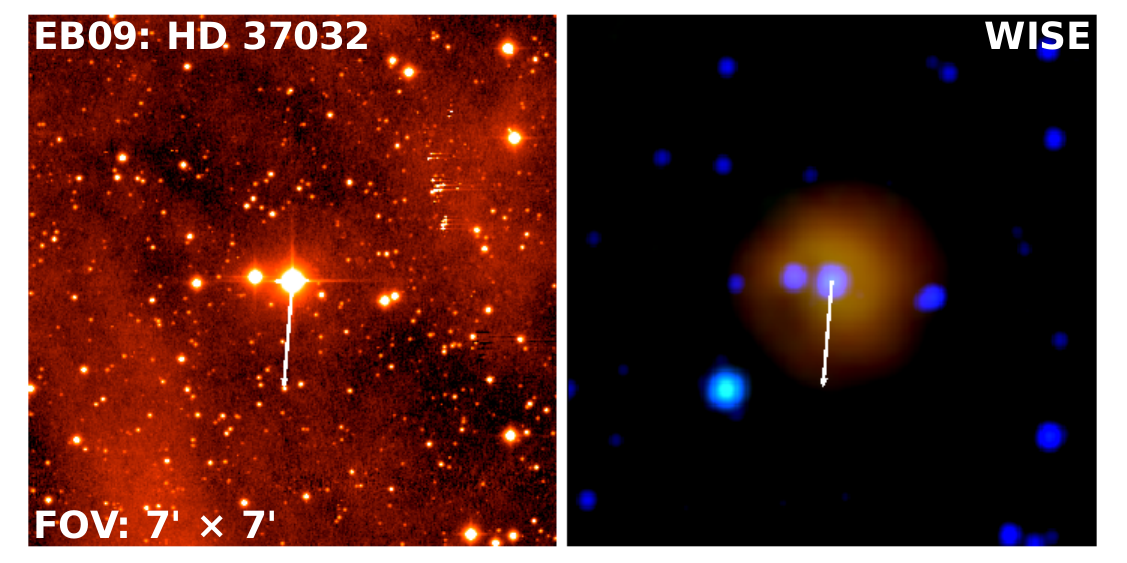} 
\includegraphics[width=0.489\linewidth]{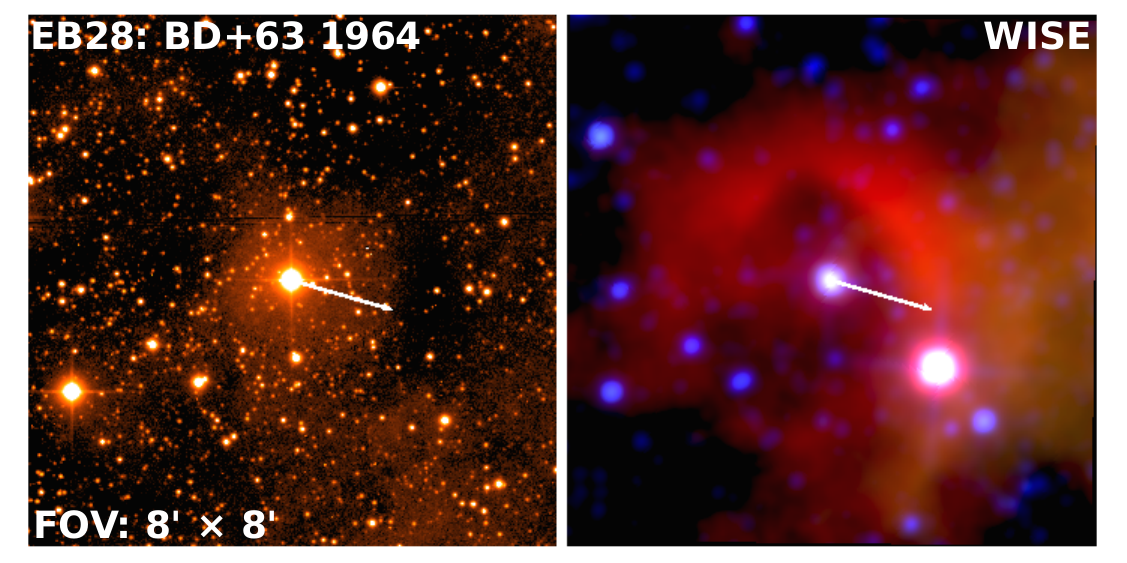} 
\caption{As Fig.\,\ref{DE:DK154andPerek}, but the H$\alpha$ images are taken from the IPHAS archive.}
\label{DE:IPHAS}
\end{figure*}

\twocolumn

\section{SHASSA-Based Flux Calibration}
\label{app:calibration}
This appendix provides a step-by-step account of the flux calibration procedure introduced in Sect.~\ref{sec:calibration}. We describe the cross-calibration of our science images against 
the absolutely calibrated SHASSA survey, the derivation of conversion factors from detector counts to H$\alpha$ surface brightness, and the estimation of the associated uncertainties.

The science images are recorded in detector counts (ADU), whereas SHASSA provides absolutely calibrated H$\alpha$ intensities in deci-Rayleighs. Prior to calibration, isolated artifacts in the science images were suppressed to prevent them from biasing the flux comparison. Extreme pixel values were identified using a robust dispersion estimate derived from the interquartile range,
\begin{equation}
  \sigma_{\mathrm{IQR}} = \frac{Q_{3} - Q_{1}}{1.349}, 
\end{equation}
where $Q_1$ and $Q_3$ denote the 25th and 75th percentiles of the pixel-value distribution. Pixels lying more than $10\,\sigma_{\mathrm{IQR}}$ from the image median were replaced by clipped values, removing residual cosmic rays, hot pixels, and other non-physical outliers while preserving the diffuse nebular emission. Invalid pixels in the SHASSA maps, identified through the survey quality flags, were masked and excluded from all subsequent analysis.

The SHASSA survey has an effective angular resolution of $\sim96''$ and a pixel size of $\sim48''$, substantially coarser than the DK154 ($0.396''$ pixel$^{-1}$), T31 ($1.1''$ pixel$^{-1}$), and SBT ($3.13''$ pixel$^{-1}$) observations. To enable a meaningful flux comparison, the science images were degraded to the SHASSA resolution. Since direct convolution at the native image scale would be computationally expensive, each science image was first rebinned to a pixel scale of $\sim20''$--$25''$ using flux-conserving block summation. The corresponding rebinning factors were approximately 60 for DK154, 8 for T31, and 6 for SBT. The rebinned images were then convolved with a Gaussian kernel whose width was determined from
\begin{equation}
 \sigma_{\mathrm{kernel}} = \sqrt{\sigma_{\mathrm{SHASSA}}^{2} - \sigma_{\mathrm{sci}}^{2}},   
\end{equation}
where $\sigma=\mathrm{FWHM}/2.355$ and $\sigma_{\mathrm{sci}}$ denotes the native image quality of the science frame and FWHM is the full-width at half-maximum of the Gaussian.

The SHASSA image was reprojected onto the WCS of the science image using bilinear interpolation and converted from deci-Rayleighs to Rayleighs. Both datasets were subsequently placed on the same coarse grid, with the reprojected SHASSA image rebinned using the same block size as the science image. This procedure ensured identical spatial sampling and allowed a direct pixel-by-pixel comparison of the two datasets over common sky areas.

A validity mask was then constructed to select reliable calibration pixels. Only pixels with finite positive values in both images were retained. Additional constraints were applied to exclude detector edges, vignetted regions, saturated pixels, and areas outside a circular region centred on the bow shock. The radius of this region was adjusted for each target to provide a representative sample of the local diffuse H$\alpha$ background while avoiding image artefacts.

The calibration factor $C$ (Rayleigh\,ADU$^{-1}$) was obtained from a linear least-squares fit of the SHASSA surface brightnesses ($y$, in Rayleighs) against the corresponding science-image counts ($x$, in ADU). Because zero detector counts should correspond to zero physical flux, the fit was constrained to pass through the origin, $y = C\,x$, yielding
\begin{equation}
    C = \frac{\sum x_i y_i}{\sum x_i^2}.
\end{equation}

An initial fit was performed using all valid pixels. Residuals about this relation were then examined and pixels deviating by more than $3\sigma$ were rejected. The fit was repeated on the clipped sample to obtain the final calibration factor. The formal least-squares uncertainty depends primarily on the number of fitted pixels and may underestimate the true error when the dominant scatter arises from real astrophysical differences between the SHASSA and science images. We therefore adopted a more conservative uncertainty based on the fractional RMS scatter of the residuals,
\begin{equation}
    \sigma_{C} =C\,\frac{\sqrt{\frac{1}{N}\sum_{i=1}^{N} r_i^2}}{\langle y \rangle},
\end{equation}
which captures the full pixel-to-pixel dispersion between the two datasets. The resulting calibration factors and their uncertainties are listed in Table~\ref{tab:calibration}.

\begin{table}[ht!]
\caption{SHASSA-derived calibration factors converting detector counts (ADU) into H$\alpha$ surface brightness (Rayleighs). Relative uncertainties are estimated from the RMS scatter of the fit residuals.}
\label{tab:calibration}
\centering
\footnotesize
\begin{tabular}{lcc}
\hline \hline
ID & $C$ (Rayleigh / ADU) & $\sigma_{C}$ (\%) \\
\hline
EB12 & 53.92 $\pm$ 24.23 & 44.9 \\
EB15  & 0.41 $\pm$ 0.03 & 7.4 \\
EB21  & 9.18 $\pm$ 1.91 & 20.8 \\
EB23  & 0.29 $\pm$ 0.002 & 0.6 \\
BB4   & 1.55 $\pm$ 0.80 & 51.7 \\
BB5   & 0.86 $\pm$ 0.08 & 8.8 \\
K065  & 0.36 $\pm$ 0.08  & 23.2 \\
K383  & 0.34 $\pm$ 0.06  & 18.1 \\
K385  & 0.25 $\pm$ 0.03  & 12.0 \\
K692 &  0.25 $\pm$ 0.05 & 18.0 \\
\hline
\end{tabular}
\end{table}

For each bow-shock detection, the calibrated H$\alpha$ surface brightness was measured using circular aperture photometry. Aperture radii were chosen to fit entirely within the nebular emission and ranged from $5''$ to $40''$ depending on the angular size of the bow shock. For a given image, the aperture size was kept constant. Multiple apertures were placed visually across the bow-shock emitting region to sample the nebula, and a number of apertures were placed in adjacent blank-sky regions or in the surrounding H\,{\sc ii} emission to characterise the local background.

Within each aperture, the robust dispersion $\sigma_{\rm IQR}$ was computed from the interquartile range and pixel values exceeding $3\sigma_{\rm IQR}$ from the aperture median were clipped to exclude stars, cosmic rays, and hot pixels. The mean of the clipped pixels was then taken as the aperture value. The net flux $F_{\rm net}$ was obtained by subtracting the mean sky level $F_{\rm sky}$ from the mean nebular level $F_{\rm apex}$, and the corresponding measurement uncertainty was estimated from the standard deviation of the aperture values,
\begin{equation}
    F_{\rm net} = F_{\rm apex} - F_{\rm sky} \quad \mathrm{and} \quad
\sigma_{\rm net}^2 = \sigma_{\rm meas}^2 + \sigma_{\rm sky}^2,
\end{equation}
where $\sigma_{\rm meas}$ is the scatter among the nebular apertures and $\sigma_{\rm sky}$ is the scatter among the sky apertures. The net flux was converted to Rayleighs using the calibration factor $C$ from Table~\ref{tab:calibration}, $\Sigma = C \times F_{\rm net}$, and the total uncertainty was computed by combining the calibration and measurement errors in quadrature,
\begin{equation}
    \sigma_\Sigma^2 = F_{\rm net}^2 \cdot \sigma_C^2 + C^2 \cdot 
\left(\sigma_{\rm meas}^2 + \sigma_{\rm sky}^2\right).
\end{equation}

\end{appendix}

\end{document}